\documentclass[12pt]{article}
\usepackage{amsmath,amsthm}
\usepackage{eucal}
\usepackage{eufrak}

\newcommand{\ar}{\renewcommand{\arraystretch}{1}} 
\DeclareMathAlphabet{\bb}{U}{msb}{m}{n}
\gdef\C{\bb C}

\gdef\dS{\bb S}
\gdef\R{\bb R}
\gdef\K{\bb K}

\DeclareMathOperator{\spin}{{\bf Spin}}

\DeclareMathOperator{\fD}{\mathfrak{D}}

\DeclareMathOperator{\Sym}{Sym}
\DeclareMathOperator{\arc}{arccos}

\DeclareMathOperator{\sign}{sign}

\newcommand{\s}{\!}

\newcommand{\iddots}{.\hspace{0.5mm}\raisebox{1mm}{.}\hspace{0.5mm}
\raisebox{2mm}{.}}

\newcommand{\re}{\mbox{\rm Re}\,}
\newcommand{\im}{\mbox{\rm Im}\,}

\newcommand{\cA}{\mathcal{A}}

\newcommand{\sA}{{\sf A}}
\newcommand{\sB}{{\sf B}}
\newcommand{\sI}{{\sf I}}

\newcommand{\sX}{{\sf X}}
\newcommand{\sY}{{\sf Y}}

\newcommand{\bx}{{\bf x}}
\newcommand{\by}{{\bf y}}
\newcommand{\bz}{{\bf z}}

\newcommand{\fM}{\mathfrak{M}}
\newcommand{\fG}{\mathfrak{G}}
\newcommand{\fC}{\mathfrak{C}}

\newcommand{\fP}{\mathfrak{P}}

\newcommand{\fT}{\mathfrak{M}}
\newcommand{\fS}{\mathfrak{S}}

\newcommand{\fg}{\mathfrak{g}}
\newcommand{\fz}{\mathfrak{z}}

\newcommand{\cl}{C\kern -0.2em \ell}

\newcommand{\e}{\mbox{\bf e}}

\newcommand{\hypergeom}[5]{\mbox{$
_#1 F_#2\left.
\!\!
\left(
\!\!\!\!
\begin{array}{c}
\multicolumn{1}{c}{\begin{array}{c}
#3
\end{array}}\\[1mm]
\multicolumn{1}{c}{\begin{array}{c}
#4
\end{array}}\end{array}
\!\!\!\!
\right|\displaystyle{#5}\right)
$}
}
\newcommand{\ld}{\left[}
\newcommand{\rd}{\right]}
\newcommand{\lf}{\left\{}
\newcommand{\rf}{\right\}}
\newcommand{\tg}{\tan}
\newcommand{\ch}{\cosh}
\newcommand{\sh}{\sinh}
\newcommand{\tnh}{\tanh}
\newcommand{\cth}{\coth}
\newcommand{\ctg}{\cot}
\newtheorem{theorem}{Theorem}

\begin{document}
\title{Hyperspherical Functions and Harmonic Analysis on the Lorentz Group}
\author{V.~V. Varlamov\\
{\small Department of Mathematics, Siberia State University of Industry,}\\
{\small Kirova 42, Novokuznetsk 654007, Russia.}}
\date{}
\maketitle
\begin{abstract}
Matrix elements of spinor and principal series representations of the
Lorentz group are studied in the basis of complex angular momentum
(helicity basis). It is shown that matrix elements are expressed via
hyperspherical functions (relativistic spherical functions). In essence,
the hyperspherical functions present itself a four-dimensional
(with respect to a pseudo-euclidean metrics of Minkowski spacetime)
generalization of the usual three-dimensional spherical functions.
An explicit form of the hyperspherical functions is given. The hypespherical
functions of the spinor representations are represented by a product of
generalized spherical functions and Jacobi functions. It is shown that
zonal hyperspherical functions are expressed via the Appell functions.
The associated hyperspherical functions are defined as the functions
on a two-dimensional complex sphere. Integral representations, addition
theorems, symmetry and recurrence relations for hyperspherical functions
are given. In case of the principal and supplementary series representations
of the Lorentz group, the matrix elements are expressed via the
functions represented by a product of spherical and conical functions.
The hyperspherical functions of the principal series representations
allow one to apply methods of harmonic analysis on the Lorentz group.
Different forms of expansions of square integrable functions on the
Lorentz group are studied. By way of example, an expansion of the wave
function, representing the Dirac field $(1/2,0)\oplus(0,1/2)$, is given.
\end{abstract}
\section{Introduction}
Numerous applications of three-dimensional spherical functions in different
areas of mathematical and theoretical physics are well known.
In XX century, under the influence of relativity theory, which represents
the world as some four-dimensional pseudo-euclidean manifold, it is
appearred a necessity to generalize Laplace three-dimensional spherical
functions on the four-dimensional case. Four-dimensional spherical functions
of an euclidean space was first studied by Fock \cite{Foc35} for the
solution of the hydrogen atom problem in momentum representation.
They have the form
\begin{equation}\label{Fock}
\psi_{nlm}(\alpha,\theta,\varphi)=
\frac{i^l\sin^l\alpha}{\sqrt{n^2(n^2-1^2)\cdots(n^2-l^2)}}
\frac{d^{l+1}\cos n\alpha}{d(\cos\alpha)^{l+1}}Y_{lm}(\theta,\varphi),
\end{equation}
where $\alpha$, $\theta$, $\varphi$ are angles of the four-dimensional
radius-vector in the euclidean space, $n=0,1,2,\ldots$. These functions
are eigenfunctions of a square of the four-dimensional angular momentum
operator
$L^2$ ($L^2$ is an angular part, in the sense of $\alpha$, $\theta$, $\varphi$, 
of the Laplace operator).

In 1956, Dolginov \cite{Dol56} (see also \cite{DT59,DM59,Esk59}) considered
an analytic continuation of the Fock four-dimensional spherical functions
(\ref{Fock}):
\begin{equation}\label{Dolginov}
\psi_{Nlm}(\alpha,\theta,\varphi)=
\frac{\sh^l\alpha}{\sqrt{N^2(N^2-1^2)\cdots(N^2-l^2)}}
\frac{d^{l+1}\ch N\alpha}{d\ch^{l+1}\alpha}Y_{lm}(\theta,\varphi),
\end{equation}
where $0\leq\alpha\leq\infty$, $N=0,1,2,\ldots$. In turn, the functions
(\ref{Dolginov}), called in the works \cite{Dol56,DT59,DM59} as
{\it relativistic spherical functions}, depend on angles of the radius-vector
in the four-dimensional spacetime. At this point, if we replace in
(\ref{Dolginov}) $N$ by $in$, where $n$ is a real number and
$0\leq n\leq\infty$, then we obtain basis functions of an irreducible
unitary infinite-dimensional representation of the Lorentz group.

In 1952, Gel'fand and Z. Shapiro \cite{GS52} (see also \cite{GMS}) showed
that matrix elements of three-dimensional rotation group are expressed via
spherical functions (a relationship between special functions and group
representations was first discovered by E. Cartan in 1929 \cite{Car29}).
The group theoretical interpretation of the theory of special functions was
intensively studied by Vilenkin and Klimyk \cite{Vil68,VK}. 
At present, it is widely accepted that the theory of special functions is
a `functional aspect' of the group representation theory.

The relationship between matrix elements of the three-dimensional rotation
group and spherical functions prompted a new way to definition of the
four-dimensional analog of the spherical functions. Namely, these functions
should be defined in terms of matrix elements of the Lorentz group.
In sixties it is appearred series of the works devoted to this topics
\cite{Esk61,Gol61,Str65,Str67,ST67,VD67,Kol70}.
The works of Str\"{o}m \cite{Str65,Str67} have been obtained the most influence.
Str\"{o}m used the fact, previously established by Naimark
\cite{Nai58}, that any Lorentz matrix can be represented in the form
\begin{equation}\label{MLorentz}
\fg=u_1(\eta,\psi,0)\fg(a)u_2(\varphi_1,\theta,\varphi_2),
\end{equation}
where $0\leq\eta<4\pi$, $0\leq\varphi_1,\varphi_2<2\pi$,
$0\leq\psi,\theta\leq\pi$, $0\leq a<\infty$, and  $u_1$, $u_2$ belong to the
group $SU(2)$ (an universal covering of the three-dimensional rotation group),
$\fg(a)$ is an element of the one-parameter noncompact subgroup of
$SL(2,\C)$. The decomposition (\ref{MLorentz}) induces a following form
for the matrix elements of the Lorentz group:
\[
D^{\nu\rho}_{JM,J^\prime M^\prime}(\fg)=
\sum^{\min(J,J^\prime)}_{\lambda=-\min(J,J^\prime)}
D^J_{M\lambda}(\eta,\psi,0)
D^{\nu\rho}_{JJ^\prime\lambda}(a)
D^{J^\prime}_{\lambda M^\prime}(\varphi_1,\theta,\varphi_2),
\]
where $D^J_{M\lambda}(\eta,\psi,0)$, 
$D^{J^\prime}_{\lambda M^\prime}(\varphi_1,\theta,\varphi_2)$ are matrix
elements of $SU(2)$, and the functions
$D^{\nu\rho}_{JJ^\prime\lambda}(a)$ of the principal series, calculated by
Str\"{o}m in the Gel'fand-Naimark basis, have the form
\begin{multline}
D^{\nu\rho}_{JJ^\prime\lambda}(a)=
\sum^{\min(J-\nu,J-\lambda)}_{d=\max(0,-\nu-\lambda)}
\sum^{\min(J^\prime-\nu,J^\prime-\lambda)}_{d^\prime=\max(0,-\nu-\lambda)}
B^{\nu\rho}_{JJ^\prime\lambda dd^\prime}
e^{-a(2d^\prime+\lambda+\nu+1-\frac{1}{2}i\rho)}\times\\
\times\hypergeom{2}{1}{1+J^\prime-\frac{1}{2}i\rho,\nu+\lambda+d+d^\prime+1}
{J+J^\prime+2}{1-e^{-2a}},\nonumber
\end{multline}
where
\begin{multline}
B^{\nu\rho}_{JJ^\prime\lambda dd^\prime}=\alpha^{\nu\rho^\ast}_J
\alpha^{\nu\rho}_{J^\prime}(-1)^{J+J^\prime-2\lambda+d+d^\prime}
\frac{1}{(J+J^\prime+1)!}\times\\
\ld(2J+1)(J-\nu)!(J+\nu)!(J-\lambda)!(J+\lambda)!(2J^\prime+1)
(J^\prime-\nu)!(J^\prime+\nu)!(J^\prime-\lambda)!
(J^\prime+\lambda)!\rd^{\frac{1}{2}}\times\\
\frac{(\nu+\lambda+d+d^\prime)!(J+J^\prime-\nu-\lambda-d-d^\prime)!}
{d!(J-\lambda-d)!(J-\nu-d)!(\lambda+\nu+d)!d^\prime!
(J^\prime-\lambda-d^\prime)!(J^\prime-\nu-d^\prime)!
(\lambda+\nu+d^\prime)!}\nonumber
\end{multline}
and
\[
\alpha^{\nu\rho}_J=\prod^J_{s=|\nu|}
\frac{-2s+i\rho}{(4s^2+\rho^2)^{\frac{1}{2}}}=
\frac{\Gamma(-|\nu|+\frac{1}{2}i\rho+1)\Gamma(-J+\frac{1}{2}i\rho)}
{\Gamma(-|\nu|+\frac{1}{2}i\rho+1)\Gamma(-J+\frac{1}{2}i\rho)}.
\]

Matrix elements in the form proposed by Str\"{o}m and also in other forms
\cite{ST67,VD67}, which used the decomposition (\ref{MLorentz}),
are very complicate. An expansion problem of relativistic amplitudes
requires the most simple form of the matrix elements for irreducible
representations of the Lorentz group. Moreover, for the matrix elements in
the Str\"{o}m form or in the Sciarrino-Toller form and etc., the question
about four-dimensional generalization of the Laplace spherical functions
remains unclear. It should be noted that Dolginov-Toptygin relativistic
spherical functions present the most degenerate form 
of the matrix elements of the Lorentz group. Smorodinsky and Huszar
\cite{Hus70,HS70,SH70,Hus85,Hus88} found more simple method for
definition (except the decomposition (\ref{MLorentz}))
of the matrix elements of the Lorentz group by means of a
complexification of the three-dimensional rotation group and solution of the
equation on eigenvalues of the Casimir operators
(see also \cite{AG64}). In general case, obtained functions
(later on called by Huszar \cite{Hus85,Hus88} as {\it spherical functions
of the Lorentz group}) are products of the two hypergeometric functions.
Matrix elements of spinor representations of the Lorentz group have been
obtained by the author \cite{Var022} via the complexification of a
fundamental representation of the group $SU(2)$. The main advantage of such a
definition is the most simple form of matrix elements expressed via a
{\it hyperspherical functions}, which is a product of the generalized
spherical functions
$P^l_{mn}$ and the Jacobi function $\fP^l_{mn}$. As is known \cite{GS52},
the matrix elements of $SU(2)$ are defined by the functions
$P^l_{mn}$, and matrix elements of the group $QU(2)$ of quasiunitary
matrices of the second order, which is isomorphic to the group
$SL(2,\R)$\footnote{Other designation of this group is
$SU(1,1)$ known also as three-dimensional Lorentz group, representations of
which was studied by Bargmann \cite{Bar47}.}, are expressed via the
functions $\fP^l_{mn}$. The hyperspherical functions present the
four-dimensional analog of the three-dimensional spherical 
functions\footnote{The hyperspherical functions (or hyperspherical
harmonics) are known in mathematics for a long time 
(see, for example, \cite{Bat2}). These functions are generalizations
of the three-dimensional spherical functions on the case of $n$-dimensional
euclidean spaces. For that reason we retain this name (hyperspherical
functions) for the case of pseudo-euclidean spaces.}. In essence, these
functions present itself a new class of special functions related to the
class of hypergeometric functions.

\section{Helicity Basis}\index{basis!helicity}
Let $\fg\rightarrow T_{\fg}$ be an arbitrary linear representation of the
proper orthochronous Lorentz group $\fG_+$ and let
$\sA_i(t)=T_{a_i(t)}$ be an 
infinitesimal operator\index{operator!infinitesimal} corresponding the rotation
$a_i(t)\in\fG_+$. Analogously, we have $\sB_i(t)=T_{b_i(t)}$, where
$b_i(t)\in\fG_+$ is a hyperbolic rotation. The operators $\sA_i$ and
$\sB_i$ satisfy the following commutation 
relations:
\begin{equation}\label{Com1}
\left.\begin{array}{lll}
\ld\sA_1,\sA_2\rd=\sA_3, & \ld\sA_2,\sA_3\rd=\sA_1, &
\ld\sA_3,\sA_1\rd=\sA_2,\\[0.1cm]
\ld\sB_1,\sB_2\rd=-\sA_3, & \ld\sB_2,\sB_3\rd=-\sA_1, &
\ld\sB_3,\sB_1\rd=-\sA_2,\\[0.1cm]
\ld\sA_1,\sB_1\rd=0, & \ld\sA_2,\sB_2\rd=0, &
\ld\sA_3,\sB_3\rd=0,\\[0.1cm]
\ld\sA_1,\sB_2\rd=\sB_3, & \ld\sA_1,\sB_3\rd=-\sB_2, & \\[0.1cm]
\ld\sA_2,\sB_3\rd=\sB_1, & \ld\sA_2,\sB_1\rd=-\sB_3, & \\[0.1cm]
\ld\sA_3,\sB_1\rd=\sB_2, & \ld\sA_3,\sB_2\rd=-\sB_1. &
\end{array}\right\}
\end{equation}
Denoting $\sI^{23}=\sA_1$, $\sI^{31}=\sA_2$,
$\sI^{12}=\sA_3$, and $\sI^{01}=\sB_1$, $\sI^{02}=\sB_2$, $\sI^{03}=\sB_3$,
we can write the relations (\ref{Com1}) in a more compact form:
\[
\ld\sI^{\mu\nu},\sI^{\lambda\rho}\rd=\delta_{\mu\rho}\sI^{\lambda\nu}+
\delta_{\nu\lambda}\sI^{\mu\rho}-\delta_{\nu\rho}\sI^{\mu\lambda}-
\delta_{\mu\lambda}\sI^{\nu\rho}.
\]
Let us consider the operators
\begin{gather}
\sX_l=\frac{1}{2}i(\sA_l+i\sB_l),\quad\sY_l=\frac{1}{2}i(\sA_l-i\sB_l),
\label{SL25}\\
(l=1,2,3).\nonumber
\end{gather}
Using the relations (\ref{Com1}), we find that
\begin{equation}\label{Com2}
\ld\sX_k,\sX_l\rd=i\varepsilon_{klm}\sX_m,\quad
\ld\sY_l,\sY_m\rd=i\varepsilon_{lmn}\sY_n,\quad
\ld\sX_l,\sY_m\rd=0.
\end{equation}
Further, introducing generators of the form
\begin{equation}\label{SL26}
\left.\begin{array}{cc}
\sX_+=\sX_1+i\sX_2, & \sX_-=\sX_1-i\sX_2,\\[0.1cm]
\sY_+=\sY_1+i\sY_2, & \sY_-=\sY_1-i\sY_2,
\end{array}\right\}
\end{equation}
we see that in virtue of commutativity of the relations (\ref{Com2}) a
space of an irreducible finite--dimensional representation of the group
$\fG_+$ can be spanned on the totality of $(2l+1)(2\dot{l}+1)$ basis
vectors $\mid l,m;\dot{l},\dot{m}\rangle$, where $l,m,\dot{l},\dot{m}$
are integer or half--integer numbers, $-l\leq m\leq l$,
$-\dot{l}\leq \dot{m}\leq \dot{l}$. Therefore,
\begin{eqnarray}
&&\sX_-\mid l,m;\dot{l},\dot{m}\rangle=
\sqrt{(l+m)(l-m+1)}\mid l,m-1,\dot{l},\dot{m}\rangle
\;\;(m>-l),\nonumber\\
&&\sX_+\mid l,m;\dot{l},\dot{m}\rangle=
\sqrt{(l-m)(l+m+1)}\mid l,m+1;\dot{l},\dot{m}\rangle
\;\;(m<l),\nonumber\\
&&\sX_3\mid l,m;\dot{l},\dot{m}\rangle=
m\mid l,m;\dot{l},\dot{m}\rangle,\nonumber\\
&&\sY_-\mid l,m;\dot{l},\dot{m}\rangle=
\sqrt{(\dot{l}+\dot{m})(\dot{l}-\dot{m}+1)}\mid l,m;\dot{l},\dot{m}-1
\rangle\;\;(\dot{m}>-\dot{l}),\nonumber\\
&&\sY_+\mid l,m;\dot{l},\dot{m}\rangle=
\sqrt{(\dot{l}-\dot{m})(\dot{l}+\dot{m}+1)}\mid l,m;\dot{l},\dot{m}+1
\rangle\;\;(\dot{m}<\dot{l}),\nonumber\\
&&\sY_3\mid l,m;\dot{l},\dot{m}\rangle=
\dot{m}\mid l,m;\dot{l},\dot{m}\rangle.\label{Waerden}
\end{eqnarray}
From the relations (\ref{Com2}) it follows that each of the sets of 
infinitisimal operators $\sX$ and $\sY$ generates the group $SU(2)$ and these
two groups commute with each other. Thus, from the relations (\ref{Com2})
and (\ref{Waerden}) it follows that the group $\fG_+$, in essence,
is equivalent to the group $SU(2)\otimes SU(2)$. In contrast to the
Gel'fand--Naimark representation\index{representation!Gel'fand-Naimark}
for the Lorentz group \cite{GMS,Nai58},
which does not find a broad application in physics,
a representation (\ref{Waerden}) is a most useful in theoretical physics
(see, for example, \cite{AB,Sch61,RF,Ryd85}). This representation for the
Lorentz group was first given by Van der Waerden in his brilliant book
\cite{Wa32}.\index{representation!Van der Waerden}
It should be noted here that the representation basis, defined by the
formulae (\ref{SL25})--(\ref{Waerden}), has an evident physical meaning.
For example, in the case of $(1,0)\oplus(0,1)$--representation space
there is an analogy with the photon spin states. Namely, the operators
$\sX$ and $\sY$ correspond to the right and left polarization states of the
photon. For that reason we will call the canonical basis consisting of the
vectors $\mid lm;\dot{l}\dot{m}\rangle$ as
{\it a helicity basis}.\index{basis!helicity}

As is known, a double covering of the proper orthochronous Lorentz group $\fG_+$,
the group $SL(2,\C)$, is isomorphic to the Clifford--Lipschitz group
$\spin_+(1,3)$, which, in turn, is completely defined within a
biquaternion algebra $\C_2$, since
\[\ar
\spin_+(1,3)\simeq\left\{\begin{pmatrix} \alpha & \beta \\ \gamma & \delta
\end{pmatrix}\in\C_2:\;\;\det\begin{pmatrix}\alpha & \beta \\ \gamma & \delta
\end{pmatrix}=1\right\}=SL(2,\C).
\]\begin{sloppypar}\noindent
Thus, a fundamental representation of the group $\fG_+$ is realized in a
spinspace $\dS_2$. The spinspace $\dS_2$ is a complexification of the
minimal left ideal of the algebra $\C_2$: $\dS_2=\C\otimes I_{2,0}=\C\otimes
\cl_{2,0}e_{20}$ or $\dS_2=\C\otimes I_{1,1}=\C\otimes\cl_{1,1}e_{11}$
($\C\otimes I_{0,2}=\C\otimes\cl_{0,2}e_{02}$), where $\cl_{p,q}$
($p+q=2$) is a real subalgebra of $\C_2$, $I_{p,q}$ is the minimal left ideal
of the algebra $\cl_{p,q}$, $e_{pq}$ is a primitive idempotent.
\end{sloppypar}
Linear
transformations of `vectors' (spinors and cospinors) of the spinspaces
$\dS_2$ and $\dot{\dS}_2$ have the form
\begin{equation}\label{6.5}\ar
\begin{array}{ccc}
\begin{array}{ccc}
{}^\prime\xi^1&=&\alpha\xi^1+\beta\xi^2,\\
{}^\prime\xi^2&=&\gamma\xi^1+\delta\xi^2,
\end{array} & \phantom{ccc} &
\begin{array}{ccc}
{}^\prime\xi^{\dot{1}}&=&\dot{\alpha}\xi^{\dot{1}}+\dot{\beta}\xi^{\dot{2}},\\
{}^\prime\xi^{\dot{2}}&=&\dot{\gamma}\xi^{\dot{1}}+\dot{\delta}\xi^{\dot{2}},
\end{array}\\
\sigma=\ar\begin{pmatrix}
\alpha & \beta\\
\gamma & \delta
\end{pmatrix} & \phantom{ccc} &
\dot{\sigma}=\begin{pmatrix}
\dot{\alpha} & \dot{\beta}\\
\dot{\gamma} & \dot{\delta}
\end{pmatrix}.
\end{array}
\end{equation}
Transformations (\ref{6.5}) form the group $SL(2,\C)$.
The expressions (\ref{6.5}) compose a base of the 2--spinor
van der Waerden formalism \cite{Wa29,Rum36}, in which the spaces
$\dS_2$ and $\dot{\dS}_2$ are called correspondingly spaces of
{\it undotted and dotted spinors}. The each of the spaces
$\dS_2$ and $\dot{\dS}_2$ is homeomorphic to an extended complex plane
$\C\cup\infty$ representing an absolute (the set of infinitely distant points)
of a Lobatchevskii space $S^{1,2}$. At this point, a group of fractional
linear transformations of the plane $\C\cup\infty$ is isomorphic to a motion
group of $S^{1,2}$ \cite{Roz55}. Besides, in accordance with
\cite{Kot27}, the Lobatchevskii space $S^{1,2}$ is an absolute\index{absolute} of the
Minkowski world $\R^{1,3}$ and, therefore, the group of fractional linear
transformations of the plane $\C\cup\infty$ (motion group of
$S^{1,2}$) twice covers a `rotation group' of the space--time $\R^{1,3}$,
that is the proper Lorentz group.

The tensor product $\C_2\otimes\C_2\otimes\cdots\otimes\C_2\simeq\C_{2k}$
of the $k$ algebras $\C_2$ induces a tensor product of the $k$
spinspaces $\dS_2$:
\[
\dS_2\otimes\dS_2\otimes\cdots\otimes\dS_2=\dS_{2^k}.
\]
Vectors of the spinspace\index{spinspace}
$\dS_{2^k}$ (or elements of the minimal left
ideal of $\C_{2k}$) are spintensors of the following form
\begin{equation}\label{6.16}
\xi^{\alpha_1\alpha_2\cdots\alpha_k}=\sum
\xi^{\alpha_1}\otimes\xi^{\alpha_2}\otimes\cdots\otimes\xi^{\alpha_k},
\end{equation}
where summation is produced on all the index collections
$(\alpha_1\ldots\alpha_k)$, $\alpha_i=1,2$. In virtue of (\ref{6.5}) for
each spinor $\xi^{\alpha_i}$ from (\ref{6.16}) we have a transformation rule
${}^\prime\xi^{\alpha^\prime_i}=
\sigma^{\alpha^\prime_i}_{\alpha_i}\xi^{\alpha_i}$. Therefore, in general
case we obtain
\begin{equation}\label{6.17}
{}^\prime\xi^{\alpha^\prime_1\alpha^\prime_2\cdots\alpha^\prime_k}=\sum
\sigma^{\alpha^\prime_1}_{\alpha_1}\sigma^{\alpha^\prime_2}_{\alpha_2}\cdots
\sigma^{\alpha^\prime_k}_{\alpha_k}\xi^{\alpha_1\alpha_2\cdots\alpha_k}.
\end{equation}
A representation (\ref{6.17}) is called
{\it undotted spintensor representation of the proper Lorentz group of the
rank $k$}.\index{representation!spintensor!undotted}

Further, let $\overset{\ast}{\C}_2$ be 
a biquaternion algebra,\index{algebra!biquaternion} the
coefficients of which are complex conjugate. Let us show that the algebra
$\overset{\ast}{\C}_2$ can be obtained from $\C_2$ under action of the
automorphism $\cA\rightarrow\cA^\star$ or antiautomorphism
$\cA\rightarrow\widetilde{\cA}$. Indeed, in virtue of an isomorphism
$\C_2\simeq\cl_{3,0}$ a general element
\[
\cA=a^0\e_0+\sum^3_{i=1}a^i\e_i+\sum^3_{i=1}\sum^3_{j=1}a^{ij}\e_{ij}+
a^{123}\e_{123}
\]
of the algebra $\cl_{3,0}$ can be written in the form
\begin{equation}\label{6.17'}
\cA=(a^0+\omega a^{123})\e_0+(a^1+\omega a^{23})\e_1+(a^2+\omega a^{31})\e_2
+(a^3+\omega a^{12})\e_3,
\end{equation}
where $\omega=\e_{123}$. Since $\omega$ belongs to 
a center\index{center} of the algebra
$\cl_{3,0}$ (commutes with all the basis elements) and $\omega^2=-1$, then
we can to suppose $\omega\equiv i$. The action of the automorphism $\star$
on the homogeneous element $\cA$ of a degree $k$ is defined by a formula
$\cA^\star=(-1)^k\cA$. In accordance with this, the action of the
automorphism $\cA\rightarrow\cA^\star$, where $\cA$ is the element
(\ref{6.17'}), has a form
\begin{equation}\label{In1}
\cA\longrightarrow\cA^\star=-(a^0-\omega a^{123})\e_0-(a^1-\omega a^{23})\e_1
-(a^2-\omega a^{31})\e_2-(a^3-\omega a^{12})\e_3.
\end{equation}
Therefore, $\star:\,\C_2\rightarrow -\overset{\ast}{\C}_2$. Correspondingly,
the action of the antiautomorphism $\cA\rightarrow\widetilde{\cA}$ on the
homogeneous element $\cA$ of a degree $k$ is defined by a formula
$\widetilde{\cA}=(-1)^{\frac{k(k-1)}{2}}\cA$. Thus, for the element
(\ref{6.17'}) we obtain
\begin{equation}\label{In2}
\cA\longrightarrow\widetilde{\cA}=(a^0-\omega a^{123})\e_0+
(a^1-\omega a^{23})\e_1+(a^2-\omega a^{31})\e_2+(a^3-\omega a^{12})\e_3,
\end{equation}
that is, $\widetilde{\phantom{cc}}:\,\C_2\rightarrow
\overset{\ast}{\C}_2$.
This allows us to define an algebraic analog of the Wigner's representation
doubling: $\C_2\oplus\overset{\ast}{\C}_2$. 
Further, from (\ref{6.17'})
it follows that $\cA=\cA_1+\omega\cA_2=(a^0\e_0+a^1\e_1+a^2\e_2+a^3\e_3)+
\omega(a^{123}\e_0+a^{23}\e_1+a^{31}\e_2+a^{12}\e_3)$. In general case,
in virtue of an isomorphism $\C_{2k}\simeq\cl_{p,q}$, where $\cl_{p,q}$ is a
real Clifford algebra with a division ring $\K\simeq\C$, $p-q\equiv 3,7
\pmod{8}$, we have for a general element of $\cl_{p,q}$ an expression
$\cA=\cA_1+\omega\cA_2$, here $\omega^2=\e^2_{12\ldots p+q}=-1$ and, therefore,
$\omega\equiv i$. Thus, from $\C_{2k}$ under action of the automorphism
$\cA\rightarrow\cA^\star$ we obtain a general algebraic doubling
\begin{equation}\label{D}
\C_{2k}\oplus\overset{\ast}{\C}_{2k}.
\end{equation}

Correspondingly, a tensor product\index{product!tensor}
$\overset{\ast}{\C}_2\otimes\overset{\ast}{\C}_2\otimes\cdots\otimes
\overset{\ast}{\C}_2\simeq\overset{\ast}{\C}_{2r}$ of $r$ algebras
$\overset{\ast}{\C}_2$ induces a tensor product of $r$ spinspaces
$\dot{\dS}_2$:
\[
\dot{\dS}_2\otimes\dot{\dS}_2\otimes\cdots\otimes\dot{\dS}_2=\dot{\dS}_{2^r}.
\]
The vectors of the spinspace $\dot{\dS}_{2^r}$ have the form
\begin{equation}\label{6.18}
\xi^{\dot{\alpha}_1\dot{\alpha}_2\cdots\dot{\alpha}_r}=\sum
\xi^{\dot{\alpha}_1}\otimes\xi^{\dot{\alpha}_2}\otimes\cdots\otimes
\xi^{\dot{\alpha}_r},
\end{equation}
where the each cospinor $\xi^{\dot{\alpha}_i}$ from (\ref{6.18}), in virtue of
(\ref{6.5}), is transformed by the rule ${}^\prime\xi^{\dot{\alpha}^\prime_i}=
\sigma^{\dot{\alpha}^\prime_i}_{\dot{\alpha}_i}\xi^{\dot{\alpha}_i}$.
Therefore,
\begin{equation}\label{6.19}
{}^\prime\xi^{\dot{\alpha}^\prime_1\dot{\alpha}^\prime_2\cdots
\dot{\alpha}^\prime_r}=\sum\sigma^{\dot{\alpha}^\prime_1}_{\dot{\alpha}_1}
\sigma^{\dot{\alpha}^\prime_2}_{\dot{\alpha}_2}\cdots
\sigma^{\dot{\alpha}^\prime_r}_{\dot{\alpha}_r}
\xi^{\dot{\alpha}_1\dot{\alpha}_2\cdots\dot{\alpha}_r}.
\end{equation}\begin{sloppypar}\noindent
A representation (\ref{6.19}) is called
{\it a dotted spintensor representation of the proper Lorentz group of the
rank $r$}.\index{representation!spintensor!dotted}\end{sloppypar}

In general case we have a tensor product of $k$ algebras $\C_2$ and
$r$ algebras $\overset{\ast}{\C}_2$:
\[
\C_2\otimes\C_2\otimes\cdots\otimes\C_2\otimes
\overset{\ast}{\C}_2\otimes
\overset{\ast}{\C}_2\otimes\cdots\otimes\overset{\ast}{\C}_2\simeq
\C_{2k}\otimes\overset{\ast}{\C}_{2r},
\]
which induces a spinspace
\begin{equation}\label{Spint}
\dS_2\otimes\dS_2\otimes\cdots\otimes\dS_2\otimes\dot{\dS}_2\otimes
\dot{\dS}_2\otimes\cdots\otimes\dot{\dS}_2=\dS_{2^{k+r}}
\end{equation}
with the vectors
\begin{equation}\label{Vect}
\xi^{\alpha_1\alpha_2\cdots\alpha_k\dot{\alpha}_1\dot{\alpha}_2\cdots
\dot{\alpha}_r}=\sum
\xi^{\alpha_1}\otimes\xi^{\alpha_2}\otimes\cdots\otimes\xi^{\alpha_k}\otimes
\xi^{\dot{\alpha}_1}\otimes\xi^{\dot{\alpha}_2}\otimes\cdots\otimes
\xi^{\dot{\alpha}_r}.
\end{equation}
In this case we have a natural unification of the representations
(\ref{6.17}) and (\ref{6.19}):
\begin{equation}\label{6.20}
{}^\prime\xi^{\alpha^\prime_1\alpha^\prime_2\cdots\alpha^\prime_k
\dot{\alpha}^\prime_1\dot{\alpha}^\prime_2\cdots\dot{\alpha}^\prime_r}=\sum
\sigma^{\alpha^\prime_1}_{\alpha_1}\sigma^{\alpha^\prime_2}_{\alpha_2}\cdots
\sigma^{\alpha^\prime_k}_{\alpha_k}\sigma^{\dot{\alpha}^\prime_1}_{
\dot{\alpha}_1}\sigma^{\dot{\alpha}^\prime_2}_{\dot{\alpha}_2}\cdots
\sigma^{\dot{\alpha}^\prime_r}_{\dot{\alpha}_r}
\xi^{\alpha_1\alpha_2\cdots\alpha_k\dot{\alpha}_1\dot{\alpha}_2\cdots
\dot{\alpha}_r}.
\end{equation}
So, a representation (\ref{6.20}) is called
{\it a spintensor representation of the proper Lorentz group of the
rank $(k,r)$}.\index{representation!spintensor}
\begin{sloppypar}
In general case, the representations, defined by the formulas
(\ref{6.17}), (\ref{6.19}) and
(\ref{6.20}), are reducible, that is, there exists the possibility of decomposition 
of the initial spinspace $\dS_{2^{k+r}}$ (correspondingly, spinspaces
$\dS_{2^k}$ and $\dS_{2^r}$) into a direct sum of invariant (with respect to
transformations of the group $\fG_+$) spinspaces
$\dS_{2^{\nu_1}}\oplus\dS_{2^{\nu_2}}\oplus\cdots\oplus\dS_{2^{\nu_s}}$,
where $\nu_1+\nu_2+\ldots+\nu_s=k+r$.
The algebras $\C_2$ ($\overset{\ast}{\C}_2$) and the spinspaces
$\dS_2$ ($\dot{\dS}_2$) correspond to fundamental represesentations
$\boldsymbol{\tau}_{\frac{1}{2},0}$ ($\boldsymbol{\tau}_{0,\frac{1}{2}}$)
of the Lorentz group $\fG_+$.
In general case the spinspace (\ref{Spint}) is reducible, that is,
there exists a decomposition of the original spinspace $\dS_{2^{k+r}}$
into a direct sum of irreducible subspaces with respect to a
representation\end{sloppypar}
\begin{equation}\label{Rept}
\underbrace{\boldsymbol{\tau}_{\frac{1}{2},0}\otimes
\boldsymbol{\tau}_{\frac{1}{2},0}\otimes\cdots\otimes
\boldsymbol{\tau}_{\frac{1}{2},0}}_{k\;\text{times}}\otimes
\underbrace{\boldsymbol{\tau}_{0,\frac{1}{2}}\otimes
\boldsymbol{\tau}_{0,\frac{1}{2}}\otimes\cdots\otimes
\boldsymbol{\tau}_{0,\frac{1}{2}}}_{r\;\text{times}}.
\end{equation}
The full representation space $\dS_{2^{k+r}}$ contains both symmetric and
antisymmetric spintensors (\ref{Vect}). 

The decomposition of the spinspace with respect to $SL(2,\C)$
is a simplest case of the Weyl scheme. Every irreducible representation of
the group $SL(2,\C)$ is defined by the 
Young tableau\index{tableau!Young} consisting of only one
row. 
Thus, all the possible irreducible representations of $SL(2,\C)$
correspond to the following Young tableaux:
\[
\unitlength=0.35mm
\begin{picture}(150.00,30.00)
\put(0,0){\line(1,0){10}}
\put(0,10){\line(1,0){10}}
\put(0,0){\line(0,1){10}}
\put(10,0){\line(0,1){10}}
\put(12,0){,}
\put(27,0){\line(1,0){10}}
\put(27,10){\line(1,0){10}}
\put(27,0){\line(0,1){10}}
\put(37,0){\line(0,1){10}}
\put(37,0){\line(1,0){10}}
\put(37,10){\line(1,0){10}}
\put(47,0){\line(0,1){10}}
\put(49,0){,}
\put(59,0){$\ldots$}
\put(75,0){,}
\put(85,0){\line(1,0){10}}
\put(85,10){\line(1,0){10}}
\put(85,0){\line(0,1){10}}
\put(95,0){\line(0,1){10}}
\put(95,0){\line(1,0){10}}
\put(95,10){\line(1,0){10}}
\put(105,0){\line(0,1){10}}
\put(105,0){\line(1,0){20}}
\put(105,10){\line(1,0){20}}
\put(108.5,4.5){$\ldots$}
\put(125,0){\line(0,1){10}}
\put(125,0){\line(1,0){10}}
\put(125,10){\line(1,0){10}}
\put(135,0){\line(0,1){10}}
\put(137,0){,}
\put(147,0){$\ldots$}
\end{picture}
\]
For that reason the representation $\boldsymbol{\tau}_{\frac{m}{2},0}$
is realized in the space
$\Sym_{(m,0)}$ of all symmetric spintensors of the rank $m$. Dimension of
$\Sym_{(m,0)}$ is equal to $m+1$.

In turn, every element of the spinspace (\ref{Spint}), related with
the representation (\ref{Rept}), corresponds to an element of
$\dS_{2^k}\otimes\dS_{2^r}$ (representations
$\boldsymbol{\tau}_{\frac{k}{2},0}\otimes\boldsymbol{\tau}_{\frac{r}{2},0}$
and
$\boldsymbol{\tau}_{\frac{k}{2},0}\otimes\boldsymbol{\tau}_{0,\frac{r}{2}}$
are equivalent). This equivalence can be described as follows
\begin{equation}\label{Equiv}
\varphi\otimes\psi\longrightarrow\varphi\otimes\psi I,
\end{equation}
where $\varphi,\psi\in\dS_{2^k}$, $\psi I\in\dS_{2^r}$ and
\[
I=\lambda\begin{pmatrix}
0 & & -1\\
  &\iddots&\\
(-1)_{\frac{r+k}{2}} && 0
\end{pmatrix}
\]
is the matrix of a bilinear form\index{form!bilinear}
(this matrix is symmetric if
$\frac{r+k}{2}\equiv 0\pmod{2}$ and skewsymmetric if 
$\frac{r+k}{2}\equiv 1\pmod{2}$). In such a way, the representation
(\ref{Rept}) is realized in 
a symmetric space\index{space!symmetric} $\Sym_{(k,r)}$ of
dimension $(k+1)(r+1)$ (or $(2l+1)(2\dot{l}+1)$ if suppose $l=k/2$ and
$\dot{l}=r/2$). The decomposition of (\ref{Rept}) is given by a Clebsh-Gordan
formula\index{Clebsh-Gordan formula}
\[
\boldsymbol{\tau}_{l_1\dot{l}_1}\otimes\boldsymbol{\tau}_{l_2\dot{l}_2}=
\sum_{|l_1-l_2|\leq k\leq l_1+l_2;|\dot{l}_1-\dot{l}_2|\leq \dot{k}\leq
\dot{l}_1+\dot{l}_2}\boldsymbol{\tau}_{k\dot{k}}.
\]
where the each $\boldsymbol{\tau}_{k\dot{k}}$ acts in the space
$\Sym_{(k,\dot{k})}$. In turn, every space $\Sym_{(k,r)}$
can be represented by 
a space of polynomials\index{space!of polynomials}
\begin{gather}
p(z_0,z_1,\bar{z}_0,\bar{z}_1)=\sum_{\substack{(\alpha_1,\ldots,\alpha_k)\\
(\dot{\alpha}_1,\ldots,\dot{\alpha}_r)}}\frac{1}{k!\,r!}
a^{\alpha_1\cdots\alpha_k\dot{\alpha}_1\cdots\dot{\alpha}_r}
z_{\alpha_1}\cdots z_{\alpha_k}\bar{z}_{\dot{\alpha}_1}\cdots
\bar{z}_{\dot{\alpha}_r}.\label{SF}\\
(\alpha_i,\dot{\alpha}_i=0,1)\nonumber
\end{gather}
where the numbers 
$a^{\alpha_1\cdots\alpha_k\dot{\alpha}_1\cdots\dot{\alpha}_r}$
are unaffected at the permutations of indices. 
The expressions (\ref{SF}) can be understood as {\it functions on the
Lorentz group}.\index{function!on the Lorentz group}
Later on, we will find 
an Euler parametrization\index{Euler parametrization}
of these functions (see section \ref{Sec:5.3}). When the coefficients
$a^{\alpha_1\cdots\alpha_k\dot{\alpha}_1\cdots\dot{\alpha}_r}$ in
(\ref{SF}) are depend on the variables $x^\alpha$ ($\alpha=0,1,2,3$),
then we have
\begin{gather}
p(x,z,\bar{z})=\sum_{\substack{(\alpha_1,\ldots,\alpha_k)\\
(\dot{\alpha}_1,\ldots,\dot{\alpha}_r)}}\frac{1}{k!\,r!}
a^{\alpha_1\cdots\alpha_k\dot{\alpha}_1\cdots\dot{\alpha}_r}(x)
z_{\alpha_1}\cdots z_{\alpha_k}\bar{z}_{\dot{\alpha}_1}\cdots
\bar{z}_{\dot{\alpha}_r}.\label{SF2}\\
(\alpha_i,\dot{\alpha}_i=0,1)\nonumber
\end{gather}
The functions (\ref{SF2}) should be considered as {\it the functions on
the Poincar\'{e} group}.\index{function!on the Poincar\'{e} group}
Some applications of these functions contained
in \cite{GS01}. Representations of 
the Poincar\'{e} group\index{group!Poincar\'{e}}
$SL(2,\C)\odot T(4)$ are realized via the functions (\ref{SF2}), here
$T(4)$ is a subgroup of 4-dimensional translations.

Infinitesimal operators of $\fG_+$ in the helicity basis have a very simple
form
\begin{eqnarray}
\sA_1\mid l,m;\do{l},\dot{m}\rangle&=&-\frac{i}{2}\boldsymbol{\alpha}^l_m
\mid l,m-1;\dot{l},\dot{m}\rangle-
\frac{i}{2}\boldsymbol{\alpha}^l_{m+1}\mid l,m+1;\dot{l}\dot{m}\rangle,
\nonumber\\
\sA_2\mid l,m;\dot{l},\dot{m}\rangle&=&\frac{1}{2}\boldsymbol{\alpha}^l_m
\mid l,m-1;\dot{l},\dot{m}\rangle-
\frac{1}{2}\boldsymbol{\alpha}^l_{m+1}
\mid l,m+1;\dot{l},\dot{m}\rangle,\label{OpA}\\
\sA_3\mid l,m;\dot{l},\dot{m}\rangle&=&-im
\mid l,m;\dot{l},\dot{m}\rangle,\nonumber
\end{eqnarray}
\begin{eqnarray}
\sB_1\mid l,m;\dot{l},\dot{m}\rangle&=&-\frac{1}{2}\boldsymbol{\alpha}^l_m
\mid l,m-1;\dot{l},\dot{m}\rangle-
\frac{1}{2}\boldsymbol{\alpha}^l_{m+1}
\mid l,m+1;\dot{l},\dot{m}\rangle,\nonumber\\
\sB_2\mid l,m;\dot{l},\dot{m}\rangle&=&-\frac{i}{2}\boldsymbol{\alpha}^l_m
\mid l,m-1;\dot{l},\dot{m}\rangle+
\frac{i}{2}\boldsymbol{\alpha}^l_{m+1}
\mid l,m+1;\dot{l},\dot{m}\rangle,\label{OpB}\\
\sB_3\mid l,m;\dot{l},\dot{m}\rangle&=&-m\mid l,m;\dot{l},\dot{m}\rangle,
\nonumber
\end{eqnarray}
\begin{eqnarray}
\widetilde{\sA}_1\mid l,m;\dot{l},\dot{m}\rangle&=&
-\frac{i}{2}\boldsymbol{\alpha}^{\dot{l}}_{\dot{m}}
\mid l,m;\dot{l},\dot{m}-1\rangle-
\frac{i}{2}\boldsymbol{\alpha}^{\dot{l}}_{\dot{m}+1}
\mid l,m;\dot{l},\dot{m}+1\rangle,\nonumber\\
\widetilde{\sA}_2\mid l,m;\dot{l},\dot{m}\rangle&=&
\frac{1}{2}\boldsymbol{\alpha}^{\dot{l}}_{\dot{m}}
\mid l,m;\dot{l},\dot{m}-1\rangle-
\frac{1}{2}\boldsymbol{\alpha}^{\dot{l}}_{\dot{m}+1}
\mid l,m;\dot{l},\dot{m}+1\rangle,\label{DopA}\\
\widetilde{\sA}_3\mid l,m;\dot{l},\dot{m}\rangle&=&
-i\dot{m}\mid l,m;\dot{l},\dot{m}\rangle,\nonumber
\end{eqnarray}
\begin{eqnarray}
\widetilde{\sB}_1\mid l,m;\dot{l},\dot{m}\rangle&=&
\frac{1}{2}\boldsymbol{\alpha}^{\dot{l}}_{\dot{m}}
\mid l,m;\dot{l},\dot{m}-1\rangle+
\frac{1}{2}\boldsymbol{\alpha}^{\dot{l}}_{\dot{m}+1}
\mid l,m;\dot{l},\dot{m}+1\rangle,\nonumber\\
\widetilde{\sB}_2\mid l,m;\dot{l},\dot{m}\rangle&=&
\frac{i}{2}\boldsymbol{\alpha}^{\dot{l}}_{\dot{m}}
\mid l,m;\dot{l},\dot{m}-1\rangle-
\frac{i}{2}\boldsymbol{\alpha}^{\dot{l}}_{\dot{m}+1}
\mid l,m;\dot{l},\dot{m}+1\rangle,\label{DopB}\\
\widetilde{\sB}_3\mid l,m;\dot{l},\dot{m}\rangle&=&
-\dot{m}\mid l,m;\dot{l},\dot{m}\rangle,\nonumber
\end{eqnarray}
where
\[
\boldsymbol{\alpha}^l_m=\sqrt{(l+m)(l-m+1)}.
\]
In the matrix notation for the operators $\sA_i$ we have
\begin{equation}\label{AI1}
\sA^j_{1}=-\frac{i}{2}
\ar\begin{bmatrix}
0 & \boldsymbol{\alpha}_{-l_j+1} & 0 & \dots & 0 & 0\\
\boldsymbol{\alpha}_{-l_j+1} &0&\boldsymbol{\alpha}_{-l_j+2} & \dots & 0 & 0\\
0 & \boldsymbol{\alpha}_{-l_j+2} & 0 & \dots & 0 & 0\\
\hdotsfor[3]{6}\\
\hdotsfor[3]{6}\\
0 & 0 & 0 & \dots & 0 & \boldsymbol{\alpha}_{l_j}\\
0 & 0 & 0 & \dots & \boldsymbol{\alpha}_{l_j} & 0
\end{bmatrix},
\end{equation}
\begin{equation}\label{AI2}
\sA^j_{2}=\frac{1}{2}
\ar\begin{bmatrix}
0 & \boldsymbol{\alpha}_{-l_j+1} & 0 & \dots & 0 & 0\\
-\boldsymbol{\alpha}_{-l_j+1}&0&\boldsymbol{\alpha}_{-l_j+2} & \dots & 0 & 0\\
0 & -\boldsymbol{\alpha}_{-l_j+2} & 0 & \dots & 0 & 0\\
\hdotsfor[3]{6}\\
\hdotsfor[3]{6}\\
0 & 0 & 0 & \dots & 0 & \boldsymbol{\alpha}_{l_j}\\
0 & 0 & 0 & \dots & -\boldsymbol{\alpha}_{l_j} & 0
\end{bmatrix},
\end{equation}
\begin{equation}\label{AI3}
\sA^j_{3}=
\ar\begin{bmatrix}
il_j & 0 & 0 & \dots & 0 & 0\\
0 & i(l_j-1) & 0 & \dots & 0 & 0\\
0 & 0 & i(l_j-2) & \dots & 0 & 0\\
\hdotsfor[3]{6}\\
\hdotsfor[3]{6}\\
0 & 0 & 0 & \dots & -i(l_j-1) & 0\\
0 & 0 & 0 & \dots & 0 & -il_j
\end{bmatrix},
\end{equation}
and so on.
\subsection{Gel'fand-Naimark basis}\index{basis!Gel'fand-Naimark}
\label{Subsec:5.2.1}
There exists another representation basis for the Lorentz group:
\begin{eqnarray}
H_{3}\xi_{lm} &=&m\xi_{lm},\nonumber\\
H_{+}\xi_{lm} &=&\sqrt{(l+m+1)(l-m)}\xi_{l,m+1},\nonumber\\
H_{-}\xi_{lm} &=&\sqrt{(l+m)(l-m+1)}\xi_{l,m-1},\nonumber\\
F_{3}\xi_{lm} &=&C_{l}\sqrt{l^{2}-m^{2}}\xi_{l-1,m}-A_{l}m\xi_{l,m}-\nonumber \\
&&\hspace{1.8cm}-C_{l+1}\sqrt{(l+1)^{2}-m^{2}}\xi_{l+1,m},\nonumber\\
F_{+}\xi_{lm} &=&C_{l}\sqrt{(l-m)(l-m-1)}\xi_{l-1,m+1}-\nonumber\\
&&\hspace{1.3cm}-A_{l}\sqrt{(l-m)(l+m+1)}\xi_{l,m+1}+\nonumber \\
&&\hspace{1.8cm}+C_{l+1}\sqrt{(l+m+1)(l+m+2)}\xi_{l+1,m+1},\nonumber\\
F_{-}\xi_{lm} &=&-C_{l}\sqrt{(l+m)(l+m-1)}\xi_{l-1,m-1}-\nonumber\\
&&\hspace{1.3cm}-A_{l}\sqrt{(l+m)(l-m+1)}\xi_{l,m-1}-\nonumber\\
&&\hspace{1.8cm}-C_{l+1}\sqrt{(l-m+1)(l-m+2)}\xi_{l+1,m-1},\nonumber
\end{eqnarray}
\[
A_{l}=\frac{il_{0}l_{1}}{l(l+1)},\quad
C_{l}=\frac{i}{l}\sqrt{\frac{(l^{2}-l^{2}_{0})(l^{2}-l^{2}_{1})}
{4l^{2}-1}},
\]
$$m=-l,-l+1,\ldots,l-1,l,$$
$$l=l_{0}\,,l_{0}+1,\ldots,$$
where $l_{0}$ is positive integer or half-integer number, $l_{1}$ is an
arbitrary complex number. 
These formulas 
define a finite--dimensional representation of the group $\fG_+$ when
$l^2_1=(l_0+p)^2$, $p$ is some natural number.
In the case $l^2_1\neq(l_0+p)^2$ we have an
infinite--dimensional representation of $\fG_+$.
The operators $H_{3},H_{+},H_{-},F_{3},F_{+},F_{-}$
are
\begin{eqnarray}
&&H_+=i\sA_1-\sA_2,\quad H_-=i\sA_1+\sA_2,\quad H_3=i\sA_3,\nonumber\\
&&F_+=i\sB_1-\sB_2,\quad F_-=i\sB_1+\sB_2,\quad F_3=i\sB_3.\nonumber
\end{eqnarray}
This basis was first given by Gel'fand in 1944 (see also
\cite{Har47,GY48,Nai58}). The following relations between generators
$\sY_\pm$, $\sX_\pm$, $\sY_3$, $\sX_3$ and $H_\pm$, $F_\pm$, $H_3$, $F_3$
define a relationship between the helicity (Van der Waerden) and
Gel'fand-Naimark basises:
\[
{\renewcommand{\arraystretch}{1.7}
\begin{array}{ccc}
\sY_+&=&-\dfrac{1}{2}(F_++iH_+),\\
\sY_-&=&-\dfrac{1}{2}(F_-+iH_-),\\
\sY_3&=&-\dfrac{1}{2}(F_3+iH_3),
\end{array}\quad
\begin{array}{ccc}
\sX_+&=&\dfrac{1}{2}(F_+-iH_+),\\
\sX_-&=&\dfrac{1}{2}(F_--iH_-),\\
\sX_3&=&\dfrac{1}{2}(F_3-iH_3).
\end{array}.
}
\]
The relation between the numbers $l_0$, $l_1$ and the number $k$ of the
factors $\C_2$ in the product $\C_2\otimes\C_2\otimes\cdots\otimes\C_2$ is
given by a following formula
\[
(l_0,l_1)=\left(\frac{k}{2},\frac{k}{2}+1\right),
\]
whence it immediately follows that $k=l_0+l_1-1$. Thus, we have
{\it a complex representation\index{representation!complex}
$\fC^{l_0+l_1-1,0}$ of the proper Lorentz
group\index{group!Lorentz} $\fG_+$ in 
the spinspace\index{spinspace} $\dS_{2^k}$}.
In accordance with \cite{GMS} a representation conjugated to
$\fC^{l_0+l_1-1,0}$ is defined by a pair
\[
(l_0,l_1)=\left(-\frac{r}{2},\,\frac{r}{2}+1\right),
\]
that is, this representation has a form $\fC^{0,l_0-l_1+1}$. In turn,
a representation conjugated to 
fundamental representation\index{representation!fundamental} $\fC^{1,0}$ is
$\fC^{0,-1}$. 
As is known \cite{GMS}, if an irreducible representation of the proper Lorentz
group $\fG_+$ is defined by the pair $(l_0,l_1)$, then a conjugated
representation is also irreducible and defined by a pair $\pm(l_0,-l_1)$.

\subsection{One-parameter subgroups}\index{subgroup!one-parameter}
The representation of the group $\fG_+$ in the space $\Sym(k,r)$ has a form
\begin{eqnarray}
T_gq(\fz,\bar{\fz})&=&\frac{1}{z^k_1\,\bar{z}^r_1}T_g\left[
z^k_1\bar{z}^r_1q\left(\frac{z_0}{z_1},\frac{\bar{z}_0}{\bar{z}_1}\right)
\right]= \nonumber\\
&=&(\gamma\fz+\delta)^k(\overset{\ast}{\gamma}\bar{\fz}+
\overset{\ast}{\delta})^rq\left(\frac{\alpha\fz+\beta}{\gamma\fz+\delta},
\frac{\overset{\ast}{\alpha}\bar{\fz}+\overset{\ast}{\beta}}
{\overset{\ast}{\gamma}\bar{\fz}+\overset{\ast}{\delta}}\right),
\label{Rep}
\end{eqnarray}
where
\[
\fz=\frac{z_0}{z_1},\quad\bar{\fz}=\frac{\bar{z}_0}{\bar{z}_1}.
\]

It is easy to see that for the group $SU(2)\subset SL(2,\C)$ the formulae
(\ref{SF}) and (\ref{Rep}) reduce to the following
\begin{gather}
p(z_0,z_1)=\sum_{(\alpha_1,\ldots,\alpha_k)}\frac{1}{k!}
a^{\alpha_1\cdots\alpha_k}z_{\alpha_1}\cdots z_{\alpha_k},\label{SF1}\\
T_gq(\fz)=(\gamma\fz+\delta)^kq\left(\frac{\alpha\fz+\beta}
{\gamma\fz+\delta}\right),\label{Rep1}
\end{gather}\begin{sloppypar}\noindent
and the representation space $\Sym(k,r)$ reduces to $\Sym(k,0)$.
One--parameter subgroups of $SU(2)$ are defined by the matrices
\end{sloppypar}
\begin{gather}
a_1(t)=
{\renewcommand{\arraystretch}{1.3}
\begin{pmatrix}
\cos\frac{t}{2} & i\sin\frac{t}{2}\\
i\sin\frac{t}{2} & \cos\frac{t}{2}
\end{pmatrix},\quad
a_2(t)=\begin{pmatrix}
\cos\frac{t}{2} & -\sin\frac{t}{2}\\
\sin\frac{t}{2} & \cos\frac{t}{2}
\end{pmatrix}},\nonumber\\[0.2cm]
a_3(t)=\begin{pmatrix}
e^{\frac{it}{2}} & 0\\
0 & e^{-\frac{it}{2}}
\end{pmatrix}.\label{PS1}
\end{gather}
An arbitrary matrix $u\in SU(2)$ written via 
Euler angles\index{Euler angles} has a form
\begin{equation}\label{SU2}\ar
u=\begin{pmatrix}
\alpha & \beta\\
-\overset{\ast}{\beta} & \overset{\ast}{\alpha}
\end{pmatrix}=
{\renewcommand{\arraystretch}{1.3}
\begin{pmatrix}
\cos\frac{\theta}{2}e^{\frac{i(\varphi+\psi)}{2}} &
i\sin\frac{\theta}{2}e^{\frac{i(\varphi-\psi)}{2}}\\
i\sin\frac{\theta}{2}e^{\frac{i(\psi-\varphi)}{2}} &
\cos\frac{\theta}{2}e^{-\frac{i(\varphi+\psi)}{2}}
\end{pmatrix}},
\end{equation}
where $0\leq\varphi<2\pi$, $0<\theta<\pi$, $-2\pi\leq\psi<2\pi$, $\det u=1$.
Hence it follows that $|\alpha|=\cos\frac{\theta}{2}$,
$|\beta|=\sin\frac{\theta}{2}$ and
\begin{eqnarray}
\cos\theta&=&2|\alpha|^2-1,\label{SU3}\\
e^{i\varphi}&=&-\frac{\alpha\beta i}{|\alpha|\,|\beta|},\label{SU4}\\
e^{\frac{i\psi}{2}}&=&\frac{\alpha e^{-\frac{i\varphi}{2}}}{|\alpha|}.
\label{SU5}
\end{eqnarray}
Diagonal matrices $\begin{pmatrix} e^{\frac{i\varphi}{2}} & 0\\ 0 &
e^{-\frac{i\varphi}{2}}\end{pmatrix}$ form one--parameter subgroup in the
group $SU(2)$. Therefore, each matrix $u\in SU(2)$ belongs to a bilateral
adjacency class containing the matrix
\[
{\renewcommand{\arraystretch}{1.3}
\begin{pmatrix}
\cos\frac{\theta}{2} & i\sin\frac{\theta}{2}\\
i\sin\frac{\theta}{2} & \cos\frac{\theta}{2}
\end{pmatrix}}.
\]
The matrix element\index{element!matrix}
$t^l_{mn}=e^{-i(m\varphi+n\psi)}\langle T_l(\theta)
\psi_n,\psi_m\rangle$ of the group $SU(2)$ in the polynomial basis
\[
\psi_n(\fz)=\frac{\fz^{l-n}}{\sqrt{\Gamma(l-n+1)\Gamma(l+n+1)}},
\quad -l\leq n\leq l,
\]
where
\[
T_l(\theta)\psi(\fz)=\left(i\sin\frac{\theta}{2}\fz+
\cos\frac{\theta}{2}\right)^{2l}\psi
\left(\frac{\cos\frac{\theta}{2}\fz+i\sin\frac{\theta}{2}}
{i\sin\frac{\theta}{2}\fz+\cos\frac{\theta}{2}}\right),
\]
has a form
\begin{multline}
t^l_{mn}(g)=e^{-i(m\varphi+n\psi)}\langle T_l(\theta)\psi_n,\psi_m\rangle
=\\[0.2cm]
\frac{e^{-i(m\varphi+n\psi)}\langle T_l(\theta)\fz^{l-n}\fz^{l-m}\rangle}
{\sqrt{\Gamma(l-m+1)\Gamma(l+m+1)\Gamma(l-n+1)\Gamma(l+n+1)}}=\\[0.2cm]
e^{-i(m\varphi+n\psi)}i^{m-n}\sqrt{\Gamma(l-m+1)\Gamma(l+m+1)\Gamma(l-n+1)
\Gamma(l+n+1)}\times\\
\cos^{2l}\frac{\theta}{2}\tg^{m-n}\frac{\theta}{2}\times\\
\sum^{\min(l-n,l+n)}_{j=\max(0,n-m)}
\frac{i^{2j}\tg^{2j}\dfrac{\theta}{2}}
{\Gamma(j+1)\Gamma(l-m-j+1)\Gamma(l+n-j+1)\Gamma(m-n+j+1)}.\label{Mat1}
\end{multline}
Further, using the formula
\begin{equation}\label{HG}
\hypergeom{2}{1}{\alpha,\beta}{\gamma}{z}
=\frac{\Gamma(\gamma)}{\Gamma(\alpha)
\Gamma(\beta)}\sum_{k\ge 0}\frac{\Gamma(\alpha+k)\Gamma(\beta+k)}
{\Gamma(\gamma+k)}\frac{z^k}{k!}
\end{equation}
we can express the matrix element (\ref{Mat1}) via the hypergeometric
function:\index{function!hypergeometric}
\begin{multline}
t^l_{mn}(g)=\frac{i^{m-n}e^{-i(m\varphi+n\psi)}}{\Gamma(m-n+1)}
\sqrt{\frac{\Gamma(l+m+1)\Gamma(l-n+1)}{\Gamma(l-m+1)\Gamma(l+n+1)}}\times\\
\cos^{2l}\frac{\theta}{2}\tg^{m-n}\frac{\theta}{2}
\hypergeom{2}{1}{m-l+1,1-l-n}{m-n+1}{i^2\tg^2\dfrac{\theta}{2}},\label{Mat2}
\end{multline}
where $m\ge n$. At $m<n$ in the right part of (\ref{Mat2}) it needs to
replace $m$ and $n$ by $-m$ and $-n$, respectively. Since $l,m$ and $n$
are finite numbers, then the hypergeometric series is interrupted.

Further, replacing in the one--parameter subgroups (\ref{PS1}) the
parameter $t$ by $-it$, we obtain
\begin{gather}
{\renewcommand{\arraystretch}{1.3}
b_1(t)=
\begin{pmatrix}
\ch\frac{t}{2} & \sh\frac{t}{2}\\
\sh\frac{t}{2} & \ch\frac{t}{2}
\end{pmatrix},\quad
b_2(t)=\begin{pmatrix}
\ch\frac{t}{2} & i\sh\frac{t}{2}\\
-i\sh\frac{t}{2} & \ch\frac{t}{2}
\end{pmatrix}},\nonumber\\[0.2cm]
b_3(t)=\begin{pmatrix}
e^{\frac{t}{2}} & 0\\
0 & e^{-\frac{t}{2}}
\end{pmatrix}.\label{PS2}
\end{gather}
These subgroups correspond to hyperbolic rotations. 
\section{Hyperspherical functions}
\label{Sec:5.3}
As is known \cite{GG78}, a root subgroup of a semisimple Lie group $O_4$
(a rotation group of the 4-dimensional space) is a normal divisor of $O_4$.
For that reason the 6-parameter group $O_4$ is semisimple, and is
represented by a direct product of the two 3-parameter unimodular groups.
By analogy with the group $O_4$, a double covering $SL(2,\C)$ of the
proper orthochronous Lorentz group $\fG_+$ (a rotation group of the
4-dimensional spacetime continuum) is semisimple, and is represented by
a direct product of the two 3-parameter special unimodular groups,
$SL(2,\C)\simeq SU(2)\otimes SU(2)$. An explicit form of this isomorphism
can be obtained by means of a complexification of the group $SU(2)$,
that is, $SL(2,\C)\simeq\mbox{\sf complex}(SU(2))\simeq
SU(2)\otimes SU(2)$ \cite{Var022}.

\subsection{Complexification of $SU(2)$}
The group $SL(2,\C)$ of all complex matrices
\[\ar
\begin{pmatrix}
\alpha & \beta\\
\gamma & \delta
\end{pmatrix}
\]
of 2-nd order with the determinant $\alpha\delta-\gamma\beta=1$, is
a {\it complexification} of the group $SU(2)$. The group $SU(2)$ is one of
the real forms of $SL(2,\C)$. The transition from $SU(2)$ to $SL(2,\C)$
is realized via the complexification of three real parameters
$\varphi,\,\theta,\,\psi$ (Euler angles). Let $\theta^c=\theta-i\tau$,
$\varphi^c=\varphi-i\epsilon$, $\psi^c=\psi-i\varepsilon$ be complex
Euler angles,\index{Euler angles!complex} where
\begin{equation}\label{CEA}
{\renewcommand{\arraystretch}{1.05}
\begin{array}{ccccc}
0 &\leq&\re\theta^c=\theta& \leq& \pi,\\
0 &\leq&\re\varphi^c=\varphi& <&2\pi,\\
-2\pi&\leq&\re\psi^c=\psi&<&2\pi,
\end{array}\quad\quad
\begin{array}{ccccc}
-\infty &<&\im\theta^c=\tau&<&+\infty,\\
-\infty&<&\im\varphi^c=\epsilon&<&+\infty,\\
-\infty&<&\im\psi^c=\varepsilon&<&+\infty.
\end{array}}
\end{equation}
Replacing in (\ref{SU2}) the angles $\varphi,\,\theta,\,\psi$ by the
complex angles $\varphi^c,\theta^c,\psi^c$, we come to the following matrix
\begin{gather}
{\renewcommand{\arraystretch}{1.3}
\mathfrak{g}=
\begin{pmatrix}
\cos\frac{\theta^c}{2}e^{\frac{i(\varphi^c+\psi^c)}{2}} &
i\sin\frac{\theta^c}{2}e^{\frac{i(\varphi^c-\psi^c)}{2}}\\
i\sin\frac{\theta^c}{2}e^{\frac{i(\psi^c-\varphi^c)}{2}} &
\cos\frac{\theta^c}{2}e^{-\frac{i(\varphi^c+\psi^c)}{2}}
\end{pmatrix}}=
\nonumber\\
{\renewcommand{\arraystretch}{1.3}
\begin{pmatrix}
\left[\cos\frac{\theta}{2}\ch\frac{\tau}{2}+
i\sin\frac{\theta}{2}\sh\frac{\tau}{2}\right]
e^{\frac{\epsilon+\varepsilon+i(\varphi+\psi)}{2}} &
\left[\cos\frac{\theta}{2}\sh\frac{\tau}{2}+
i\sin\frac{\theta}{2}\ch\frac{\tau}{2}\right]
e^{\frac{\epsilon-\varepsilon+i(\varphi-\psi)}{2}} \\
\left[\cos\frac{\theta}{2}\sh\frac{\tau}{2}+
i\sin\frac{\theta}{2}\ch\frac{\tau}{2}\right]
e^{\frac{\varepsilon-\epsilon+i(\psi-\varphi)}{2}} &
\left[\cos\frac{\theta}{2}\ch\frac{\tau}{2}+
i\sin\frac{\theta}{2}\sh\frac{\tau}{2}\right]
e^{\frac{-\epsilon-\varepsilon-i(\varphi+\psi)}{2}}
\end{pmatrix}},\label{SL1}
\end{gather}
since $\cos\frac{1}{2}(\theta-i\tau)=\cos\frac{\theta}{2}\ch\frac{\tau}{2}+
i\sin\frac{\theta}{2}\sh\frac{\tau}{2}$, and 
$\sin\frac{1}{2}(\theta-i\tau)=\sin\frac{\theta}{2}\ch\frac{\tau}{2}-
i\cos\frac{\theta}{2}\sh\frac{\tau}{2}$. It is easy to verify that the
matrix (\ref{SL1}) coincides with a matrix of the fundamental reprsentation
of the group $SL(2,\C)$ (in Euler parametrization):
\begin{multline}
\mathfrak{g}(\varphi,\,\epsilon,\,\theta,\,\tau,\,\psi,\,\varepsilon)=\\[0.2cm]
\begin{pmatrix}
e^{i\frac{\varphi}{2}} & 0\\
0 & e^{-i\frac{\varphi}{2}}
\end{pmatrix}{\renewcommand{\arraystretch}{1.1}\!\!\!\begin{pmatrix}
e^{\frac{\epsilon}{2}} & 0\\
0 & e^{-\frac{\epsilon}{2}}
\end{pmatrix}}\!\!\!\!{\renewcommand{\arraystretch}{1.3}\begin{pmatrix}
\cos\frac{\theta}{2} & i\sin\frac{\theta}{2}\\
i\sin\frac{\theta}{2} & \cos\frac{\theta}{2}
\end{pmatrix}\!\!\!\!
\begin{pmatrix}
\ch\frac{\tau}{2} & \sh\frac{\tau}{2}\\
\sh\frac{\tau}{2} & \ch\frac{\tau}{2}
\end{pmatrix}}\!\!\!{\renewcommand{\arraystretch}{1.1}\begin{pmatrix}
e^{i\frac{\psi}{2}} & 0\\
0 & e^{-i\frac{\psi}{2}}
\end{pmatrix}}\!\!\!
\begin{pmatrix}
e^{\frac{\varepsilon}{2}} & 0\\
0 & e^{-\frac{\varepsilon}{2}}
\end{pmatrix}.\label{FUN}
\end{multline}
\begin{sloppypar}
Moreover, the complexification of $SU(2)$ gives us the most simple and
direct way for calculation of matrix elements\index{element!matrix}
of the Lorentz group.
It is known that these elements have a great importance in quantum
field theory and widely used at the study of relativistic amplitudes.
\end{sloppypar}
\begin{sloppypar}
The matrix element $t^l_{mn}=e^{-m(\epsilon+i\varphi)-n(\varepsilon+
i\psi)}\langle T_l(\theta,\tau)\psi_\lambda,\psi_{\dot{\lambda}}\rangle$
of the finite--dimensional repsesentation of $SL(2,\C)$ at $l=\dot{l}$
in the polynomial basis\end{sloppypar}
\[
\psi_\lambda(\fz,\bar{\fz})=\frac{\fz^{l-n}\bar{\fz}^{l-m}}
{\sqrt{\Gamma(l-n+1)\Gamma(l+n+1)\Gamma(l-m+1)\Gamma(l+m+1)}},
\]
has a form
\begin{multline}
t^l_{mn}(\mathfrak{g})=e^{-m(\epsilon+i\varphi)-n(\varepsilon+i\psi)}
Z^l_{mn}=e^{-m(\epsilon+i\varphi)-n(\varepsilon+i\psi)}\times\\[0.2cm]
\sum^l_{k=-l}i^{m-k}
\sqrt{\Gamma(l-m+1)\Gamma(l+m+1)\Gamma(l-k+1)\Gamma(l+k+1)}\times\\
\cos^{2l}\frac{\theta}{2}\tg^{m-k}\frac{\theta}{2}\times\\[0.2cm]
\sum^{\min(l-m,l+k)}_{j=\max(0,k-m)}
\frac{i^{2j}\tg^{2j}\dfrac{\theta}{2}}
{\Gamma(j+1)\Gamma(l-m-j+1)\Gamma(l+k-j+1)\Gamma(m-k+j+1)}\times\\[0.2cm]
\sqrt{\Gamma(l-n+1)\Gamma(l+n+1)\Gamma(l-k+1)\Gamma(l+k+1)}
\ch^{2l}\frac{\tau}{2}\tnh^{n-k}\frac{\tau}{2}\times\\[0.2cm]
\sum^{\min(l-n,l+k)}_{s=\max(0,k-n)}
\frac{\tnh^{2s}\dfrac{\tau}{2}}
{\Gamma(s+1)\Gamma(l-n-s+1)\Gamma(l+k-s+1)\Gamma(n-k+s+1)}.\label{HS}
\end{multline}
We will call the functions $Z^l_{mn}$ in (\ref{HS}) as
{\it hyperspherical functions}.\index{function!hyperspherical}
Using (\ref{HG}), we can write the
hyperspherical functions $Z^l_{mn}$ via the hypergeometric series:
\begin{multline}
Z^l_{mn}=\cos^{2l}\frac{\theta}{2}\ch^{2l}\frac{\tau}{2}
\sum^l_{k=-l}i^{m-k}\tg^{m-k}\frac{\theta}{2}
\tnh^{n-k}\frac{\tau}{2}\times\\[0.2cm]
\hypergeom{2}{1}{m-l+1,1-l-k}{m-k+1}{i^2\tg^2\dfrac{\theta}{2}}
\hypergeom{2}{1}{n-l+1,1-l-k}{n-k+1}{\tnh^2\dfrac{\tau}{2}}.\label{HS1}
\end{multline}
Therefore, matrix elements can be expressed by means of the function
({\it a generalized 
hyperspherical function})\index{function!hyperspherical!generalized}
\begin{equation}\label{HS2}
\fT^l_{mn}(\mathfrak{g})=e^{-m(\epsilon+i\varphi)}Z^l_{mn}
e^{-n(\varepsilon+i\psi)},
\end{equation}
where
\begin{equation}\label{HS3}
Z^l_{mn}(\cos\theta^c)=
\sum^l_{k=-l}P^l_{mk}(\cos\theta)\mathfrak{P}^l_{kn}(\ch\tau),
\end{equation}
here $P^l_{mn}(\cos\theta)$ is a 
generalized spherical 
function\index{function!spherical!generalized} on the
group $SU(2)$ (see \cite{GMS}), and $\mathfrak{P}^l_{mn}$ is an analog of
the generalized spherical function for the group $QU(2)$ (so--called
Jacobi function\index{function!Jacobi}
\cite{Vil68}). $QU(2)$ is a group of quasiunitary
unimodular matrices of second order. As well as the group $SU(2)$, the
group $QU(2)$ is one of the real forms of $SL(2,\C)$
($QU(2)$ is noncompact).

\subsection{Appell functions}
From (\ref{HS1}) we see that the function $Z^l_{mn}$ depends on
two variables $\theta$ and $\tau$. Therefore, using Bateman factorization we can
express the hyperspherical functions $Z^l_{mn}$ via Appell functions
$F_1$--$F_4$ (hypergeometric series of two variables \cite{AK26,Bat}).
Appell functions are defined by the following expressions:
\begin{eqnarray}
F_1(\alpha,\beta,\beta^\prime;\gamma;x,y)&=&\sum^\infty_{m,n}
\frac{(\alpha)_{m+n}(\beta)_m(\beta^\prime)_n}{(\gamma)_{m+n}m!n!}
x^my^n,\nonumber\\
F_2(\alpha,\beta,\beta^\prime;\gamma,\gamma^\prime;x,y)&=&\sum^\infty_{m,n}
\frac{(\alpha)_m(\beta)_m(\beta^\prime)_n}
{(\gamma)_m(\gamma^\prime)_nm!n!}x^my^n,\nonumber\\
F_3(\alpha,\alpha^\prime,\beta,\beta^\prime;\gamma;x,y)&=&\sum^\infty_{m,n}
\frac{(\alpha)_m(\alpha^\prime)_n(\beta)_m(\beta^\prime)_n}
{(\gamma)_{m+n}m!n!}x^my^n,\nonumber\\
F_4(\alpha,\beta;\gamma,\gamma^\prime;x,y)&=&\sum^\infty_{m,n}
\frac{(\alpha)_{m+n}(\beta)_{m+n}}
{(\gamma)_m(\gamma^\prime)_nm!n!}x^my^n,\nonumber
\end{eqnarray}
where $(a)_n=\Gamma(a+n)/\Gamma(a)$. For example, for the function $F_4$
there exists a following factorization (see \cite{BC41,F4})
\begin{multline}
\hypergeom{2}{1}{\alpha,\beta}{\gamma}{x}
\hypergeom{2}{1}{\alpha,\beta}{\gamma^\prime}{y}=\\
=\sum^\infty_{r=0}\frac{(\alpha)_r(\beta)_r(\gamma+\gamma^\prime-\alpha
-\beta-1)_r}{r!(\gamma)_r(\gamma^\prime)_r}x^ry^r\times\\
\times F_4(\alpha+r,\beta+r;\gamma+r,\gamma^\prime+r;x-xy,y-xy).\nonumber
\end{multline}
Then, using this formula we can express the hyperspherical functions
$Z^l_{mn}$, defining the matrix elements of the main diagonal ($m=n$), via
the Appell function $F_4$:
\begin{multline}
Z^l_{mm}=\cos^{2l}\frac{\theta}{2}\ch^{2l}\frac{\tau}{2}
\sum^l_{k=-l}\sum_{r\ge 0}i^{m-k}(-1)^r\times\\
\frac{(m-l+1)_r(1-l-k)_r(m-k+2l-1)_r}{\Gamma(r+1)(m-k+1)^2_r}
\tg^{m-k}\frac{\theta}{2}\tnh^{m-k}\frac{\tau}{2}\tg^{2r}\frac{\theta}{2}
\tnh^{2r}\frac{\tau}{2}\times\\
F_4\left(m-l+r+1,1-l-k+r;m-k+r+1,m-k+r+1;-\tg^2\frac{\theta}{2}+\right.\\
+\left.\tg^2\frac{\theta}{2}\tnh^2\frac{\tau}{2},\tnh^2\frac{\tau}{2}+
\tg^2\frac{\theta}{2}\tnh^2\frac{\tau}{2}\right).\label{HF3}
\end{multline}
Since $l$ and $m$ are finite numbers, then the series standing in the right
part of (\ref{HF3}) is interrupted.

In turn, for the function $F_2$ there exists a following decomposition
\cite{BC41,Bat}:
\begin{multline}
\hypergeom{2}{1}{\alpha,\beta}{\gamma}{x}
\hypergeom{2}{1}{\alpha,\beta^\prime}{\gamma^\prime}{y}=\\
=\sum^\infty_{r=0}(-1)^r\frac{(\alpha)_r(\beta)_r(\beta^\prime)_r}
{r!(\gamma)_r(\gamma^\prime)_r}x^ry^r\times\\
\times F_2(\alpha+r,\beta+r,\beta^\prime+r;\gamma+r,\gamma^\prime+r;x,y).
\nonumber
\end{multline}
Thus, the hyperspherical functions of the main diagonal are expressed via
the Appell function $F_2$ as follows:
\begin{multline}
Z^l_{mm}=\cos^{2l}\frac{\theta}{2}\ch^{2l}\frac{\tau}{2}\sum^l_{k=-l}
\sum_{r\ge 0}i^{m-k}\times\\
\frac{(m-l+1)_r(1-l-k)^2_r}{\Gamma(r+1)(m-k+1)^2_r}\tg^{m-k}\frac{\theta}{2}
\tnh^{m-k}\frac{\tau}{2}\tg^{2r}\frac{\theta}{2}\tnh^{2r}\frac{\tau}{2}\times\\
F_2\left(m-l+r+1,1-l-k+r,1-l-k+r;\right.\\
\left. m-k+r+1,m-k+r+1;-\tg^2\frac{\theta}{2},
\tnh^2\frac{\tau}{2}\right).\label{HF4}
\end{multline}

In \cite{BC41}, Burchnall and Chaundy obtained 15 pairs of decompositions
containing the Appell functions and usual hypergeometric functions.
A detailed consideration of the relationship between the hyperspherical
functions $Z^l_{mn}$ and the Appell functions $F_1-F_4$ (and also
Horn functions \cite{Hor31}) comes beyond the framework of the present
work and will be considered in a future paper.
\subsection{Integral representations of $Z^l_{mn}(\theta,\tau)$}
Let us choose in the space $\fD_\chi$ of the representation $T_\chi(\fg)$ of
$SU(2)\otimes SU(2)$ the basis consisting of the functions
$\lf e^{-in\vartheta}\rf$, then
\begin{equation}\label{Representation}
T_\chi(\fg)f(e^{i\vartheta})=(\alpha e^{i\vartheta}+\gamma)^{l+o}
(\beta e^{i\vartheta}+\delta)^{l-o}f\left(\frac{\alpha e^{i\vartheta}+\gamma}
{\beta e^{i\vartheta}+\delta}\right).
\end{equation}
Therefore,
\[
T_\chi(\fg)e^{-in\vartheta}=(\alpha e^{i\vartheta}+\gamma)^{l-n-o}
(\beta e^{i\vartheta}+\delta)^{l+n+o}e^{-in\vartheta}.
\]
The matrix elements $t^\chi_{mn}(\fg)$ of the representation $T_\chi(\fg)$
in the basis $\lf e^{-in\vartheta}\rf$ are Fourier coefficients in the
decomposition of the functions $T_\chi(\fg)e^{-in\vartheta}$ on the system
$\lf e^{-im\vartheta}\rf$:
\[
T_\chi(\fg)e^{-in\vartheta}=\sum^\infty_{m=-\infty}t^\chi_{mn}(\fg)
e^{-im\vartheta}.
\]
From the formula for the Fourier coefficients we obtain
\begin{equation}\label{Integral}
t^\chi_{mn}(\fg)=\frac{1}{2\pi}\int\limits^{2\pi}_0(\alpha e^{i\vartheta}+
\gamma)^{l-n-o}(\beta e^{i\vartheta}+\delta)^{l+n+o}e^{i(m-n)\vartheta}
d\vartheta.
\end{equation}
In such a way, we have an integral representation for the matrix elements.
Making the substitution $e^{i\vartheta}=z$ we come to the following
representation:
\begin{equation}\label{Integral2}
t^\chi_{mn}(\fg)=\frac{1}{2\pi i}\oint\limits_\Gamma(\alpha z+\gamma)^{l-n-o}
(\beta z+\delta)^{l+n+o}z^{m-l+o-1}dz,
\end{equation}
where $\Gamma:\;|z|=1$ is an unit circle. In the basis 
$\lf e^{-im\vartheta}\rf$ the matrices of the operators $T_\chi(h)$ have
a very simple form, where
\[
h=\begin{pmatrix}
e^{\frac{it}{2}} & 0\\
0 & e^{-\frac{it}{2}}
\end{pmatrix}
\]
is the diagonal matrix. Then from the formula (\ref{Representation}) we
see that the operator $T_\chi(h)$ has a form
\[
T_\chi(h)f(e^{i\vartheta})=e^{-i\vartheta}f(e^{i(\vartheta+t)})
\]
and, therefore,
\[
T_\chi(h)e^{-im\vartheta}=e^{-i(m+o)t}e^{-im\vartheta}.
\]
Thus, the matrix of $T_\chi(h)$ in the basis $\lf e^{-im\vartheta}\rf$
is an infinite diagonal matrix, on the main diagonal of which we have the
numbers $e^{-i(m+o)t}$.

As it has been shown previously (see (\ref{FUN})), the element $\fg$ of the
fundamental representation of $SU(2)\otimes SU(2)$ has a form
\[
\fg(\varphi^c,\theta^c,\psi^c)=\begin{pmatrix}
e^{\frac{i\varphi^c}{2}} & 0\\
0 & e^{-\frac{i\varphi^c}{2}}
\end{pmatrix}\!{\renewcommand{\arraystretch}{1.3}\begin{pmatrix}
\cos\frac{\theta^c}{2} & i\sin\frac{\theta^c}{2}\\
i\sin\frac{\theta^c}{2} & \cos\frac{\theta^c}{2}
\end{pmatrix}}\!\begin{pmatrix}
e^{\frac{i\psi^c}{2}} & 0\\
0 & e^{-\frac{i\psi^c}{2}}
\end{pmatrix},
\]
where $\varphi^c$, $\theta^c$, $\psi^c$ are the complex Euler angles.
Since the matrix $T_\chi(\fg)$ for the diagonal matrices has been found, then
it is remain to find a matrix of the operator $T_\chi(\fg(\theta,\tau))$
corresponding to the element
\[
\fg(\theta,\tau)={\renewcommand{\arraystretch}{1.3}
\begin{pmatrix}
\cos\frac{\theta^c}{2} & i\sin\frac{\theta^c}{2}\\
i\sin\frac{\theta^c}{2} & \cos\frac{\theta^c}{2}
\end{pmatrix}=\begin{pmatrix}
\cos\frac{\theta}{2}\ch\frac{\tau}{2}+i\sin\frac{\theta}{2}\sh\frac{\tau}{2} &
\cos\frac{\theta}{2}\sh\frac{\tau}{2}+i\sin\frac{\theta}{2}\ch\frac{\tau}{2}\\
\cos\frac{\theta}{2}\sh\frac{\tau}{2}+i\sin\frac{\theta}{2}\ch\frac{\tau}{2} &
\cos\frac{\theta}{2}\ch\frac{\tau}{2}+i\sin\frac{\theta}{2}\sh\frac{\tau}{2}
\end{pmatrix}}.
\]
Then from the integral representation (\ref{Integral}) we obtain
\[
t^\chi_{mn}(\fg(\theta,\tau))=\frac{1}{2\pi}\int\limits^{2\pi}_0
\left(\cos\frac{\theta^c}{2}e^{i\vartheta}+
i\sin\frac{\theta^c}{2}\right)^{l-n-o}
\left(i\sin\frac{\theta^c}{2}e^{i\vartheta}+
\cos\frac{\theta^c}{2}\right)^{l+n+o}e^{i(m-n)\vartheta}d\vartheta.
\]
Therefore, an integral representation for the hyperspherical function
$Z^l_{mn}(\theta,\tau)$ has a form
\begin{multline}
Z^l_{mn}(\theta,\tau)=\frac{1}{2\pi}\int\limits^{2\pi}_0
\left(\cos\frac{\theta^c}{2}e^{i\vartheta}+
i\sin\frac{\theta^c}{2}\right)^{l-n}\left(i\sin\frac{\theta^c}{2}e^{i\vartheta}
+\cos\frac{\theta^c}{2}\right)^{l+n}e^{i(m-n)\vartheta}d\vartheta=\\
\frac{1}{2\pi}\int\limits^{2\pi}_0\left(\left[
\cos\frac{\theta}{2}\ch\frac{\tau}{2}+
i\sin\frac{\theta}{2}\sh\frac{\tau}{2}\right]
e^{i\vartheta}+\cos\frac{\theta}{2}\sh\frac{\tau}{2}+
i\sin\frac{\theta}{2}\ch\frac{\tau}{2}\right)^{l-n}\times\\
\left(\left[\cos\frac{\theta}{2}\sh\frac{\tau}{2}+
i\sin\frac{\theta}{2}\ch\frac{\tau}{2}\right]e^{i\vartheta}+
\cos\frac{\theta}{2}\ch\frac{\tau}{2}+
i\sin\frac{\theta}{2}\sh\frac{\tau}{2}\right)^{l+n}
e^{i(m-n)\vartheta}d\vartheta.\label{Integral3}
\end{multline}
Or, using the integral representation (\ref{Integral2}), we obtain
\begin{multline}
Z^l_{mn}(\theta,\tau)=\frac{1}{2\pi i}\oint\limits_\Gamma
\left(\cos\frac{\theta^c}{2}z+i\sin\frac{\theta^c}{2}\right)^{l-n}
\left(i\sin\frac{\theta^c}{2}z+\cos\frac{\theta^c}{2}\right)^{l+n}
z^{m-l-1}dz=\\
\frac{1}{2\pi i}\oint\limits_\Gamma
\left(\left[\cos\frac{\theta}{2}\ch\frac{\tau}{2}+
i\sin\frac{\theta}{2}\sh\frac{\tau}{2}\right]z+
\cos\frac{\theta}{2}\sh\frac{\tau}{2}+
i\sin\frac{\theta}{2}\ch\frac{\tau}{2}\right)^{l-n}\times\\
\left(\left[\cos\frac{\theta}{2}\sh\frac{\tau}{2}+
i\sin\frac{\theta}{2}\ch\frac{\tau}{2}\right]z+
\cos\frac{\theta}{2}\ch\frac{\tau}{2}+
i\sin\frac{\theta}{2}\sh\frac{\tau}{2}\right)^{l+n}z^{m-l-1}dz.
\label{Integral4}
\end{multline}
\subsection{Zonal hyperspherical functions}\label{Zonal}
Let $o=0$, then in the space $\fD_l$ there is a function which
invariant with respect to all the operators $T_l(h)$, where $h$ is
the diagonal matrix from the group $SL(2,\C)\sim SU(2)\otimes SU(2)$:
\[
h=\begin{pmatrix}
e^{\frac{it}{2}} & 0\\
0 & e^{-\frac{it}{2}}
\end{pmatrix}.
\]
For example, the basis element 1 is such a function. The matrix element
$t^l_{00}(\fg)$, corresponding to such a basis element, we will call
{\it a zonal hyperspherical function} of the representation $T_l(\fg)$
with respect to the subgroup $\Omega$ of the diagonal matrices.

So, at $o=0$ the representation $T_l(\fg)$ has a zonal hyperspherical
function satisfying the following equation
\[
t^l_{00}(h_1\fg h_2)=t^l_{00}(\fg),
\]
where $h_1,h_2\in\Omega$. Hence it immediately follows that
$t^l_{00}(\fg)$ depends only on the Euler angles $\theta$ and $\tau$.
From (\ref{HS}) we obtain
\[
t^l_{00}(\fg)=Z^l_{00}(\theta,\tau).
\]
We will denote the function $Z^l_{00}(\theta,\tau)$ via $Z_l(\theta,\tau)$.
Thus,
\[
Z_l(\theta,\tau)=t^l_{00}(0,0,\theta,\tau,0,0)=Z^l_{00}(\theta,\tau).
\]
From (\ref{HS}) it follows an explicit expression for the zonal
hypersperical function:
\begin{multline}
Z_l(\theta,\tau)=
\sum^l_{k=-l}i^{-k}
\Gamma(l+1)\sqrt{\Gamma(l-k+1)\Gamma(l+k+1)}\times\\
\cos^{2l}\frac{\theta}{2}\tg^{-k}\frac{\theta}{2}\times\\[0.2cm]
\sum^{\min(l,l+k)}_{j=\max(0,k)}
\frac{i^{2j}\tg^{2j}\dfrac{\theta}{2}}
{\Gamma(j+1)\Gamma(l-j+1)\Gamma(l+k-j+1)\Gamma(j-k+1)}\times\\[0.2cm]
\Gamma(l+1)\sqrt{\Gamma(l-k+1)\Gamma(l+k+1)}
\ch^{2l}\frac{\tau}{2}\tnh^{-k}\frac{\tau}{2}\times\\[0.2cm]
\sum^{\min(l,l+k)}_{s=\max(0,k)}
\frac{\tnh^{2s}\dfrac{\tau}{2}}
{\Gamma(s+1)\Gamma(l-s+1)\Gamma(l+k-s+1)\Gamma(s-k+1)}.\nonumber
\end{multline}
Further, from (\ref{HS1}), (\ref{HF3}) and (\ref{HF4}) we obtain
corresponding expressions for $Z_l(\theta,\tau)$ via the hypergeometric
function ${}_2F_1$ and the Appell functions $F_4$ and $F_2$:
\begin{multline}
Z^l_{mn}=\cos^{2l}\frac{\theta}{2}\ch^{2l}\frac{\tau}{2}
\sum^l_{k=-l}i^{-k}\tg^{-k}\frac{\theta}{2}
\tnh^{-k}\frac{\tau}{2}\times\\[0.2cm]
\hypergeom{2}{1}{-l+1,1-l-k}{-k+1}{i^2\tg^2\dfrac{\theta}{2}}
\hypergeom{2}{1}{-l+1,1-l-k}{-k+1}{\tnh^2\dfrac{\tau}{2}},\nonumber
\end{multline}
\begin{multline}
Z^l_{mm}=\cos^{2l}\frac{\theta}{2}\ch^{2l}\frac{\tau}{2}
\sum^l_{k=-l}\sum_{r\ge 0}i^{-k}(-1)^r\times\\
\frac{(-l+1)_r(1-l-k)_r(2l-k-1)_r}{\Gamma(r+1)(-k+1)^2_r}
\tg^{-k}\frac{\theta}{2}\tnh^{-k}\frac{\tau}{2}\tg^{2r}\frac{\theta}{2}
\tnh^{2r}\frac{\tau}{2}\times\\
F_4\left(r-l+1,1-l-k+r;r-k+1,r-k+1;-\tg^2\frac{\theta}{2}+\right.\\
+\left.\tg^2\frac{\theta}{2}\tnh^2\frac{\tau}{2},\tnh^2\frac{\tau}{2}+
\tg^2\frac{\theta}{2}\tnh^2\frac{\tau}{2}\right),\nonumber
\end{multline}
\begin{multline}
Z^l_{mm}=\cos^{2l}\frac{\theta}{2}\ch^{2l}\frac{\tau}{2}\sum^l_{k=-l}
\sum_{r\ge 0}i^{-k}\times\\
\frac{(-l+1)_r(1-l-k)^2_r}{\Gamma(r+1)(-k+1)^2_r}\tg^{-k}\frac{\theta}{2}
\tnh^{-k}\frac{\tau}{2}\tg^{2r}\frac{\theta}{2}\tnh^{2r}\frac{\tau}{2}\times\\
F_2\left(r-l+1,1-l-k+r,1-l-k+r;r-k+1,r-k+1;-\tg^2\frac{\theta}{2},
\tnh^2\frac{\tau}{2}\right).\nonumber
\end{multline}
Integral representations (\ref{Integral3}) and (\ref{Integral4}) have
in this case the following form
\begin{multline}
Z_l(\theta,\tau)=\frac{1}{2\pi}\int\limits^{2\pi}_0
\left(\cos\frac{\theta^c}{2}e^{i\vartheta}+
i\sin\frac{\theta^c}{2}\right)^{l}\left(i\sin\frac{\theta^c}{2}e^{i\vartheta}
+\cos\frac{\theta^c}{2}\right)^{l}d\vartheta=\\
\frac{1}{2\pi}\int\limits^{2\pi}_0
(\cos\theta^c+i\sin\theta^c\cos\vartheta)d\vartheta=\\
\frac{1}{2\pi}\int\limits^{2\pi}_0(\cos\theta\ch\tau+i\sin\theta\sh\tau+
i[\sin\theta\ch\tau-i\cos\theta\sh\tau]\cos\vartheta)d\vartheta,\nonumber
\end{multline}
\begin{multline}
Z_l(\theta,\tau)=\frac{1}{2\pi i}\oint\limits_\Gamma
\left(\cos\frac{\theta^c}{2}z+i\sin\frac{\theta^c}{2}\right)^l
\left(i\sin\frac{\theta^c}{2}z+\cos\frac{\theta^c}{2}\right)^lz^{-l-1}dz=\\
\frac{i^l\sin^l\theta^c}{2^{l+1}\pi i}
\oint\limits_\Gamma(z^2-2iz\ctg\theta^c+1)^lz^{-l-1}dz=\\
\frac{i^l(\sin\theta\ch\tau-i\cos\theta\sh\tau)^l}{2^{l+1}\pi i}
\oint\limits_\Gamma(z^2-2iz\frac{\ctg\theta\cth\tau+1}{\cth\tau-\ctg\theta}
+1)z^{-l-1}dz.\nonumber
\end{multline}
\subsection{Associated hyperspherical functions}\label{Associated}
Let us consider now associated hyperspherical functions of the
representation $T_\chi(\fg)$, $\chi=(l,0)$, that is, the matrix elements
$t^l_{m0}(\fg)$ standing in one column with the function $t^l_{00}(\fg)$.
In this case from (\ref{HS}) we have
\[
t^l_{m0}(\fg)=e^{-m(\epsilon+i\varphi)}Z^l_{m0}(\theta,\tau).
\]
Hence it follows that matrix elements $t^l_{m0}(\fg)$ do not depend on
the Euler angles $\varepsilon$ and $\psi$, that is, $t^l_{m0}(\fg)$
are constant on the left adjacency classes formed by the subgroup
$\Omega^c_\psi$ of the diagonal matrices
$\begin{pmatrix} e^{\frac{i\psi^c}{2}} & 0\\ 0 & e^{-\frac{i\psi^c}{2}}
\end{pmatrix}$. Therefore,
\[
t^l_{m0}(\fg h)=t^l_{m0}(\fg),\quad h\in\Omega^c_\psi.
\]
We will denote the functions $Z^{m0}(\theta,\tau)$ via $Z^m_l(\theta,\tau)$.
From (\ref{HS}) we obtain an explicit expression for the
{\it associated hyperspherical function} $Z^m_l(\theta,\tau)$:
\begin{multline}
Z^m_l(\theta,\tau)=
\sum^l_{k=-l}i^{m-k}
\sqrt{\Gamma(l-m+1)\Gamma(l+m+1)\Gamma(l-k+1)\Gamma(l+k+1)}\times\\
\cos^{2l}\frac{\theta}{2}\tg^{m-k}\frac{\theta}{2}\times\\[0.2cm]
\sum^{\min(l-m,l+k)}_{j=\max(0,k-m)}
\frac{i^{2j}\tg^{2j}\dfrac{\theta}{2}}
{\Gamma(j+1)\Gamma(l-m-j+1)\Gamma(l+k-j+1)\Gamma(m-k+j+1)}\times\\[0.2cm]
\Gamma(l+1)\sqrt{\Gamma(l-k+1)\Gamma(l+k+1)}
\ch^{2l}\frac{\tau}{2}\tnh^{-k}\frac{\tau}{2}\times\\[0.2cm]
\sum^{\min(l,l+k)}_{s=\max(0,k)}
\frac{\tnh^{2s}\dfrac{\tau}{2}}
{\Gamma(s+1)\Gamma(l-s+1)\Gamma(l+k-s+1)\Gamma(s-k+1)}.\label{AHS}
\end{multline}
Further, from (\ref{HS1}) it follows that
\begin{multline}
Z^l_{mn}=\cos^{2l}\frac{\theta}{2}\ch^{2l}\frac{\tau}{2}
\sum^l_{k=-l}i^{m-k}\tg^{m-k}\frac{\theta}{2}
\tnh^{-k}\frac{\tau}{2}\times\\[0.2cm]
\hypergeom{2}{1}{m-l+1,1-l-k}{m-k+1}{i^2\tg^2\dfrac{\theta}{2}}
\hypergeom{2}{1}{-l+1,1-l-k}{-k+1}{\tnh^2\dfrac{\tau}{2}}.\label{AHS1}
\end{multline}
Corresponding integral representations for the functions $Z^m_l(\theta,\tau)$
have the form:
\begin{multline}
Z^m_l(\theta,\tau)=\frac{1}{2\pi}\int\limits^{2\pi}_0
(\cos\theta^c+i\sin\theta^c\cos\vartheta)e^{im\vartheta}d\vartheta=\\
\frac{1}{2\pi}\int\limits^{2\pi}_0(\cos\theta\ch\tau+i\sin\theta\sh\tau+
i[\sin\theta\ch\tau-i\cos\theta\sh\tau]\cos\vartheta)e^{im\vartheta}
d\vartheta,\nonumber
\end{multline}
\begin{multline}
Z^m_l(\theta,\tau)=\frac{i^l\sin^l\theta^c}{2^{l+1}\pi i}
\oint\limits_\Gamma(z^2-2iz\ctg\theta^c+1)^lz^{m-l-1}dz=\\
\frac{i^l(\sin\theta\ch\tau-i\cos\theta\sh\tau)^l}{2^{l+1}\pi i}
\oint\limits_\Gamma(z^2-2iz\frac{\ctg\theta\cth\tau+1}{\cth\tau-\ctg\theta}
+1)z^{m-l-1}dz.\nonumber
\end{multline}
Associated hyperspherical functions admit a very elegant geometric
interpretation, namely, they are the functions on the surface of a
two-dimensional complex sphere. 
Indeed, let us construct in $\C^3$ a two--dimensional complex sphere from the
quantities $z_k=x_k+iy_k$, $\overset{\ast}{z}_k=x_k-iy_k$
as follows (see Figure 1)
\begin{equation}\label{CS}
\bz^2=z^2_1+z^2_2+z^2_3=\bx^2-\by^2+2i\bx\by=r^2
\end{equation}
and its complex conjugate (dual) sphere
\begin{equation}\label{DS}
\overset{\ast}{\bz}{}^2=\overset{\ast}{z}_1{}^2+\overset{\ast}{z}_2{}^2+
\overset{\ast}{z}_3{}^2=\bx^2-\by^2-2i\bx\by=\overset{\ast}{r}{}^2.
\end{equation}
For more details about the two-dimensional complex sphere see
\cite{Hus70,HS70,SH70}.
It is well-known that both quantities $\bx^2-\by^2$, $\bx\by$ are
invariant with respect to the Lorentz transformations, since a surface of
the complex sphere is invariant 
(Casimir operators\index{operator!Casimir} of the Lorentz group are
constructed from such quantities, see also (\ref{KO})).
It is easy to see that three--dimensional complex space $\C^3$ is
isometric to a real space $\R^{3,3}$ with a basis
$\left\{i\e_1,i\e_2,i\e_3,\e_4,\e_5,\e_6\right\}$. At this point
a metric tensor of $\R^{3,3}$ has the form
\[
g_{ab}=\begin{pmatrix}
-1 & 0 & 0 & 0 & 0 & 0\\
0  &-1 & 0 & 0 & 0 & 0\\
0  & 0 &-1 & 0 & 0 & 0\\
0  & 0 & 0 & 1 & 0 & 0\\
0  & 0 & 0 & 0 & 1 & 0\\
0  & 0 & 0 & 0 & 0 & 1
\end{pmatrix}.
\] 
Hence it
immediately follows that $\C^3$ is isometric to 
the bivector space\index{space!bivector} $\R^6$.
\begin{figure}
\[
\unitlength=0.3mm
\begin{picture}(100.00,175.00)(0,25)
\put(50,50){\vector(0,1){51}}
\put(50,150){\vector(0,-1){48}}
\put(47,49.5){$\bullet$}
\put(47,150){$\bullet$}
\put(54,75){$r^\ast$}
\put(54,125){$r$}
\put(58,13){$\bullet$}
\put(58,183){$\bullet$}
\put(60,15){\vector(1,1){23}}
\put(60,15){\vector(1,-1){23}}
\put(60,185){\vector(1,1){23}}
\put(60,185){\vector(1,-1){23}}
\put(46,10){$P^\prime$}
\put(46,180){$P$}
\put(-95,50){$\mathfrak{M}_{\dot{l}}^{\dot{m}}
(\dot{\varphi}^c,\dot{\theta}^c,0)$}
\put(-95,150){$\mathfrak{M}_{l}^{m}(\varphi^c,\theta^c,0)$}
\put(85,-12){$\boldsymbol{e}_{\dot{\varphi}^c}$}
\put(80,210){$\boldsymbol{e}_{\varphi^c}$}
\put(80,30){$\boldsymbol{e}_{\dot{\theta}^c}$}
\put(80,170){$\boldsymbol{e}_{\theta^c}$}
\put(105,18){$\dot{\varphi}^c=\varphi+i\epsilon$}
\put(105,2){$\dot{\theta}^c=\theta+i\tau$}
\put(105,198){$\varphi^c=\varphi-i\epsilon$}
\put(105,182){$\theta^c=\theta-i\tau$}
\put(50,34){$\cdot$}
\put(49.5,34){$\cdot$}
\put(49,34){$\cdot$}
\put(48.5,34){$\cdot$}
\put(48,34.01){$\cdot$}
\put(47.5,34.02){$\cdot$}
\put(47,34.03){$\cdot$}
\put(46.5,34.04){$\cdot$}
\put(46,34.05){$\cdot$}
\put(45.5,34.06){$\cdot$}
\put(45,34.08){$\cdot$}
\put(44.5,34.10){$\cdot$}
\put(44,34.11){$\cdot$}
\put(43.5,34.13){$\cdot$}
\put(43,34.16){$\cdot$}
\put(42.5,34.18){$\cdot$}
\put(42,34.20){$\cdot$}
\put(41.5,34.23){$\cdot$}
\put(41,34.26){$\cdot$}
\put(40.5,34.29){$\cdot$}
\put(40,34.32){$\cdot$}
\put(39.5,34.36){$\cdot$}
\put(39,34.39){$\cdot$}
\put(38.5,34.43){$\cdot$}
\put(38,34.47){$\cdot$}
\put(37.5,34.51){$\cdot$}
\put(37,34.55){$\cdot$}
\put(36.5,34.59){$\cdot$}
\put(36,34.64){$\cdot$}
\put(35.5,34.69){$\cdot$}
\put(35,34.74){$\cdot$}
\put(34.5,34.79){$\cdot$}
\put(34,34.84){$\cdot$}
\put(33.5,34.90){$\cdot$}
\put(33,34.95){$\cdot$}
\put(32.5,35.01){$\cdot$}
\put(32,35.07){$\cdot$}
\put(31.5,35.13){$\cdot$}
\put(31,35.20){$\cdot$}
\put(30.5,35.27){$\cdot$}
\put(30,35.33){$\cdot$}
\put(29.5,35.40){$\cdot$}
\put(29,35.48){$\cdot$}
\put(28.5,35.55){$\cdot$}
\put(28,35.63){$\cdot$}
\put(27.5,35.71){$\cdot$}
\put(27,35.79){$\cdot$}
\put(26.5,35.88){$\cdot$}
\put(26,35.96){$\cdot$}
\put(25.5,36.05){$\cdot$}
\put(25,36.14){$\cdot$}
\put(24.5,36.24){$\cdot$}
\put(24,36.33){$\cdot$}
\put(23.5,36.43){$\cdot$}
\put(23,36.53){$\cdot$}
\put(22.5,36.64){$\cdot$}
\put(22,36.74){$\cdot$}
\put(21.5,36.85){$\cdot$}
\put(21,36.97){$\cdot$}
\put(20.5,37.08){$\cdot$}
\put(20,37.20){$\cdot$}
\put(19.5,37.32){$\cdot$}
\put(19,37.45){$\cdot$}
\put(18.5,37.57){$\cdot$}
\put(18,37.70){$\cdot$}
\put(17.5,37.84){$\cdot$}
\put(17,37.98){$\cdot$}
\put(16.5,38.12){$\cdot$}
\put(16,38.27){$\cdot$}
\put(15.5,38.42){$\cdot$}
\put(15,38.57){$\cdot$}
\put(14.5,38.73){$\cdot$}
\put(14,38.90){$\cdot$}
\put(13.5,39.06){$\cdot$}
\put(13,39.24){$\cdot$}
\put(12.5,39.42){$\cdot$}
\put(12,39.60){$\cdot$}
\put(11.5,39.79){$\cdot$}
\put(11,39.99){$\cdot$}
\put(10.5,40.19){$\cdot$}
\put(10,40.40){$\cdot$}
\put(9.5,40.62){$\cdot$}
\put(9,40.84){$\cdot$}
\put(8.5,41.07){$\cdot$}
\put(8,41.32){$\cdot$}
\put(7.5,41.57){$\cdot$}
\put(7,41.83){$\cdot$}
\put(6.5,42.11){$\cdot$}
\put(6,42.40){$\cdot$}
\put(5.5,42.70){$\cdot$}
\put(5,43.02){$\cdot$}
\put(4.5,43.36){$\cdot$}
\put(4,43.73){$\cdot$}
\put(3.5,44.12){$\cdot$}
\put(3,44.54){$\cdot$}
\put(2.5,45.00){$\cdot$}
\put(2,45.52){$\cdot$}
\put(1.5,46.11){$\cdot$}
\put(1,46.82){$\cdot$}
\put(0.5,47.74){$\cdot$}
\put(0,50){$\cdot$}
\put(50,34){$\cdot$}
\put(50.5,34){$\cdot$}
\put(51,34){$\cdot$}
\put(51.5,34){$\cdot$}
\put(52,34.01){$\cdot$}
\put(52.5,34.02){$\cdot$}
\put(53,34.03){$\cdot$}
\put(53.5,34.04){$\cdot$}
\put(54,34.05){$\cdot$}
\put(54.5,34.06){$\cdot$}
\put(55,34.08){$\cdot$}
\put(55.5,34.10){$\cdot$}
\put(56,34.11){$\cdot$}
\put(56.5,34.13){$\cdot$}
\put(57,34.16){$\cdot$}
\put(57.5,34.18){$\cdot$}
\put(58,34.20){$\cdot$}
\put(58.5,34.23){$\cdot$}
\put(59,34.26){$\cdot$}
\put(59.5,34.29){$\cdot$}
\put(60,34.32){$\cdot$}
\put(60.5,34.36){$\cdot$}
\put(61,34.39){$\cdot$}
\put(61.5,34.43){$\cdot$}
\put(62,34.47){$\cdot$}
\put(62.5,34.51){$\cdot$}
\put(63,34.55){$\cdot$}
\put(63.5,34.59){$\cdot$}
\put(64,34.64){$\cdot$}
\put(64.5,34.69){$\cdot$}
\put(65,34.74){$\cdot$}
\put(65.5,34.79){$\cdot$}
\put(66,34.84){$\cdot$}
\put(66.5,34.90){$\cdot$}
\put(67,34.95){$\cdot$}
\put(67.5,35.01){$\cdot$}
\put(68,35.07){$\cdot$}
\put(68.5,35.13){$\cdot$}
\put(69,35.20){$\cdot$}
\put(69.5,35.27){$\cdot$}
\put(70,35.33){$\cdot$}
\put(70.5,35.40){$\cdot$}
\put(71,35.48){$\cdot$}
\put(71.5,35.55){$\cdot$}
\put(72,35.63){$\cdot$}
\put(72.5,35.71){$\cdot$}
\put(73,35.79){$\cdot$}
\put(73.5,35.88){$\cdot$}
\put(74,35.96){$\cdot$}
\put(74.5,36.05){$\cdot$}
\put(75,36.14){$\cdot$}
\put(75.5,36.24){$\cdot$}
\put(76,36.33){$\cdot$}
\put(76.5,36.43){$\cdot$}
\put(77,36.53){$\cdot$}
\put(77.5,36.64){$\cdot$}
\put(78,36.74){$\cdot$}
\put(78.5,36.85){$\cdot$}
\put(79,36.97){$\cdot$}
\put(79.5,37.08){$\cdot$}
\put(80,37.20){$\cdot$}
\put(80.5,37.32){$\cdot$}
\put(81,37.45){$\cdot$}
\put(81.5,37.57){$\cdot$}
\put(82,37.70){$\cdot$}
\put(82.5,37.84){$\cdot$}
\put(83,37.98){$\cdot$}
\put(83.5,38.12){$\cdot$}
\put(84,38.27){$\cdot$}
\put(84.5,38.42){$\cdot$}
\put(85,38.57){$\cdot$}
\put(85.5,38.73){$\cdot$}
\put(86,38.90){$\cdot$}
\put(86.5,39.06){$\cdot$}
\put(87,39.24){$\cdot$}
\put(87.5,39.42){$\cdot$}
\put(88,39.60){$\cdot$}
\put(88.5,39.79){$\cdot$}
\put(89,39.99){$\cdot$}
\put(89.5,40.19){$\cdot$}
\put(90,40.40){$\cdot$}
\put(90.5,40.62){$\cdot$}
\put(91,40.84){$\cdot$}
\put(91.5,41.07){$\cdot$}
\put(92,41.32){$\cdot$}
\put(92.5,41.57){$\cdot$}
\put(93,41.83){$\cdot$}
\put(93.5,42.11){$\cdot$}
\put(94,42.40){$\cdot$}
\put(94.5,42.70){$\cdot$}
\put(95,43.02){$\cdot$}
\put(95.5,43.36){$\cdot$}
\put(96,43.73){$\cdot$}
\put(96.5,44.12){$\cdot$}
\put(97,44.54){$\cdot$}
\put(97.5,45.00){$\cdot$}
\put(98,45.52){$\cdot$}
\put(98.5,46.11){$\cdot$}
\put(99,46.82){$\cdot$}
\put(99.5,47.74){$\cdot$}
\put(100,50){$\cdot$}

\put(50,134){$\cdot$}
\put(49.5,134){$\cdot$}
\put(49,134){$\cdot$}
\put(48.5,134){$\cdot$}
\put(48,134.01){$\cdot$}
\put(47.5,134.02){$\cdot$}
\put(47,134.03){$\cdot$}
\put(46.5,134.04){$\cdot$}
\put(46,134.05){$\cdot$}
\put(45.5,134.06){$\cdot$}
\put(45,134.08){$\cdot$}
\put(44.5,134.10){$\cdot$}
\put(44,134.11){$\cdot$}
\put(43.5,134.13){$\cdot$}
\put(43,134.16){$\cdot$}
\put(42.5,134.18){$\cdot$}
\put(42,134.20){$\cdot$}
\put(41.5,134.23){$\cdot$}
\put(41,134.26){$\cdot$}
\put(40.5,134.29){$\cdot$}
\put(40,134.32){$\cdot$}
\put(39.5,134.36){$\cdot$}
\put(39,134.39){$\cdot$}
\put(38.5,134.43){$\cdot$}
\put(38,134.47){$\cdot$}
\put(37.5,134.51){$\cdot$}
\put(37,134.55){$\cdot$}
\put(36.5,134.59){$\cdot$}
\put(36,134.64){$\cdot$}
\put(35.5,134.69){$\cdot$}
\put(35,134.74){$\cdot$}
\put(34.5,134.79){$\cdot$}
\put(34,134.84){$\cdot$}
\put(33.5,134.90){$\cdot$}
\put(33,134.95){$\cdot$}
\put(32.5,135.01){$\cdot$}
\put(32,135.07){$\cdot$}
\put(31.5,135.13){$\cdot$}
\put(31,135.20){$\cdot$}
\put(30.5,135.27){$\cdot$}
\put(30,135.33){$\cdot$}
\put(29.5,135.40){$\cdot$}
\put(29,135.48){$\cdot$}
\put(28.5,135.55){$\cdot$}
\put(28,135.63){$\cdot$}
\put(27.5,135.71){$\cdot$}
\put(27,135.79){$\cdot$}
\put(26.5,135.88){$\cdot$}
\put(26,135.96){$\cdot$}
\put(25.5,136.05){$\cdot$}
\put(25,136.14){$\cdot$}
\put(24.5,136.24){$\cdot$}
\put(24,136.33){$\cdot$}
\put(23.5,136.43){$\cdot$}
\put(23,136.53){$\cdot$}
\put(22.5,136.64){$\cdot$}
\put(22,136.74){$\cdot$}
\put(21.5,136.85){$\cdot$}
\put(21,136.97){$\cdot$}
\put(20.5,137.08){$\cdot$}
\put(20,137.20){$\cdot$}
\put(19.5,137.32){$\cdot$}
\put(19,137.45){$\cdot$}
\put(18.5,137.57){$\cdot$}
\put(18,137.70){$\cdot$}
\put(17.5,137.84){$\cdot$}
\put(17,137.98){$\cdot$}
\put(16.5,138.12){$\cdot$}
\put(16,138.27){$\cdot$}
\put(15.5,138.42){$\cdot$}
\put(15,138.57){$\cdot$}
\put(14.5,138.73){$\cdot$}
\put(14,138.90){$\cdot$}
\put(13.5,139.06){$\cdot$}
\put(13,139.24){$\cdot$}
\put(12.5,139.42){$\cdot$}
\put(12,139.60){$\cdot$}
\put(11.5,139.79){$\cdot$}
\put(11,139.99){$\cdot$}
\put(10.5,140.19){$\cdot$}
\put(10,140.40){$\cdot$}
\put(9.5,140.62){$\cdot$}
\put(9,140.84){$\cdot$}
\put(8.5,141.07){$\cdot$}
\put(8,141.32){$\cdot$}
\put(7.5,141.57){$\cdot$}
\put(7,141.83){$\cdot$}
\put(6.5,142.11){$\cdot$}
\put(6,142.40){$\cdot$}
\put(5.5,142.70){$\cdot$}
\put(5,143.02){$\cdot$}
\put(4.5,143.36){$\cdot$}
\put(4,143.73){$\cdot$}
\put(3.5,144.12){$\cdot$}
\put(3,144.54){$\cdot$}
\put(2.5,145.00){$\cdot$}
\put(2,145.52){$\cdot$}
\put(1.5,146.11){$\cdot$}
\put(1,146.82){$\cdot$}
\put(0.5,147.74){$\cdot$}
\put(0,150){$\cdot$}
\put(50,134){$\cdot$}
\put(50.5,134){$\cdot$}
\put(51,134){$\cdot$}
\put(51.5,134){$\cdot$}
\put(52,134.01){$\cdot$}
\put(52.5,134.02){$\cdot$}
\put(53,134.03){$\cdot$}
\put(53.5,134.04){$\cdot$}
\put(54,134.05){$\cdot$}
\put(54.5,134.06){$\cdot$}
\put(55,134.08){$\cdot$}
\put(55.5,134.10){$\cdot$}
\put(56,134.11){$\cdot$}
\put(56.5,134.13){$\cdot$}
\put(57,134.16){$\cdot$}
\put(57.5,134.18){$\cdot$}
\put(58,134.20){$\cdot$}
\put(58.5,134.23){$\cdot$}
\put(59,134.26){$\cdot$}
\put(59.5,134.29){$\cdot$}
\put(60,134.32){$\cdot$}
\put(60.5,134.36){$\cdot$}
\put(61,134.39){$\cdot$}
\put(61.5,134.43){$\cdot$}
\put(62,134.47){$\cdot$}
\put(62.5,134.51){$\cdot$}
\put(63,134.55){$\cdot$}
\put(63.5,134.59){$\cdot$}
\put(64,134.64){$\cdot$}
\put(64.5,134.69){$\cdot$}
\put(65,134.74){$\cdot$}
\put(65.5,134.79){$\cdot$}
\put(66,134.84){$\cdot$}
\put(66.5,134.90){$\cdot$}
\put(67,134.95){$\cdot$}
\put(67.5,135.01){$\cdot$}
\put(68,135.07){$\cdot$}
\put(68.5,135.13){$\cdot$}
\put(69,135.20){$\cdot$}
\put(69.5,135.27){$\cdot$}
\put(70,135.33){$\cdot$}
\put(70.5,135.40){$\cdot$}
\put(71,135.48){$\cdot$}
\put(71.5,135.55){$\cdot$}
\put(72,135.63){$\cdot$}
\put(72.5,135.71){$\cdot$}
\put(73,135.79){$\cdot$}
\put(73.5,135.88){$\cdot$}
\put(74,135.96){$\cdot$}
\put(74.5,136.05){$\cdot$}
\put(75,136.14){$\cdot$}
\put(75.5,136.24){$\cdot$}
\put(76,136.33){$\cdot$}
\put(76.5,136.43){$\cdot$}
\put(77,136.53){$\cdot$}
\put(77.5,136.64){$\cdot$}
\put(78,136.74){$\cdot$}
\put(78.5,136.85){$\cdot$}
\put(79,136.97){$\cdot$}
\put(79.5,137.08){$\cdot$}
\put(80,137.20){$\cdot$}
\put(80.5,137.32){$\cdot$}
\put(81,137.45){$\cdot$}
\put(81.5,137.57){$\cdot$}
\put(82,137.70){$\cdot$}
\put(82.5,137.84){$\cdot$}
\put(83,137.98){$\cdot$}
\put(83.5,138.12){$\cdot$}
\put(84,138.27){$\cdot$}
\put(84.5,138.42){$\cdot$}
\put(85,138.57){$\cdot$}
\put(85.5,138.73){$\cdot$}
\put(86,138.90){$\cdot$}
\put(86.5,139.06){$\cdot$}
\put(87,139.24){$\cdot$}
\put(87.5,139.42){$\cdot$}
\put(88,139.60){$\cdot$}
\put(88.5,139.79){$\cdot$}
\put(89,139.99){$\cdot$}
\put(89.5,140.19){$\cdot$}
\put(90,140.40){$\cdot$}
\put(90.5,140.62){$\cdot$}
\put(91,140.84){$\cdot$}
\put(91.5,141.07){$\cdot$}
\put(92,141.32){$\cdot$}
\put(92.5,141.57){$\cdot$}
\put(93,141.83){$\cdot$}
\put(93.5,142.11){$\cdot$}
\put(94,142.40){$\cdot$}
\put(94.5,142.70){$\cdot$}
\put(95,143.02){$\cdot$}
\put(95.5,143.36){$\cdot$}
\put(96,143.73){$\cdot$}
\put(96.5,144.12){$\cdot$}
\put(97,144.54){$\cdot$}
\put(97.5,145.00){$\cdot$}
\put(98,145.52){$\cdot$}
\put(98.5,146.11){$\cdot$}
\put(99,146.82){$\cdot$}
\put(99.5,147.74){$\cdot$}
\put(100,150){$\cdot$}
\put(50,66){$\cdot$}
\put(49.5,65.99){$\cdot$}
\put(49,65.99){$\cdot$}
\put(48.5,65.99){$\cdot$}
\put(48,65.98){$\cdot$}
\put(45,65.91){$\cdot$}
\put(44.5,65.90){$\cdot$}
\put(44,65.88){$\cdot$}
\put(43.5,65.86){$\cdot$}
\put(43,65.84){$\cdot$}
\put(40,65.68){$\cdot$}
\put(39.5,65.64){$\cdot$}
\put(39,65.60){$\cdot$}
\put(38.5,65.57){$\cdot$}
\put(38,65.53){$\cdot$}
\put(35,65.26){$\cdot$}
\put(34.5,65.21){$\cdot$}
\put(34,65.16){$\cdot$}
\put(33.5,65.10){$\cdot$}
\put(33,65.04){$\cdot$}
\put(30,64.66){$\cdot$}
\put(29.5,64.59){$\cdot$}
\put(29,64.52){$\cdot$}
\put(28.5,64.44){$\cdot$}
\put(28,64.37){$\cdot$}
\put(25,63.86){$\cdot$}
\put(24.5,63.76){$\cdot$}
\put(24,63.66){$\cdot$}
\put(23.5,63.57){$\cdot$}
\put(23,63.47){$\cdot$}
\put(20,62.80){$\cdot$}
\put(19.5,62.68){$\cdot$}
\put(19,62.55){$\cdot$}
\put(18.5,62.42){$\cdot$}
\put(18,62.29){$\cdot$}
\put(15,61.43){$\cdot$}
\put(14.5,61.27){$\cdot$}
\put(14,61.10){$\cdot$}
\put(13.5,60.93){$\cdot$}
\put(13,60.76){$\cdot$}
\put(10,59.60){$\cdot$}
\put(9.5,59.38){$\cdot$}
\put(9,59.16){$\cdot$}
\put(8.5,58.92){$\cdot$}
\put(8,58.68){$\cdot$}
\put(5,56.97){$\cdot$}
\put(4.5,56.63){$\cdot$}
\put(4,56.27){$\cdot$}
\put(3.5,55.88){$\cdot$}
\put(3,55.46){$\cdot$}
\put(53,65.98){$\cdot$}
\put(53.5,65.97){$\cdot$}
\put(54,65.96){$\cdot$}
\put(54.5,65.95){$\cdot$}
\put(55,65.93){$\cdot$}
\put(58,65.82){$\cdot$}
\put(58.5,65.79){$\cdot$}
\put(59,65.77){$\cdot$}
\put(59.5,65.74){$\cdot$}
\put(60,65.70){$\cdot$}
\put(63,65.49){$\cdot$}
\put(63.5,65.45){$\cdot$}
\put(64,65.40){$\cdot$}
\put(64.5,65.36){$\cdot$}
\put(65,65.31){$\cdot$}
\put(68,64.99){$\cdot$}
\put(68.5,64.93){$\cdot$}
\put(69,64.86){$\cdot$}
\put(69.5,64.79){$\cdot$}
\put(70,64.73){$\cdot$}
\put(73,64.29){$\cdot$}
\put(73.5,64.20){$\cdot$}
\put(74,64.12){$\cdot$}
\put(74.5,64.04){$\cdot$}
\put(75,63.95){$\cdot$}
\put(78,63.36){$\cdot$}
\put(78.5,63.25){$\cdot$}
\put(79,63.15){$\cdot$}
\put(79.5,63.03){$\cdot$}
\put(80,62.92){$\cdot$}
\put(83,62.16){$\cdot$}
\put(83.5,62.02){$\cdot$}
\put(84,61.88){$\cdot$}
\put(84.5,61.73){$\cdot$}
\put(85,61.58){$\cdot$}
\put(88,60.58){$\cdot$}
\put(88.5,60.40){$\cdot$}
\put(89,60.20){$\cdot$}
\put(89.5,60.01){$\cdot$}
\put(90,59.81){$\cdot$}
\put(93,58.43){$\cdot$}
\put(93.5,58.16){$\cdot$}
\put(94,57.88){$\cdot$}
\put(94.5,57.59){$\cdot$}
\put(95,57.29){$\cdot$}
\put(98,54.99){$\cdot$}
\put(98.5,54.48){$\cdot$}
\put(99,53.89){$\cdot$}
\put(99.5,53.18){$\cdot$}
\put(100,52.26){$\cdot$}
\put(50,166){$\cdot$}
\put(49.5,165.99){$\cdot$}
\put(49,165.99){$\cdot$}
\put(48.5,165.99){$\cdot$}
\put(48,165.98){$\cdot$}
\put(45,165.91){$\cdot$}
\put(44.5,165.90){$\cdot$}
\put(44,165.88){$\cdot$}
\put(43.5,165.86){$\cdot$}
\put(43,165.84){$\cdot$}
\put(40,165.68){$\cdot$}
\put(39.5,165.64){$\cdot$}
\put(39,165.60){$\cdot$}
\put(38.5,165.57){$\cdot$}
\put(38,165.53){$\cdot$}
\put(35,165.26){$\cdot$}
\put(34.5,165.21){$\cdot$}
\put(34,165.16){$\cdot$}
\put(33.5,165.10){$\cdot$}
\put(33,165.04){$\cdot$}
\put(30,164.66){$\cdot$}
\put(29.5,164.59){$\cdot$}
\put(29,164.52){$\cdot$}
\put(28.5,164.44){$\cdot$}
\put(28,164.37){$\cdot$}
\put(25,163.86){$\cdot$}
\put(24.5,163.76){$\cdot$}
\put(24,163.66){$\cdot$}
\put(23.5,163.57){$\cdot$}
\put(23,163.47){$\cdot$}
\put(20,162.80){$\cdot$}
\put(19.5,162.68){$\cdot$}
\put(19,162.55){$\cdot$}
\put(18.5,162.42){$\cdot$}
\put(18,162.29){$\cdot$}
\put(15,161.43){$\cdot$}
\put(14.5,161.27){$\cdot$}
\put(14,161.10){$\cdot$}
\put(13.5,160.93){$\cdot$}
\put(13,160.76){$\cdot$}
\put(10,159.60){$\cdot$}
\put(9.5,159.38){$\cdot$}
\put(9,159.16){$\cdot$}
\put(8.5,158.92){$\cdot$}
\put(8,158.68){$\cdot$}
\put(5,156.97){$\cdot$}
\put(4.5,156.63){$\cdot$}
\put(4,156.27){$\cdot$}
\put(3.5,155.88){$\cdot$}
\put(3,155.46){$\cdot$}
\put(53,165.98){$\cdot$}
\put(53.5,165.97){$\cdot$}
\put(54,165.96){$\cdot$}
\put(54.5,165.95){$\cdot$}
\put(55,165.93){$\cdot$}
\put(58,165.82){$\cdot$}
\put(58.5,165.79){$\cdot$}
\put(59,165.77){$\cdot$}
\put(59.5,165.74){$\cdot$}
\put(60,165.70){$\cdot$}
\put(63,165.49){$\cdot$}
\put(63.5,165.45){$\cdot$}
\put(64,165.40){$\cdot$}
\put(64.5,165.36){$\cdot$}
\put(65,165.31){$\cdot$}
\put(68,164.99){$\cdot$}
\put(68.5,164.93){$\cdot$}
\put(69,164.86){$\cdot$}
\put(69.5,164.79){$\cdot$}
\put(70,164.73){$\cdot$}
\put(73,164.29){$\cdot$}
\put(73.5,164.20){$\cdot$}
\put(74,164.12){$\cdot$}
\put(74.5,164.04){$\cdot$}
\put(75,163.95){$\cdot$}
\put(78,163.36){$\cdot$}
\put(78.5,163.25){$\cdot$}
\put(79,163.15){$\cdot$}
\put(79.5,163.03){$\cdot$}
\put(80,162.92){$\cdot$}
\put(83,162.16){$\cdot$}
\put(83.5,162.02){$\cdot$}
\put(84,161.88){$\cdot$}
\put(84.5,161.73){$\cdot$}
\put(85,161.58){$\cdot$}
\put(88,160.58){$\cdot$}
\put(88.5,160.40){$\cdot$}
\put(89,160.20){$\cdot$}
\put(89.5,160.01){$\cdot$}
\put(90,159.81){$\cdot$}
\put(93,158.43){$\cdot$}
\put(93.5,158.16){$\cdot$}
\put(94,157.88){$\cdot$}
\put(94.5,157.59){$\cdot$}
\put(95,157.29){$\cdot$}
\put(98,154.99){$\cdot$}
\put(98.5,154.48){$\cdot$}
\put(99,153.89){$\cdot$}
\put(99.5,153.18){$\cdot$}
\put(100,152.26){$\cdot$}
\put(50,100){$\cdot$}
\put(49.5,99.99){$\cdot$}
\put(49,99.98){$\cdot$}
\put(48.5,99.97){$\cdot$}
\put(48,99.96){$\cdot$}
\put(47.5,99.94){$\cdot$}
\put(47,99.90){$\cdot$}
\put(46.5,99.88){$\cdot$}
\put(46,99.84){$\cdot$}
\put(45.5,99.80){$\cdot$}
\put(45,99.75){$\cdot$}
\put(44.5,99.70){$\cdot$}
\put(44,99.64){$\cdot$}
\put(43.5,99.58){$\cdot$}
\put(43,99.51){$\cdot$}
\put(42.5,99.43){$\cdot$}
\put(42,99.35){$\cdot$}
\put(41.5,99.27){$\cdot$}
\put(41,99.18){$\cdot$}
\put(40.5,99.09){$\cdot$}
\put(40,98.99){$\cdot$}
\put(39.5,98.88){$\cdot$}
\put(39,98.77){$\cdot$}
\put(38.5,98.66){$\cdot$}
\put(38,98.54){$\cdot$}
\put(37.5,98.41){$\cdot$}
\put(37,98.28){$\cdot$}
\put(36.5,98.14){$\cdot$}
\put(36,98.00){$\cdot$}
\put(35.5,97.85){$\cdot$}
\put(35,97.70){$\cdot$}
\put(34.5,97.54){$\cdot$}
\put(34,97.37){$\cdot$}
\put(33.5,97.20){$\cdot$}
\put(33,97.02){$\cdot$}
\put(32.5,96.84){$\cdot$}
\put(32,96.65){$\cdot$}
\put(31.5,96.45){$\cdot$}
\put(31,96.25){$\cdot$}
\put(30.5,96.04){$\cdot$}
\put(30,95.82){$\cdot$}
\put(29.5,95.60){$\cdot$}
\put(29,95.38){$\cdot$}
\put(28.5,95.14){$\cdot$}
\put(28,94.90){$\cdot$}
\put(27.5,94.65){$\cdot$}
\put(27,94.39){$\cdot$}
\put(26.5,94.13){$\cdot$}
\put(26,93.86){$\cdot$}
\put(25.5,93.59){$\cdot$}
\put(25,93.30){$\cdot$}
\put(24.5,93.01){$\cdot$}
\put(24,92.71){$\cdot$}
\put(23.5,92.40){$\cdot$}
\put(23,92.08){$\cdot$}
\put(22.5,91.76){$\cdot$}
\put(22,91.42){$\cdot$}
\put(21.5,91.08){$\cdot$}
\put(21,90.73){$\cdot$}
\put(20.5,90.37){$\cdot$}
\put(20,90.00){$\cdot$}
\put(19.5,89.62){$\cdot$}
\put(19,89.23){$\cdot$}
\put(18.5,88.83){$\cdot$}
\put(18,88.42){$\cdot$}
\put(17.5,87.99){$\cdot$}
\put(17,87.56){$\cdot$}
\put(16.5,87.12){$\cdot$}
\put(16,86.66){$\cdot$}
\put(15.5,86.19){$\cdot$}
\put(15,85.70){$\cdot$}
\put(14.5,85.21){$\cdot$}
\put(14,84.70){$\cdot$}
\put(13.5,84.17){$\cdot$}
\put(13,83.63){$\cdot$}
\put(12.5,83.07){$\cdot$}
\put(12,82.49){$\cdot$}
\put(11.5,81.90){$\cdot$}
\put(11,81.29){$\cdot$}
\put(10.5,80.65){$\cdot$}
\put(10,80.00){$\cdot$}
\put(9.5,79.32){$\cdot$}
\put(9,78.62){$\cdot$}
\put(8.5,77.88){$\cdot$}
\put(8,77.13){$\cdot$}
\put(7.5,76.34){$\cdot$}
\put(7,75.51){$\cdot$}
\put(6.5,74.65){$\cdot$}
\put(6,73.75){$\cdot$}
\put(5.5,72.80){$\cdot$}
\put(5,71.79){$\cdot$}
\put(4.5,70.73){$\cdot$}
\put(4,69.59){$\cdot$}
\put(3.5,68.38){$\cdot$}
\put(3,67.06){$\cdot$}
\put(2.5,65.61){$\cdot$}
\put(2,64.00){$\cdot$}
\put(1.5,62.15){$\cdot$}
\put(1,59.95){$\cdot$}
\put(0.5,57.05){$\cdot$}
\put(0,50){$\cdot$}
\put(50,100){$\cdot$}
\put(50.5,99.99){$\cdot$}
\put(51,99.98){$\cdot$}
\put(51.5,99.97){$\cdot$}
\put(52,99.96){$\cdot$}
\put(52.5,99.94){$\cdot$}
\put(53,99.90){$\cdot$}
\put(53.5,99.88){$\cdot$}
\put(54,99.84){$\cdot$}
\put(54.5,99.80){$\cdot$}
\put(55,99.75){$\cdot$}
\put(55.5,99.70){$\cdot$}
\put(56,99.64){$\cdot$}
\put(56.5,99.58){$\cdot$}
\put(57,99.51){$\cdot$}
\put(57.5,99.43){$\cdot$}
\put(58,99.35){$\cdot$}
\put(58.5,99.27){$\cdot$}
\put(59,99.18){$\cdot$}
\put(59.5,99.09){$\cdot$}
\put(60,98.99){$\cdot$}
\put(60.5,98.88){$\cdot$}
\put(61,98.77){$\cdot$}
\put(61.5,98.66){$\cdot$}
\put(62,98.54){$\cdot$}
\put(62.5,98.41){$\cdot$}
\put(63,98.28){$\cdot$}
\put(63.5,98.14){$\cdot$}
\put(64,98.00){$\cdot$}
\put(64.5,97.85){$\cdot$}
\put(65,97.70){$\cdot$}
\put(65.5,97.54){$\cdot$}
\put(66,97.37){$\cdot$}
\put(66.5,97.20){$\cdot$}
\put(67,97.02){$\cdot$}
\put(67.5,96.84){$\cdot$}
\put(68,96.65){$\cdot$}
\put(68.5,96.45){$\cdot$}
\put(69,96.25){$\cdot$}
\put(69.5,96.04){$\cdot$}
\put(70,95.82){$\cdot$}
\put(70.5,95.60){$\cdot$}
\put(71,95.38){$\cdot$}
\put(71.5,95.14){$\cdot$}
\put(72,94.90){$\cdot$}
\put(72.5,94.65){$\cdot$}
\put(73,94.39){$\cdot$}
\put(73.5,94.13){$\cdot$}
\put(74,93.86){$\cdot$}
\put(74.5,93.59){$\cdot$}
\put(75,93.30){$\cdot$}
\put(75.5,93.01){$\cdot$}
\put(76,92.71){$\cdot$}
\put(76.5,92.40){$\cdot$}
\put(77,92.08){$\cdot$}
\put(77.5,91.76){$\cdot$}
\put(78,91.42){$\cdot$}
\put(78.5,91.08){$\cdot$}
\put(79,90.73){$\cdot$}
\put(79.5,90.37){$\cdot$}
\put(80,90.00){$\cdot$}
\put(80.5,89.62){$\cdot$}
\put(81,89.23){$\cdot$}
\put(81.5,88.83){$\cdot$}
\put(82,88.42){$\cdot$}
\put(82.5,87.99){$\cdot$}
\put(83,87.56){$\cdot$}
\put(83.5,87.12){$\cdot$}
\put(84,86.66){$\cdot$}
\put(84.5,86.19){$\cdot$}
\put(85,85.70){$\cdot$}
\put(85.5,85.21){$\cdot$}
\put(86,84.70){$\cdot$}
\put(86.5,84.17){$\cdot$}
\put(87,83.63){$\cdot$}
\put(87.5,83.07){$\cdot$}
\put(88,82.49){$\cdot$}
\put(88.5,81.90){$\cdot$}
\put(89,81.29){$\cdot$}
\put(89.5,80.65){$\cdot$}
\put(90,80.00){$\cdot$}
\put(90.5,79.32){$\cdot$}
\put(91,78.62){$\cdot$}
\put(91.5,77.88){$\cdot$}
\put(92,77.13){$\cdot$}
\put(92.5,76.34){$\cdot$}
\put(93,75.51){$\cdot$}
\put(93.5,74.65){$\cdot$}
\put(94,73.75){$\cdot$}
\put(94.5,72.80){$\cdot$}
\put(95,71.79){$\cdot$}
\put(95.5,70.73){$\cdot$}
\put(96,69.59){$\cdot$}
\put(96.5,68.38){$\cdot$}
\put(97,67.06){$\cdot$}
\put(97.5,65.61){$\cdot$}
\put(98,64.00){$\cdot$}
\put(98.5,62.15){$\cdot$}
\put(99,59.95){$\cdot$}
\put(99.5,57.05){$\cdot$}
\put(100,50){$\cdot$}
\put(50,200){$\cdot$}
\put(49.5,199.99){$\cdot$}
\put(49,199.98){$\cdot$}
\put(48.5,199.97){$\cdot$}
\put(48,199.96){$\cdot$}
\put(47.5,199.94){$\cdot$}
\put(47,199.90){$\cdot$}
\put(46.5,199.88){$\cdot$}
\put(46,199.84){$\cdot$}
\put(45.5,199.80){$\cdot$}
\put(45,199.75){$\cdot$}
\put(44.5,199.70){$\cdot$}
\put(44,199.64){$\cdot$}
\put(43.5,199.58){$\cdot$}
\put(43,199.51){$\cdot$}
\put(42.5,199.43){$\cdot$}
\put(42,199.35){$\cdot$}
\put(41.5,199.27){$\cdot$}
\put(41,199.18){$\cdot$}
\put(40.5,199.09){$\cdot$}
\put(40,198.99){$\cdot$}
\put(39.5,198.88){$\cdot$}
\put(39,198.77){$\cdot$}
\put(38.5,198.66){$\cdot$}
\put(38,198.54){$\cdot$}
\put(37.5,198.41){$\cdot$}
\put(37,198.28){$\cdot$}
\put(36.5,198.14){$\cdot$}
\put(36,198.00){$\cdot$}
\put(35.5,197.85){$\cdot$}
\put(35,197.70){$\cdot$}
\put(34.5,197.54){$\cdot$}
\put(34,197.37){$\cdot$}
\put(33.5,197.20){$\cdot$}
\put(33,197.02){$\cdot$}
\put(32.5,196.84){$\cdot$}
\put(32,196.65){$\cdot$}
\put(31.5,196.45){$\cdot$}
\put(31,196.25){$\cdot$}
\put(30.5,196.04){$\cdot$}
\put(30,195.82){$\cdot$}
\put(29.5,195.60){$\cdot$}
\put(29,195.38){$\cdot$}
\put(28.5,195.14){$\cdot$}
\put(28,194.90){$\cdot$}
\put(27.5,194.65){$\cdot$}
\put(27,194.39){$\cdot$}
\put(26.5,194.13){$\cdot$}
\put(26,193.86){$\cdot$}
\put(25.5,193.59){$\cdot$}
\put(25,193.30){$\cdot$}
\put(24.5,193.01){$\cdot$}
\put(24,192.71){$\cdot$}
\put(23.5,192.40){$\cdot$}
\put(23,192.08){$\cdot$}
\put(22.5,191.76){$\cdot$}
\put(22,191.42){$\cdot$}
\put(21.5,191.08){$\cdot$}
\put(21,190.73){$\cdot$}
\put(20.5,190.37){$\cdot$}
\put(20,190.00){$\cdot$}
\put(19.5,189.62){$\cdot$}
\put(19,189.23){$\cdot$}
\put(18.5,188.83){$\cdot$}
\put(18,188.42){$\cdot$}
\put(17.5,187.99){$\cdot$}
\put(17,187.56){$\cdot$}
\put(16.5,187.12){$\cdot$}
\put(16,186.66){$\cdot$}
\put(15.5,186.19){$\cdot$}
\put(15,185.70){$\cdot$}
\put(14.5,185.21){$\cdot$}
\put(14,184.70){$\cdot$}
\put(13.5,184.17){$\cdot$}
\put(13,183.63){$\cdot$}
\put(12.5,183.07){$\cdot$}
\put(12,182.49){$\cdot$}
\put(11.5,181.90){$\cdot$}
\put(11,181.29){$\cdot$}
\put(10.5,180.65){$\cdot$}
\put(10,180.00){$\cdot$}
\put(9.5,179.32){$\cdot$}
\put(9,178.62){$\cdot$}
\put(8.5,177.88){$\cdot$}
\put(8,177.13){$\cdot$}
\put(7.5,176.34){$\cdot$}
\put(7,175.51){$\cdot$}
\put(6.5,174.65){$\cdot$}
\put(6,173.75){$\cdot$}
\put(5.5,172.80){$\cdot$}
\put(5,171.79){$\cdot$}
\put(4.5,170.73){$\cdot$}
\put(4,169.59){$\cdot$}
\put(3.5,168.38){$\cdot$}
\put(3,167.06){$\cdot$}
\put(2.5,165.61){$\cdot$}
\put(2,164.00){$\cdot$}
\put(1.5,162.15){$\cdot$}
\put(1,159.95){$\cdot$}
\put(0.5,157.05){$\cdot$}
\put(0,150){$\cdot$}
\put(50,200){$\cdot$}
\put(50.5,199.99){$\cdot$}
\put(51,199.98){$\cdot$}
\put(51.5,199.97){$\cdot$}
\put(52,199.96){$\cdot$}
\put(52.5,199.94){$\cdot$}
\put(53,199.90){$\cdot$}
\put(53.5,199.88){$\cdot$}
\put(54,199.84){$\cdot$}
\put(54.5,199.80){$\cdot$}
\put(55,199.75){$\cdot$}
\put(55.5,199.70){$\cdot$}
\put(56,199.64){$\cdot$}
\put(56.5,199.58){$\cdot$}
\put(57,199.51){$\cdot$}
\put(57.5,199.43){$\cdot$}
\put(58,199.35){$\cdot$}
\put(58.5,199.27){$\cdot$}
\put(59,199.18){$\cdot$}
\put(59.5,199.09){$\cdot$}
\put(60,198.99){$\cdot$}
\put(60.5,198.88){$\cdot$}
\put(61,198.77){$\cdot$}
\put(61.5,198.66){$\cdot$}
\put(62,198.54){$\cdot$}
\put(62.5,198.41){$\cdot$}
\put(63,198.28){$\cdot$}
\put(63.5,198.14){$\cdot$}
\put(64,198.00){$\cdot$}
\put(64.5,197.85){$\cdot$}
\put(65,197.70){$\cdot$}
\put(65.5,197.54){$\cdot$}
\put(66,197.37){$\cdot$}
\put(66.5,197.20){$\cdot$}
\put(67,197.02){$\cdot$}
\put(67.5,196.84){$\cdot$}
\put(68,196.65){$\cdot$}
\put(68.5,196.45){$\cdot$}
\put(69,196.25){$\cdot$}
\put(69.5,196.04){$\cdot$}
\put(70,195.82){$\cdot$}
\put(70.5,195.60){$\cdot$}
\put(71,195.38){$\cdot$}
\put(71.5,195.14){$\cdot$}
\put(72,194.90){$\cdot$}
\put(72.5,194.65){$\cdot$}
\put(73,194.39){$\cdot$}
\put(73.5,194.13){$\cdot$}
\put(74,193.86){$\cdot$}
\put(74.5,193.59){$\cdot$}
\put(75,193.30){$\cdot$}
\put(75.5,193.01){$\cdot$}
\put(76,192.71){$\cdot$}
\put(76.5,192.40){$\cdot$}
\put(77,192.08){$\cdot$}
\put(77.5,191.76){$\cdot$}
\put(78,191.42){$\cdot$}
\put(78.5,191.08){$\cdot$}
\put(79,190.73){$\cdot$}
\put(79.5,190.37){$\cdot$}
\put(80,190.00){$\cdot$}
\put(80.5,189.62){$\cdot$}
\put(81,189.23){$\cdot$}
\put(81.5,188.83){$\cdot$}
\put(82,188.42){$\cdot$}
\put(82.5,187.99){$\cdot$}
\put(83,187.56){$\cdot$}
\put(83.5,187.12){$\cdot$}
\put(84,186.66){$\cdot$}
\put(84.5,186.19){$\cdot$}
\put(85,185.70){$\cdot$}
\put(85.5,185.21){$\cdot$}
\put(86,184.70){$\cdot$}
\put(86.5,184.17){$\cdot$}
\put(87,183.63){$\cdot$}
\put(87.5,183.07){$\cdot$}
\put(88,182.49){$\cdot$}
\put(88.5,181.90){$\cdot$}
\put(89,181.29){$\cdot$}
\put(89.5,180.65){$\cdot$}
\put(90,180.00){$\cdot$}
\put(90.5,179.32){$\cdot$}
\put(91,178.62){$\cdot$}
\put(91.5,177.88){$\cdot$}
\put(92,177.13){$\cdot$}
\put(92.5,176.34){$\cdot$}
\put(93,175.51){$\cdot$}
\put(93.5,174.65){$\cdot$}
\put(94,173.75){$\cdot$}
\put(94.5,172.80){$\cdot$}
\put(95,171.79){$\cdot$}
\put(95.5,170.73){$\cdot$}
\put(96,169.59){$\cdot$}
\put(96.5,168.38){$\cdot$}
\put(97,167.06){$\cdot$}
\put(97.5,165.61){$\cdot$}
\put(98,164.00){$\cdot$}
\put(98.5,162.15){$\cdot$}
\put(99,159.95){$\cdot$}
\put(99.5,157.05){$\cdot$}
\put(100,150){$\cdot$}
\put(50,0){$\cdot$}
\put(49.5,0){$\cdot$}
\put(49,0.01){$\cdot$}
\put(48.5,0.02){$\cdot$}
\put(48,0.04){$\cdot$}
\put(47.5,0.06){$\cdot$}
\put(47,0.09){$\cdot$}
\put(46.5,0.12){$\cdot$}
\put(46,0.16){$\cdot$}
\put(45.5,0.2){$\cdot$}
\put(45,0.25){$\cdot$}
\put(44.5,0.3){$\cdot$}
\put(44,0.36){$\cdot$}
\put(43.5,0.42){$\cdot$}
\put(43,0.49){$\cdot$}
\put(42.5,0.56){$\cdot$}
\put(42,0.64){$\cdot$}
\put(41.5,0.73){$\cdot$}
\put(41,0.82){$\cdot$}
\put(40.5,0.91){$\cdot$}
\put(40,1.01){$\cdot$}
\put(39.5,1.11){$\cdot$}
\put(39,1.22){$\cdot$}
\put(38.5,1.34){$\cdot$}
\put(38,1.46){$\cdot$}
\put(37.5,1.59){$\cdot$}
\put(37,1.72){$\cdot$}
\put(36.5,1.86){$\cdot$}
\put(36,2.0){$\cdot$}
\put(35.5,2.15){$\cdot$}
\put(35,2.3){$\cdot$}
\put(34.5,2.46){$\cdot$}
\put(34,2.63){$\cdot$}
\put(33.5,2.8){$\cdot$}
\put(33,2.98){$\cdot$}
\put(32.5,3.16){$\cdot$}
\put(32,3.35){$\cdot$}
\put(31.5,3.55){$\cdot$}
\put(31,3.75){$\cdot$}
\put(30.5,3.96){$\cdot$}
\put(30,4.17){$\cdot$}
\put(29.5,4.39){$\cdot$}
\put(29,4.62){$\cdot$}
\put(28.5,4.86){$\cdot$}
\put(28,5.1){$\cdot$}
\put(27.5,5.35){$\cdot$}
\put(27,5.6){$\cdot$}
\put(26.5,5.87){$\cdot$}
\put(26,6.14){$\cdot$}
\put(25.5,6.41){$\cdot$}
\put(25,6.7){$\cdot$}
\put(24.5,6.99){$\cdot$}
\put(24,7.29){$\cdot$}
\put(23.5,7.6){$\cdot$}
\put(23,7.9){$\cdot$}
\put(22.5,8.24){$\cdot$}
\put(22,8.57){$\cdot$}
\put(21.5,8.92){$\cdot$}
\put(21,9.27){$\cdot$}
\put(20.5,9.63){$\cdot$}
\put(20,10.0){$\cdot$}
\put(19.5,10.38){$\cdot$}
\put(19,10.77){$\cdot$}
\put(18.5,11.17){$\cdot$}
\put(18,11.58){$\cdot$}
\put(17.5,12.0){$\cdot$}
\put(17,12.44){$\cdot$}
\put(16.5,12.88){$\cdot$}
\put(16,13.34){$\cdot$}
\put(15.5,13.81){$\cdot$}
\put(15,14.29){$\cdot$}
\put(14.5,14.79){$\cdot$}
\put(14,15.3){$\cdot$}
\put(13.5,15.83){$\cdot$}
\put(13,16.37){$\cdot$}
\put(12.5,16.93){$\cdot$}
\put(12,17.5){$\cdot$}
\put(11.5,18.1){$\cdot$}
\put(11,18.71){$\cdot$}
\put(10.5,19.34){$\cdot$}
\put(10,20.00){$\cdot$}
\put(9.5,20.68){$\cdot$}
\put(9,21.38){$\cdot$}
\put(8.5,22.11){$\cdot$}
\put(8,22.87){$\cdot$}
\put(7.5,23.66){$\cdot$}
\put(7,24.48){$\cdot$}
\put(6.5,25.35){$\cdot$}
\put(6,26.25){$\cdot$}
\put(5.5,27.2){$\cdot$}
\put(5,28.2){$\cdot$}
\put(4.5,29.27){$\cdot$}
\put(4,30.4){$\cdot$}
\put(3.5,31.62){$\cdot$}
\put(3,32.94){$\cdot$}
\put(2.5,34.39){$\cdot$}
\put(2,36.00){$\cdot$}
\put(1.5,37.84){$\cdot$}
\put(1,40.05){$\cdot$}
\put(0.5,42.95){$\cdot$}
\put(0,50){$\cdot$}
\put(50,0){$\cdot$}
\put(50.5,0){$\cdot$}
\put(51,0.01){$\cdot$}
\put(51.5,0.02){$\cdot$}
\put(52,0.04){$\cdot$}
\put(52.5,0.06){$\cdot$}
\put(53,0.09){$\cdot$}
\put(53.5,0.12){$\cdot$}
\put(54,0.16){$\cdot$}
\put(54.5,0.2){$\cdot$}
\put(55,0.25){$\cdot$}
\put(55.5,0.3){$\cdot$}
\put(56,0.36){$\cdot$}
\put(56.5,0.42){$\cdot$}
\put(57,0.49){$\cdot$}
\put(57.5,0.56){$\cdot$}
\put(58,0.64){$\cdot$}
\put(58.5,0.73){$\cdot$}
\put(59,0.82){$\cdot$}
\put(59.5,0.91){$\cdot$}
\put(60,1.01){$\cdot$}
\put(60.5,1.11){$\cdot$}
\put(61,1.22){$\cdot$}
\put(61.5,1.34){$\cdot$}
\put(62,1.46){$\cdot$}
\put(62.5,1.59){$\cdot$}
\put(63,1.72){$\cdot$}
\put(63.5,1.86){$\cdot$}
\put(64,2.0){$\cdot$}
\put(64.5,2.15){$\cdot$}
\put(65,2.3){$\cdot$}
\put(65.5,2.46){$\cdot$}
\put(66,2.63){$\cdot$}
\put(66.5,2.8){$\cdot$}
\put(67,2.98){$\cdot$}
\put(67.5,3.16){$\cdot$}
\put(68,3.35){$\cdot$}
\put(68.5,3.55){$\cdot$}
\put(69,3.75){$\cdot$}
\put(69.5,3.96){$\cdot$}
\put(70,4.17){$\cdot$}
\put(70.5,4.39){$\cdot$}
\put(71,4.62){$\cdot$}
\put(71.5,4.86){$\cdot$}
\put(72,5.1){$\cdot$}
\put(72.5,5.35){$\cdot$}
\put(73,5.6){$\cdot$}
\put(73.5,5.87){$\cdot$}
\put(74,6.14){$\cdot$}
\put(74.5,6.41){$\cdot$}
\put(75,6.7){$\cdot$}
\put(75.5,6.99){$\cdot$}
\put(76,7.29){$\cdot$}
\put(76.5,7.6){$\cdot$}
\put(77,7.9){$\cdot$}
\put(77.5,8.24){$\cdot$}
\put(78,8.57){$\cdot$}
\put(78.5,8.92){$\cdot$}
\put(79,9.27){$\cdot$}
\put(79.5,9.63){$\cdot$}
\put(80,10.0){$\cdot$}
\put(80.5,10.38){$\cdot$}
\put(81,10.77){$\cdot$}
\put(81.5,11.17){$\cdot$}
\put(82,11.58){$\cdot$}
\put(82.5,12.0){$\cdot$}
\put(83,12.44){$\cdot$}
\put(83.5,12.88){$\cdot$}
\put(84,13.34){$\cdot$}
\put(84.5,13.81){$\cdot$}
\put(85,14.29){$\cdot$}
\put(85.5,14.79){$\cdot$}
\put(86,15.3){$\cdot$}
\put(86.5,15.83){$\cdot$}
\put(87,16.37){$\cdot$}
\put(87.5,16.93){$\cdot$}
\put(88,17.5){$\cdot$}
\put(88.5,18.1){$\cdot$}
\put(89,18.71){$\cdot$}
\put(89.5,19.34){$\cdot$}
\put(90,20.00){$\cdot$}
\put(90.5,20.68){$\cdot$}
\put(91,21.38){$\cdot$}
\put(91.5,22.11){$\cdot$}
\put(92,22.87){$\cdot$}
\put(92.5,23.66){$\cdot$}
\put(93,24.48){$\cdot$}
\put(93.5,25.35){$\cdot$}
\put(94,26.25){$\cdot$}
\put(94.5,27.2){$\cdot$}
\put(95,28.2){$\cdot$}
\put(95.5,29.27){$\cdot$}
\put(96,30.4){$\cdot$}
\put(96.5,31.62){$\cdot$}
\put(97,32.94){$\cdot$}
\put(97.5,34.39){$\cdot$}
\put(98,36.00){$\cdot$}
\put(98.5,37.84){$\cdot$}
\put(99,40.05){$\cdot$}
\put(99.5,42.95){$\cdot$}
\put(100,50){$\cdot$}
\put(50,100){$\cdot$}
\put(49.5,100){$\cdot$}
\put(49,100.01){$\cdot$}
\put(48.5,100.02){$\cdot$}
\put(48,100.04){$\cdot$}
\put(47.5,100.06){$\cdot$}
\put(47,100.09){$\cdot$}
\put(46.5,100.12){$\cdot$}
\put(46,100.16){$\cdot$}
\put(45.5,100.2){$\cdot$}
\put(45,100.25){$\cdot$}
\put(44.5,100.3){$\cdot$}
\put(44,100.36){$\cdot$}
\put(43.5,100.42){$\cdot$}
\put(43,100.49){$\cdot$}
\put(42.5,100.56){$\cdot$}
\put(42,100.64){$\cdot$}
\put(41.5,100.73){$\cdot$}
\put(41,100.82){$\cdot$}
\put(40.5,100.91){$\cdot$}
\put(40,101.01){$\cdot$}
\put(39.5,101.11){$\cdot$}
\put(39,101.22){$\cdot$}
\put(38.5,101.34){$\cdot$}
\put(38,101.46){$\cdot$}
\put(37.5,101.59){$\cdot$}
\put(37,101.72){$\cdot$}
\put(36.5,101.86){$\cdot$}
\put(36,102.0){$\cdot$}
\put(35.5,102.15){$\cdot$}
\put(35,102.3){$\cdot$}
\put(34.5,102.46){$\cdot$}
\put(34,102.63){$\cdot$}
\put(33.5,102.8){$\cdot$}
\put(33,102.98){$\cdot$}
\put(32.5,103.16){$\cdot$}
\put(32,103.35){$\cdot$}
\put(31.5,103.55){$\cdot$}
\put(31,103.75){$\cdot$}
\put(30.5,103.96){$\cdot$}
\put(30,104.17){$\cdot$}
\put(29.5,104.39){$\cdot$}
\put(29,104.62){$\cdot$}
\put(28.5,104.86){$\cdot$}
\put(28,105.1){$\cdot$}
\put(27.5,105.35){$\cdot$}
\put(27,105.6){$\cdot$}
\put(26.5,105.87){$\cdot$}
\put(26,106.14){$\cdot$}
\put(25.5,106.41){$\cdot$}
\put(25,106.7){$\cdot$}
\put(24.5,106.99){$\cdot$}
\put(24,107.29){$\cdot$}
\put(23.5,107.6){$\cdot$}
\put(23,107.9){$\cdot$}
\put(22.5,108.24){$\cdot$}
\put(22,108.57){$\cdot$}
\put(21.5,108.92){$\cdot$}
\put(21,109.27){$\cdot$}
\put(20.5,109.63){$\cdot$}
\put(20,110.0){$\cdot$}
\put(19.5,110.38){$\cdot$}
\put(19,110.77){$\cdot$}
\put(18.5,111.17){$\cdot$}
\put(18,111.58){$\cdot$}
\put(17.5,112.0){$\cdot$}
\put(17,112.44){$\cdot$}
\put(16.5,112.88){$\cdot$}
\put(16,113.34){$\cdot$}
\put(15.5,113.81){$\cdot$}
\put(15,114.29){$\cdot$}
\put(14.5,114.79){$\cdot$}
\put(14,115.3){$\cdot$}
\put(13.5,115.83){$\cdot$}
\put(13,116.37){$\cdot$}
\put(12.5,116.93){$\cdot$}
\put(12,117.5){$\cdot$}
\put(11.5,118.1){$\cdot$}
\put(11,118.71){$\cdot$}
\put(10.5,119.34){$\cdot$}
\put(10,120.00){$\cdot$}
\put(9.5,120.68){$\cdot$}
\put(9,121.38){$\cdot$}
\put(8.5,122.11){$\cdot$}
\put(8,122.87){$\cdot$}
\put(7.5,123.66){$\cdot$}
\put(7,124.48){$\cdot$}
\put(6.5,125.35){$\cdot$}
\put(6,126.25){$\cdot$}
\put(5.5,127.2){$\cdot$}
\put(5,128.2){$\cdot$}
\put(4.5,129.27){$\cdot$}
\put(4,130.4){$\cdot$}
\put(3.5,131.62){$\cdot$}
\put(3,132.94){$\cdot$}
\put(2.5,134.39){$\cdot$}
\put(2,136.00){$\cdot$}
\put(1.5,137.84){$\cdot$}
\put(1,140.05){$\cdot$}
\put(0.5,142.95){$\cdot$}
\put(0,150){$\cdot$}
\put(50,100){$\cdot$}
\put(50.5,100){$\cdot$}
\put(51,100.01){$\cdot$}
\put(51.5,100.02){$\cdot$}
\put(52,100.04){$\cdot$}
\put(52.5,100.06){$\cdot$}
\put(53,100.09){$\cdot$}
\put(53.5,100.12){$\cdot$}
\put(54,100.16){$\cdot$}
\put(54.5,100.2){$\cdot$}
\put(55,100.25){$\cdot$}
\put(55.5,100.3){$\cdot$}
\put(56,100.36){$\cdot$}
\put(56.5,100.42){$\cdot$}
\put(57,100.49){$\cdot$}
\put(57.5,100.56){$\cdot$}
\put(58,100.64){$\cdot$}
\put(58.5,100.73){$\cdot$}
\put(59,100.82){$\cdot$}
\put(59.5,100.91){$\cdot$}
\put(60,101.01){$\cdot$}
\put(60.5,101.11){$\cdot$}
\put(61,101.22){$\cdot$}
\put(61.5,101.34){$\cdot$}
\put(62,101.46){$\cdot$}
\put(62.5,101.59){$\cdot$}
\put(63,101.72){$\cdot$}
\put(63.5,101.86){$\cdot$}
\put(64,102.0){$\cdot$}
\put(64.5,102.15){$\cdot$}
\put(65,102.3){$\cdot$}
\put(65.5,102.46){$\cdot$}
\put(66,102.63){$\cdot$}
\put(66.5,102.8){$\cdot$}
\put(67,102.98){$\cdot$}
\put(67.5,103.16){$\cdot$}
\put(68,103.35){$\cdot$}
\put(68.5,103.55){$\cdot$}
\put(69,103.75){$\cdot$}
\put(69.5,103.96){$\cdot$}
\put(70,104.17){$\cdot$}
\put(70.5,104.39){$\cdot$}
\put(71,104.62){$\cdot$}
\put(71.5,104.86){$\cdot$}
\put(72,105.1){$\cdot$}
\put(72.5,105.35){$\cdot$}
\put(73,105.6){$\cdot$}
\put(73.5,105.87){$\cdot$}
\put(74,106.14){$\cdot$}
\put(74.5,106.41){$\cdot$}
\put(75,106.7){$\cdot$}
\put(75.5,106.99){$\cdot$}
\put(76,107.29){$\cdot$}
\put(76.5,107.6){$\cdot$}
\put(77,107.9){$\cdot$}
\put(77.5,108.24){$\cdot$}
\put(78,108.57){$\cdot$}
\put(78.5,108.92){$\cdot$}
\put(79,109.27){$\cdot$}
\put(79.5,109.63){$\cdot$}
\put(80,110.0){$\cdot$}
\put(80.5,110.38){$\cdot$}
\put(81,110.77){$\cdot$}
\put(81.5,111.17){$\cdot$}
\put(82,111.58){$\cdot$}
\put(82.5,112.0){$\cdot$}
\put(83,112.44){$\cdot$}
\put(83.5,112.88){$\cdot$}
\put(84,113.34){$\cdot$}
\put(84.5,113.81){$\cdot$}
\put(85,114.29){$\cdot$}
\put(85.5,114.79){$\cdot$}
\put(86,115.3){$\cdot$}
\put(86.5,115.83){$\cdot$}
\put(87,116.37){$\cdot$}
\put(87.5,116.93){$\cdot$}
\put(88,117.5){$\cdot$}
\put(88.5,118.1){$\cdot$}
\put(89,118.71){$\cdot$}
\put(89.5,119.34){$\cdot$}
\put(90,120.00){$\cdot$}
\put(90.5,120.68){$\cdot$}
\put(91,121.38){$\cdot$}
\put(91.5,122.11){$\cdot$}
\put(92,122.87){$\cdot$}
\put(92.5,123.66){$\cdot$}
\put(93,124.48){$\cdot$}
\put(93.5,125.35){$\cdot$}
\put(94,126.25){$\cdot$}
\put(94.5,127.2){$\cdot$}
\put(95,128.2){$\cdot$}
\put(95.5,129.27){$\cdot$}
\put(96,130.4){$\cdot$}
\put(96.5,131.62){$\cdot$}
\put(97,132.94){$\cdot$}
\put(97.5,134.39){$\cdot$}
\put(98,136.00){$\cdot$}
\put(98.5,137.84){$\cdot$}
\put(99,140.05){$\cdot$}
\put(99.5,142.95){$\cdot$}
\put(100,150){$\cdot$}
\put(0,50){$\cdot$}
\put(0.01,50.5){$\cdot$}
\put(0.01,51){$\cdot$}
\put(0.02,51.5){$\cdot$}
\put(0.04,52){$\cdot$}
\put(0.06,52.5){$\cdot$}
\put(0.09,53){$\cdot$}
\put(0.12,53.5){$\cdot$}
\put(0.16,54){$\cdot$}
\put(0.2,54.5){$\cdot$}
\put(0.25,55){$\cdot$}
\put(0.3,55.5){$\cdot$}
\put(0.36,56){$\cdot$}
\put(0.42,56.5){$\cdot$}
\put(0.49,57){$\cdot$}
\put(0.56,57.5){$\cdot$}
\put(0.64,58){$\cdot$}
\put(0.78,58.5){$\cdot$}
\put(0.82,59){$\cdot$}
\put(0.91,59.5){$\cdot$}
\put(1.01,60){$\cdot$}
\put(1.11,60.5){$\cdot$}
\put(1.22,61){$\cdot$}
\put(1.34,61.5){$\cdot$}
\put(1.46,62){$\cdot$}
\put(1.59,62.5){$\cdot$}
\put(1.72,63){$\cdot$}
\put(1.86,63.5){$\cdot$}
\put(2,64){$\cdot$}
\put(2.15,64.5){$\cdot$}
\put(2.3,65){$\cdot$}
\put(2.46,65.5){$\cdot$}
\put(2.63,66){$\cdot$}
\put(2.6,66.5){$\cdot$}
\put(2.98,67){$\cdot$}
\put(3.16,67.5){$\cdot$}
\put(3.35,68){$\cdot$}
\put(3.54,68.5){$\cdot$}
\put(3.75,69){$\cdot$}
\put(3.96,69.5){$\cdot$}
\put(4.17,70){$\cdot$}
\put(4.39,70.5){$\cdot$}
\put(4.62,71){$\cdot$}
\put(4.86,71.5){$\cdot$}
\put(5.1,72){$\cdot$}
\put(5.35,72.5){$\cdot$}
\put(5.6,73){$\cdot$}
\put(5.87,73.5){$\cdot$}
\put(6.14,74){$\cdot$}
\put(6.41,74.5){$\cdot$}
\put(6.7,75){$\cdot$}

\put(0,50){$\cdot$}
\put(0.01,49.5){$\cdot$}
\put(0.01,49){$\cdot$}
\put(0.02,48.5){$\cdot$}
\put(0.04,48){$\cdot$}
\put(0.06,47.5){$\cdot$}
\put(0.09,47){$\cdot$}
\put(0.12,46.5){$\cdot$}
\put(0.16,46){$\cdot$}
\put(0.2,45.5){$\cdot$}
\put(0.25,45){$\cdot$}
\put(0.3,44.5){$\cdot$}
\put(0.36,44){$\cdot$}
\put(0.42,43.5){$\cdot$}
\put(0.49,43){$\cdot$}
\put(0.56,42.5){$\cdot$}
\put(0.64,42){$\cdot$}
\put(0.78,41.5){$\cdot$}
\put(0.82,41){$\cdot$}
\put(0.91,40.5){$\cdot$}
\put(1.01,40){$\cdot$}
\put(1.11,39.5){$\cdot$}
\put(1.22,39){$\cdot$}
\put(1.34,38.5){$\cdot$}
\put(1.46,38){$\cdot$}
\put(1.59,37.5){$\cdot$}
\put(1.72,37){$\cdot$}
\put(1.86,36.5){$\cdot$}
\put(2,36){$\cdot$}
\put(2.15,35.5){$\cdot$}
\put(2.3,35){$\cdot$}
\put(2.46,34.5){$\cdot$}
\put(2.63,34){$\cdot$}
\put(2.6,33.5){$\cdot$}
\put(2.98,33){$\cdot$}
\put(3.16,32.5){$\cdot$}
\put(3.35,32){$\cdot$}
\put(3.54,31.5){$\cdot$}
\put(3.75,31){$\cdot$}
\put(3.96,30.5){$\cdot$}
\put(4.17,30){$\cdot$}
\put(4.39,29.5){$\cdot$}
\put(4.62,29){$\cdot$}
\put(4.86,28.5){$\cdot$}
\put(5.1,28){$\cdot$}
\put(5.35,27.5){$\cdot$}
\put(5.6,27){$\cdot$}
\put(5.87,26.5){$\cdot$}
\put(6.14,26){$\cdot$}
\put(6.41,25.5){$\cdot$}
\put(6.7,25){$\cdot$}
\put(0,150){$\cdot$}
\put(0.01,150.5){$\cdot$}
\put(0.01,151){$\cdot$}
\put(0.02,151.5){$\cdot$}
\put(0.04,152){$\cdot$}
\put(0.06,152.5){$\cdot$}
\put(0.09,153){$\cdot$}
\put(0.12,153.5){$\cdot$}
\put(0.16,154){$\cdot$}
\put(0.2,154.5){$\cdot$}
\put(0.25,155){$\cdot$}
\put(0.3,155.5){$\cdot$}
\put(0.36,156){$\cdot$}
\put(0.42,156.5){$\cdot$}
\put(0.49,157){$\cdot$}
\put(0.56,157.5){$\cdot$}
\put(0.64,158){$\cdot$}
\put(0.78,158.5){$\cdot$}
\put(0.82,159){$\cdot$}
\put(0.91,159.5){$\cdot$}
\put(1.01,160){$\cdot$}
\put(1.11,160.5){$\cdot$}
\put(1.22,161){$\cdot$}
\put(1.34,161.5){$\cdot$}
\put(1.46,162){$\cdot$}
\put(1.59,162.5){$\cdot$}
\put(1.72,163){$\cdot$}
\put(1.86,163.5){$\cdot$}
\put(2,164){$\cdot$}
\put(2.15,164.5){$\cdot$}
\put(2.3,165){$\cdot$}
\put(2.46,165.5){$\cdot$}
\put(2.63,166){$\cdot$}
\put(2.6,166.5){$\cdot$}
\put(2.98,167){$\cdot$}
\put(3.16,167.5){$\cdot$}
\put(3.35,168){$\cdot$}
\put(3.54,168.5){$\cdot$}
\put(3.75,169){$\cdot$}
\put(3.96,169.5){$\cdot$}
\put(4.17,170){$\cdot$}
\put(4.39,170.5){$\cdot$}
\put(4.62,171){$\cdot$}
\put(4.86,171.5){$\cdot$}
\put(5.1,172){$\cdot$}
\put(5.35,172.5){$\cdot$}
\put(5.6,173){$\cdot$}
\put(5.87,173.5){$\cdot$}
\put(6.14,174){$\cdot$}
\put(6.41,174.5){$\cdot$}
\put(6.7,175){$\cdot$}

\put(0,150){$\cdot$}
\put(0.01,149.5){$\cdot$}
\put(0.01,149){$\cdot$}
\put(0.02,148.5){$\cdot$}
\put(0.04,148){$\cdot$}
\put(0.06,147.5){$\cdot$}
\put(0.09,147){$\cdot$}
\put(0.12,146.5){$\cdot$}
\put(0.16,146){$\cdot$}
\put(0.2,145.5){$\cdot$}
\put(0.25,145){$\cdot$}
\put(0.3,144.5){$\cdot$}
\put(0.36,144){$\cdot$}
\put(0.42,143.5){$\cdot$}
\put(0.49,143){$\cdot$}
\put(0.56,142.5){$\cdot$}
\put(0.64,142){$\cdot$}
\put(0.78,141.5){$\cdot$}
\put(0.82,141){$\cdot$}
\put(0.91,140.5){$\cdot$}
\put(1.01,140){$\cdot$}
\put(1.11,139.5){$\cdot$}
\put(1.22,139){$\cdot$}
\put(1.34,138.5){$\cdot$}
\put(1.46,138){$\cdot$}
\put(1.59,137.5){$\cdot$}
\put(1.72,137){$\cdot$}
\put(1.86,136.5){$\cdot$}
\put(2,136){$\cdot$}
\put(2.15,135.5){$\cdot$}
\put(2.3,135){$\cdot$}
\put(2.46,134.5){$\cdot$}
\put(2.63,134){$\cdot$}
\put(2.6,133.5){$\cdot$}
\put(2.98,133){$\cdot$}
\put(3.16,132.5){$\cdot$}
\put(3.35,132){$\cdot$}
\put(3.54,131.5){$\cdot$}
\put(3.75,131){$\cdot$}
\put(3.96,130.5){$\cdot$}
\put(4.17,130){$\cdot$}
\put(4.39,129.5){$\cdot$}
\put(4.62,129){$\cdot$}
\put(4.86,128.5){$\cdot$}
\put(5.1,128){$\cdot$}
\put(5.35,127.5){$\cdot$}
\put(5.6,127){$\cdot$}
\put(5.87,126.5){$\cdot$}
\put(6.14,126){$\cdot$}
\put(6.41,125.5){$\cdot$}
\put(6.7,125){$\cdot$}

\put(100,50){$\cdot$}
\put(99.99,50.5){$\cdot$}
\put(99.99,51){$\cdot$}
\put(99.98,51.5){$\cdot$}
\put(99.96,52){$\cdot$}
\put(99.94,52.5){$\cdot$}
\put(99.91,53){$\cdot$}
\put(99.88,53.5){$\cdot$}
\put(99.84,54){$\cdot$}
\put(99.8,54.5){$\cdot$}
\put(99.75,55){$\cdot$}
\put(99.7,55.5){$\cdot$}
\put(99.64,56){$\cdot$}
\put(99.58,56.5){$\cdot$}
\put(99.51,57){$\cdot$}
\put(99.44,57.5){$\cdot$}
\put(99.36,58){$\cdot$}
\put(99.22,58.5){$\cdot$}
\put(99.18,59){$\cdot$}
\put(99.09,59.5){$\cdot$}
\put(98.99,60){$\cdot$}
\put(98.89,60.5){$\cdot$}
\put(98.78,61){$\cdot$}
\put(98.66,61.5){$\cdot$}
\put(98.54,62){$\cdot$}
\put(98.41,62.5){$\cdot$}
\put(98.28,63){$\cdot$}
\put(98.14,63.5){$\cdot$}
\put(98,64){$\cdot$}
\put(97.85,64.5){$\cdot$}
\put(97.7,65){$\cdot$}
\put(97.54,65.5){$\cdot$}
\put(97.37,66){$\cdot$}
\put(97.4,66.5){$\cdot$}
\put(97.02,67){$\cdot$}
\put(96.84,67.5){$\cdot$}
\put(96.65,68){$\cdot$}
\put(96.46,68.5){$\cdot$}
\put(96.25,69){$\cdot$}
\put(96.04,69.5){$\cdot$}
\put(95.83,70){$\cdot$}
\put(95.61,70.5){$\cdot$}
\put(95.38,71){$\cdot$}
\put(95.14,71.5){$\cdot$}
\put(94.9,72){$\cdot$}
\put(94.65,72.5){$\cdot$}
\put(94.4,73){$\cdot$}
\put(94.13,73.5){$\cdot$}
\put(93.86,74){$\cdot$}
\put(93.59,74.5){$\cdot$}
\put(93.3,75){$\cdot$}

\put(100,50){$\cdot$}
\put(99.99,49.5){$\cdot$}
\put(99.99,49){$\cdot$}
\put(99.98,48.5){$\cdot$}
\put(99.96,48){$\cdot$}
\put(99.94,47.5){$\cdot$}
\put(99.91,47){$\cdot$}
\put(99.88,46.5){$\cdot$}
\put(99.84,46){$\cdot$}
\put(99.8,45.5){$\cdot$}
\put(99.75,45){$\cdot$}
\put(99.7,44.5){$\cdot$}
\put(99.64,44){$\cdot$}
\put(99.58,43.5){$\cdot$}
\put(99.51,43){$\cdot$}
\put(99.44,42.5){$\cdot$}
\put(99.36,42){$\cdot$}
\put(99.22,41.5){$\cdot$}
\put(99.18,41){$\cdot$}
\put(99.09,40.5){$\cdot$}
\put(98.99,40){$\cdot$}
\put(98.89,39.5){$\cdot$}
\put(98.78,39){$\cdot$}
\put(98.66,38.5){$\cdot$}
\put(98.54,38){$\cdot$}
\put(98.41,37.5){$\cdot$}
\put(98.28,37){$\cdot$}
\put(98.14,36.5){$\cdot$}
\put(98,36){$\cdot$}
\put(97.85,35.5){$\cdot$}
\put(97.7,35){$\cdot$}
\put(97.54,34.5){$\cdot$}
\put(97.37,34){$\cdot$}
\put(97.4,33.5){$\cdot$}
\put(97.02,33){$\cdot$}
\put(96.84,32.5){$\cdot$}
\put(96.65,32){$\cdot$}
\put(96.46,31.5){$\cdot$}
\put(96.25,31){$\cdot$}
\put(96.04,30.5){$\cdot$}
\put(95.83,30){$\cdot$}
\put(95.61,29.5){$\cdot$}
\put(95.38,29){$\cdot$}
\put(95.14,28.5){$\cdot$}
\put(94.9,28){$\cdot$}
\put(94.65,27.5){$\cdot$}
\put(94.4,27){$\cdot$}
\put(94.13,26.5){$\cdot$}
\put(93.86,26){$\cdot$}
\put(93.59,25.5){$\cdot$}
\put(93.3,25){$\cdot$}

\put(100,150){$\cdot$}
\put(99.99,150.5){$\cdot$}
\put(99.99,151){$\cdot$}
\put(99.98,151.5){$\cdot$}
\put(99.96,152){$\cdot$}
\put(99.94,152.5){$\cdot$}
\put(99.91,153){$\cdot$}
\put(99.88,153.5){$\cdot$}
\put(99.84,154){$\cdot$}
\put(99.8,154.5){$\cdot$}
\put(99.75,155){$\cdot$}
\put(99.7,155.5){$\cdot$}
\put(99.64,156){$\cdot$}
\put(99.58,156.5){$\cdot$}
\put(99.51,157){$\cdot$}
\put(99.44,157.5){$\cdot$}
\put(99.36,158){$\cdot$}
\put(99.22,158.5){$\cdot$}
\put(99.18,159){$\cdot$}
\put(99.09,159.5){$\cdot$}
\put(98.99,160){$\cdot$}
\put(98.89,160.5){$\cdot$}
\put(98.78,161){$\cdot$}
\put(98.66,161.5){$\cdot$}
\put(98.54,162){$\cdot$}
\put(98.41,162.5){$\cdot$}
\put(98.28,163){$\cdot$}
\put(98.14,163.5){$\cdot$}
\put(98,164){$\cdot$}
\put(97.85,164.5){$\cdot$}
\put(97.7,165){$\cdot$}
\put(97.54,165.5){$\cdot$}
\put(97.37,166){$\cdot$}
\put(97.4,166.5){$\cdot$}
\put(97.02,167){$\cdot$}
\put(96.84,167.5){$\cdot$}
\put(96.65,168){$\cdot$}
\put(96.46,168.5){$\cdot$}
\put(96.25,169){$\cdot$}
\put(96.04,169.5){$\cdot$}
\put(95.83,170){$\cdot$}
\put(95.61,170.5){$\cdot$}
\put(95.38,171){$\cdot$}
\put(95.14,171.5){$\cdot$}
\put(94.9,172){$\cdot$}
\put(94.65,172.5){$\cdot$}
\put(94.4,173){$\cdot$}
\put(94.13,173.5){$\cdot$}
\put(93.86,174){$\cdot$}
\put(93.59,174.5){$\cdot$}
\put(93.3,175){$\cdot$}

\put(100,150){$\cdot$}
\put(99.99,149.5){$\cdot$}
\put(99.99,149){$\cdot$}
\put(99.98,148.5){$\cdot$}
\put(99.96,148){$\cdot$}
\put(99.94,147.5){$\cdot$}
\put(99.91,147){$\cdot$}
\put(99.88,146.5){$\cdot$}
\put(99.84,146){$\cdot$}
\put(99.8,145.5){$\cdot$}
\put(99.75,145){$\cdot$}
\put(99.7,144.5){$\cdot$}
\put(99.64,144){$\cdot$}
\put(99.58,143.5){$\cdot$}
\put(99.51,143){$\cdot$}
\put(99.44,142.5){$\cdot$}
\put(99.36,142){$\cdot$}
\put(99.22,141.5){$\cdot$}
\put(99.18,141){$\cdot$}
\put(99.09,140.5){$\cdot$}
\put(98.99,140){$\cdot$}
\put(98.89,139.5){$\cdot$}
\put(98.78,139){$\cdot$}
\put(98.66,138.5){$\cdot$}
\put(98.54,138){$\cdot$}
\put(98.41,137.5){$\cdot$}
\put(98.28,137){$\cdot$}
\put(98.14,136.5){$\cdot$}
\put(98,136){$\cdot$}
\put(97.85,135.5){$\cdot$}
\put(97.7,135){$\cdot$}
\put(97.54,134.5){$\cdot$}
\put(97.37,134){$\cdot$}
\put(97.4,133.5){$\cdot$}
\put(97.02,133){$\cdot$}
\put(96.84,132.5){$\cdot$}
\put(96.65,132){$\cdot$}
\put(96.46,131.5){$\cdot$}
\put(96.25,131){$\cdot$}
\put(96.04,130.5){$\cdot$}
\put(95.83,130){$\cdot$}
\put(95.61,129.5){$\cdot$}
\put(95.38,129){$\cdot$}
\put(95.14,128.5){$\cdot$}
\put(94.9,128){$\cdot$}
\put(94.65,127.5){$\cdot$}
\put(94.4,127){$\cdot$}
\put(94.13,126.5){$\cdot$}
\put(93.86,126){$\cdot$}
\put(93.59,125.5){$\cdot$}
\put(93.3,125){$\cdot$}

\end{picture}
\]
\vspace{4ex}
\begin{center}
\begin{minipage}{25pc}{{\renewcommand{\baselinestretch}{1.2}\small
{\rm Figure 1}\;
Two--dimensional complex sphere $z^2_1+z^2_2+z^2_3=r^2$ in
three--dimensional complex space $\C^3$. The space $\C^3$ is isometric
to the bivector space $\R^6$. The dual (complex conjugate) sphere
$\overset{\ast}{z}_1{}^2+\overset{\ast}{z}_2{}^2+\overset{\ast}{z}_3{}^2=
\overset{\ast}{r}{}^2$ is a mirror image of the complex sphere with
respect to the hyperplane. The associated hyperspherical functions
$\fM_l^{m}(\varphi^c,\theta^c,0)$ 
($\fM_{\dot{l}}^{\dot{m}}(\dot{\varphi}^c,\dot{\theta}^c,0)$)
are defined on the surface of the complex (dual) sphere.}}
\end{minipage}
\end{center}
\label{Sphere}
\end{figure}

\subsection{Symmetry relations for the functions $Z^l_{mn}(\theta,\tau)$}
The hyperspherical functions $Z^l_{mn}(\theta,\tau)$ satisfy some
symmetry relations with respect to the indices $l$, $m$, $n$.
Let us show that these functions satisfy the relation
\begin{equation}\label{Symm1}
Z^l_{mn}(\theta,\tau)=Z^l_{-m,-n}(\theta,\tau).
\end{equation}
Consider the operator $S$ transforming $f(e^{i\vartheta})$ into
$e^{2io\vartheta}f(e^{-i\vartheta})$. It is easy to verify that 
$S$ commutes with an operator $T_\chi(\fg(\theta,\tau))$, where
\[
\fg(\theta,\tau)={\renewcommand{\arraystretch}{1.3}
\begin{pmatrix}
\cos\frac{\theta}{2}\ch\frac{\tau}{2}+i\sin\frac{\theta}{2}\sh\frac{\tau}{2} &
\cos\frac{\theta}{2}\sh\frac{\tau}{2}+i\sin\frac{\theta}{2}\ch\frac{\tau}{2}\\
\cos\frac{\theta}{2}\sh\frac{\tau}{2}+i\sin\frac{\theta}{2}\ch\frac{\tau}{2} &
\cos\frac{\theta}{2}\ch\frac{\tau}{2}+i\sin\frac{\theta}{2}\sh\frac{\tau}{2}
\end{pmatrix}},
\]
that is,
\begin{equation}\label{Symm2}
ST_\chi(\fg(\theta,\tau))=T_\chi(\fg(\theta,\tau))S.
\end{equation}
Matrix elements $s_{mn}$ of the operator $S$ in the basis 
$\{e^{-im\vartheta}\}$ equal to zero if $m+n\neq-2o$, and equal to the unit
if $m+n=-2o$. Multiplying the matrices in (\ref{Symm2}) and comparing
the corresponding elements, we obtain
\[
t^\chi_{-2o-m,n}(\fg(\theta,\tau))=t^\chi_{m,-2o-n}(\fg(\theta,\tau)).
\]
Since $t^\chi_{mn}(\fg(\theta,\tau))=Z^l_{m^\prime,n^\prime}(\theta,\tau)$,
where $m^\prime=m+o$, $n^\prime=n+o$, then
\[
Z^l_{-m-o,n+o}(\theta,\tau)=Z^l_{m+o,-n-o}(\theta,\tau).
\]
Hence it follows that
\[
Z^l_{mn}(\theta,\tau)=Z^l_{-m,-n}(\theta,\tau)
\]
for all the complex values of $l$ and all integer or half-integer
values of $m$ and $n$. Further, replacing in the expression (\ref{HS})
$m$ by $-n$, and $n$ by $-m$, we obtain
\[
Z^l_{-n,-m}(\theta,\tau)=Z^l_{m,n}(\theta,\tau)
\]
or, in virtue of the relation (\ref{Symm1}),
\begin{equation}\label{Symm3}
Z^l_{nm}(\theta,\tau)=Z^l_{mn}(\theta,\tau).
\end{equation}
Therefore, the hyperspherical functions $Z^l_{mn}(\theta,\tau)$ are
symmetric.\subsection{Matrices $T_l(\mathfrak{g})$}
Using the formula (\ref{HS}), let us find explicit expressions for the
matrices $T_l(\mathfrak{g})$ of the finite--dimensional representations of
$\fG_+$ at $l=0,\frac{1}{2},1$:
\begin{gather}
T_0(\theta,\tau)=1,\nonumber\\[0.3cm]
T_{\frac{1}{2}}(\theta,\tau)=\begin{pmatrix}
Z^{\frac{1}{2}}_{-\frac{1}{2}-\frac{1}{2}} &
Z^{\frac{1}{2}}_{\frac{1}{2}-\frac{1}{2}}\\
Z^{\frac{1}{2}}_{-\frac{1}{2}\frac{1}{2}} &
Z^{\frac{1}{2}}_{\frac{1}{2}\frac{1}{2}}
\end{pmatrix}=\nonumber\\[0.3cm]
{\renewcommand{\arraystretch}{1.3}
\begin{pmatrix}
\cos\frac{\theta}{2}\ch\frac{\tau}{2}+
i\sin\frac{\theta}{2}\sh\frac{\tau}{2} &
\cos\frac{\theta}{2}\sh\frac{\tau}{2}+
i\sin\frac{\theta}{2}\ch\frac{\tau}{2} \\
\cos\frac{\theta}{2}\sh\frac{\tau}{2}+
i\sin\frac{\theta}{2}\ch\frac{\tau}{2} &
\cos\frac{\theta}{2}\ch\frac{\tau}{2}+
i\sin\frac{\theta}{2}\sh\frac{\tau}{2} 
\end{pmatrix}},\label{T1}\\[0.4cm]
T_1(\theta,\tau)=\begin{pmatrix}
Z^1_{-1-1} & Z^1_{-10} & Z^1_{-11}\\
Z^1_{0-1} & Z^1_{00} & Z^1_{01}\\
Z^1_{1-1} & Z^1_{10} & Z^1_{11}
\end{pmatrix}=\nonumber
\end{gather}
\begin{multline}
{\renewcommand{\arraystretch}{1.8}\left(\begin{array}{cc}\s
\cos^2\tfrac{\theta}{2}\ch^2\tfrac{\tau}{2}+\tfrac{i\sin\theta\sh\tau}{2}-
\sin^2\tfrac{\theta}{2}\sh^2\tfrac{\tau}{2} &\s
\tfrac{1}{\sqrt{2}}(\cos\theta\sh\tau+i\sin\theta\ch\tau) \\
\s\tfrac{1}{\sqrt{2}}(\cos\theta\sh\tau+i\sin\theta\ch\tau) &
\s\cos\theta\ch\tau+i\sin\theta\sh\tau \\
\s\cos^2\tfrac{\theta}{2}\sh^2\tfrac{\tau}{2}+\tfrac{i\sin\theta\sh\tau}{2}-
\sin^2\tfrac{\theta}{2}\ch^2\tfrac{\tau}{2} &
\s
\tfrac{1}{\sqrt{2}}(\cos\theta\sh\tau+i\sin\theta\ch\tau) 
\end{array}\right.}\\
{\renewcommand{\arraystretch}{1.8}\left.\begin{array}{c}\s
\cos^2\tfrac{\theta}{2}\sh^2\tfrac{\tau}{2}+\tfrac{i\sin\theta\sh\tau}{2}-
\sin^2\tfrac{\theta}{2}\ch^2\tfrac{\tau}{2} \\
\s\tfrac{1}{\sqrt{2}}(\cos\theta\sh\tau+i\sin\theta\ch\tau) \\
\s\cos^2\tfrac{\theta}{2}\ch^2\tfrac{\tau}{2}+\tfrac{i\sin\theta\sh\tau}{2}-
\sin^2\tfrac{\theta}{2}\sh^2\tfrac{\tau}{2} 
\end{array}\right).}\label{T2}
\end{multline}
\subsection{Addition Theorem}\index{theorem!addition}
Let $\mathfrak{g}=\mathfrak{g}_1\mathfrak{g}_2$ be the product of two
matrices $\mathfrak{g}_1,\,\mathfrak{g}_2\in SL(2,\C)$. Let us denote
the Euler angles of the matrix $\mathfrak{g}$ via $\varphi^c,\theta^c,\psi^c$,
of the matrix $\mathfrak{g}_1$ via $\varphi^c_1,\theta^c_1,\psi^c_1$ and of the
matrix $\mathfrak{g}_2$ via $\varphi^c_2,\theta^c_2,\psi^c_2$.
Expressing now the Euler angles\index{Euler angles}
of the matrix $\mathfrak{g}$ via the
Euler angles of the factors $\mathfrak{g}_1,\,\mathfrak{g}_2$, we consider
at first the particular case $\varphi^c_1=\psi^c_1=\psi^c_2=0$:
\[
\mathfrak{g}=
{\renewcommand{\arraystretch}{1.3}
\begin{pmatrix}
\cos\frac{\theta^c_1}{2} & i\sin\frac{\theta^c_1}{2}\\
i\sin\frac{\theta^c_1}{2} & \cos\frac{\theta^c_1}{2}
\end{pmatrix}\!\!\!
\begin{pmatrix}
\cos\frac{\theta^c_2}{2}e^{\frac{i\varphi^c_2}{2}} &
i\sin\frac{\theta^c_2}{2}e^{\frac{i\varphi^c_2}{2}} \\
i\sin\frac{\theta^c_2}{2}e^{-\frac{i\varphi^c_2}{2}} &
\cos\frac{\theta^c_2}{2}e^{-\frac{i\varphi^c_2}{2}}
\end{pmatrix}}.
\]
Multiplying the matrices in the right part of this equality and using
a complex analog of the formulae (\ref{SU3})--(\ref{SU5}), we obtain
\begin{eqnarray}
\cos\theta^c&=&\cos\theta^c_1\cos\theta^c_2-
\sin\theta^c_1\sin\theta^c_2\cos\varphi^c_2,\label{SL2}\\[0.2cm]
e^{i\varphi^c}&=&\frac{\sin\theta^c_1\cos\theta^c_2+
\cos\theta^c_1\sin\theta^c_2\cos\varphi^c_2+i\sin\theta^c_2\sin\varphi^c_2}
{\sin\theta^c},\label{SL3}\\
e^{\frac{i(\varphi^c+\psi^c)}{2}}&=&\frac{\cos\frac{\theta^c_1}{2}
\cos\frac{\theta^c_2}{2}e^{\frac{i\varphi^c_2}{2}}-
\sin\frac{\theta^c_1}{2}\sin\frac{\theta^c_2}{2}e^{-\frac{i\varphi^c_2}{2}}}
{\cos\frac{\theta^c}{2}}.\label{SL4}
\end{eqnarray}
It is not difficult to obtain a general case. Indeed, in virtue of
(\ref{FUN}) the matrix $\mathfrak{g}\in SL(2,\C)$ admits a representation
\begin{eqnarray}
\mathfrak{g}(\varphi^c,\theta^c,\psi^c)&=&
\begin{pmatrix}
e^{\frac{i\varphi^c}{2}} & 0\\
0 & e^{-\frac{i\varphi^c}{2}}
\end{pmatrix}\!\!\!{\renewcommand{\arraystretch}{1.3}
\begin{pmatrix}
\cos\frac{\theta^c}{2} & i\sin\frac{\theta^c}{2}\\
i\sin\frac{\theta^c}{2} & \cos\frac{\theta^c}{2}
\end{pmatrix}}\!\!\!
\begin{pmatrix}
e^{\frac{i\psi^c}{2}} & 0\\
0 & e^{-\frac{i\psi^c}{2}}
\end{pmatrix}
\equiv\nonumber\\
&\equiv&
\mathfrak{g}(\varphi^c,0,0)\mathfrak{g}(0,\theta^c,0)\mathfrak{g}(0,0,\psi^c).
\nonumber
\end{eqnarray}
Therefore,
\begin{multline}
\mathfrak{g}(\varphi^c_1,\theta^c_1,\psi^c_1)
\mathfrak{g}(\varphi^c_2,\theta^c_2,\psi^c_2)=\\
\mathfrak{g}(\varphi^c_1,0,0)\mathfrak{g}(0,\theta^c_1,0)
\mathfrak{g}(0,0,\psi^c_1)\mathfrak{g}(\varphi^c_2,0,0)
\mathfrak{g}(0,\theta^c_2,0)\mathfrak{g}(0,0,\psi^c_2).
\end{multline}
It is obvious that
\[
\mathfrak{g}(0,0,\psi^c_1)\mathfrak{g}(\varphi^c_2,0,0)=
\mathfrak{g}(\varphi^c_2+\psi^c_1,0,0).
\]
Besides, if we multiply the matrix $\mathfrak{g}(\varphi^c,\theta^c,\psi^c)$
at the left by the matrix $\mathfrak{g}(\varphi^c_1,0,0)$, the Euler angle
$\varphi^c$ increases by $\varphi^c_1$, and other Euler angles remain
unaltered. Analogously, if we multiply at the right the matrix
$\mathfrak{g}(\varphi^c,\theta^c,\psi^c)$ by
$\mathfrak{g}(0,0,\psi^c_2)$, the angle $\psi^c$ increases by $\psi^c_2$.
Hence it follows that in general case the angle $\varphi^c_2$ should be
replaced by $\varphi^c_2+\psi^c_1$, and the angles $\varphi^c$ and
$\psi^c$ should be replaced 
by $\varphi^c-\varphi^c_1$ and $\psi^c-\psi^c_2$, that is,
\begin{eqnarray}
\cos\theta^c&=&\cos\theta^c_1\cos\theta^c_2-\sin\theta^c_1\sin\theta^c_2
\cos(\varphi^c_2+\psi^c_1),\nonumber\\[0.2cm]
e^{i(\varphi^c-\varphi^c_1)}&=&
\frac{\sin\theta^c_1\cos\theta^c_2+
\cos\theta^c_1\sin\theta^c_2\cos(\varphi^c_2+\psi^c_1)+
i\sin\theta^c_2\sin(\varphi^c_2+\psi^c_1)}{\sin\theta^c},\nonumber\\
e^{\frac{i(\varphi^c+\psi^c-\varphi^c_1-\psi^c_2)}{2}}&=&
\frac{\cos\frac{\theta^c_1}{2}\cos\frac{\theta^c_2}{2}
e^{\frac{i(\varphi^c_2+\psi^c_1)}{2}}-\sin\frac{\theta^c_1}{2}
\sin\frac{\theta^c_2}{2}e^{-\frac{i(\varphi^c_2+\psi^c_1)}{2}}}
{\cos\dfrac{\theta^c}{2}}.\label{SL7}
\end{eqnarray}

Addition Theorem for hyperspherical functions\index{function!hyperspherical}
$Z^l_{mn}$ follows from the
relation
\[
T_l(\mathfrak{g}_1\mathfrak{g}_2)=T_l(\mathfrak{g}_1)T_l(\mathfrak{g}_2).
\]
Hence it follows that
\[
t^l_{mn}(\mathfrak{g}_1\mathfrak{g}_2)=\sum^l_{k=-l}
t^l_{mk}(\mathfrak{g}_1)t^l_{kn}(\mathfrak{g}_2).
\]
Let us apply this equality to the matrices $\mathfrak{g}_1$ and
$\mathfrak{g}_2$  with the Euler angles $0,0,\theta_1,\tau_1,0,0$ and
$\varphi_2,\epsilon_2,\theta_2,\tau_2,0,0$, respectively. Using the
formula (\ref{HS}), we obtain
\begin{eqnarray}
t^l_{mk}(\mathfrak{g}_1)&=&Z^l_{mk}(\cos\theta_1,\ch\tau_1),\nonumber\\
t^l_{kn}(\mathfrak{g}_2)&=&e^{-k(\epsilon_2+i\varphi_2)}
Z^l_{kn}(\cos\theta_2,\ch\tau_2)\nonumber
\end{eqnarray}
and
\[
t^l_{mn}(\mathfrak{g}_1\mathfrak{g}_2)=e^{-m(\epsilon+i\varphi)-
n(\varepsilon+i\psi)}Z^l_{mn}(\cos\theta,\ch\tau),
\]
where $\epsilon,\varphi,\theta,\tau,\varepsilon,\psi$ are the Euler angles
of the matrix $\mathfrak{g}_1\mathfrak{g}_2$. In accordance with
(\ref{SL2})--(\ref{SL4}) these angles are expressed via the factor angles
$0,0,\theta_1,\tau_1,0,0$ and $\varphi_2,\epsilon_2,\theta_2,\tau_2,0,0$.
Thus, in this case the functions $Z^l_{mn}$ satisfy the following addition
theorem:
\begin{equation}\label{AdTheor}
e^{-m(\epsilon+i\varphi)-n(\varepsilon+i\psi)}
Z^l_{mn}(\cos\theta^c)=
\sum^l_{k=-l}e^{-k(\epsilon_2+i\varphi_2)}Z^l_{mk}(\cos\theta^c_1)
Z^l_{kn}(\cos\theta^c_2).
\end{equation}
In general case when the Euler angles are related by the formulae
(\ref{SL7}) we obtain
\begin{multline}
e^{-m[\epsilon+\epsilon_1+i(\varphi_1-\varphi)]-n[\varepsilon+\varepsilon_2-
i(\psi_2-\psi)]}Z^l_{mn}(\cos\theta^c)=\\
\sum^l_{k=-l}e^{-k[\epsilon_2+\varepsilon_1+i(\varphi_2+\psi_1)]}
Z^l_{mk}(\cos\theta^c_1)Z^l_{kn}(\cos\theta^c_2).\nonumber
\end{multline}

Addition theorems for the associated and zonal hyperspherical functions
follow as particular cases from the proved theorem for the functions
$Z^l_{mn}(\theta,\tau)$. In the subsections \ref{Zonal} and \ref{Associated}
these functions have been defined by the formulas
\[
Z_l(\cos\theta^c)=Z^l_{00}(\cos\theta^c)
\]
and
\[
Z^m_l(\cos\theta^c)=Z^l_{m0}(\cos\theta^c).
\]
Supposing $n=0$ in the formula (\ref{AdTheor}), we obtain an addition theorem
for the associated hyperspherical functions:
\[
e^{-m(\epsilon+i\varphi)}Z^m_l(\cos\theta^c)=
\sum^l_{k=-l}e^{-k(\epsilon_2+i\varphi_2)}Z^l_{mk}(\cos\theta^c_1)
Z^k_l(\cos\theta^c_2),
\]
where the angles $\varphi^c$, $\varphi^c_2$, $\theta^c$, $\theta^c_1$,
$\theta^c_2$ are related by (\ref{SL2})--(\ref{SL4}). Supposing
$m=0$, $n=0$, we obtain
\[
Z_l(\cos\theta^c)=\sum^l_{k=-l}e^{-k(\epsilon_2+i\varphi_2)}
Z^k_l(\cos\theta^c_1)Z^k_l(\cos\theta^c_2).
\]\subsection{Matrix elements of the principal series representations of the
Lorentz group}
As it has been shown in \cite{Nai58}, for the case of principal series
representations there exists an analog of the spinor representation
formula (\ref{Rep}):
\begin{equation}\label{Principal}
V_af(z)=(a_{12}z+a_{22})^{\frac{\lambda}{2}+i\frac{\rho}{2}-1}
\overline{(a_{12}z+a_{22})}^{-\frac{\lambda}{2}+i\frac{\rho}{2}-1}
f\left(\frac{a_{11}z+a_{21}}{a_{12}z+a_{22}}\right),
\end{equation}
where $f(z)$ is a measurable functions of the Hilbert space $L_2(Z)$,
satisfying the condition $\int|f(z)|^2dz<\infty$, $z=x+iy$. At this point,
the numbers
$l_0$, $l_1$ and $\lambda$, $\rho$ are related by the formulas
\begin{gather}
l_0=\left|\frac{\lambda}{2}\right|,\quad
l_1=-i(\sign\lambda)\frac{\rho}{2}\quad\text{if $m\neq 0$},\nonumber\\
l_0=0,\quad
l_1=\pm i\frac{\rho}{2}\quad\text{if $m=0$}.\nonumber
\end{gather}
A totality of all representations
$a\rightarrow V_a$, corresponding to all possible pairs
$\lambda$, $\rho$, is called a principal series of representations of the
group
$SL(2,\C)$. At this point, a comparison of (\ref{Principal}) 
with the formula (\ref{Rep}) for the spinor representation
$\fS_l$ shows that the both formulas have the same structure;
only the exponents at the factors
$(a_{12}z+a_{22})$, $\overline{(a_{12}z+a_{22})}$ and the functions $f(z)$
are different. In the case of spinor representations the functions
$f(z)$ are polynomials
$p(z,\bar{z})$ in the spaces
$\Sym_{(k,r)}$, and in the case of a representation
$\fS_{\lambda,\rho}$ of the principal series $f(z)$ are functions from
the Hilbert space $L_2(Z)$.

As is known, a representation $S_l$ of the group $SU(2)$ is realized in terms
of the functions $P^l_{mn}(\cos\theta)$. 
\begin{theorem}[{\rm Naimark \cite{Nai58}}]
The representation $S_l$ is contained in
$\fS_{\lambda,\rho}$ no more then one time. At this point, $S_l$ is
contained in
$\fS_{\lambda,\rho}$, when $\frac{\lambda}{2}$ is one from the numbers
$-l,-l+1,\ldots,l$.
\end{theorem}
Therefore, matrix elements of the principal series representations of the
Lorentz group, making infinite-dimensional matrix, have the form
(see also \cite{FNR66}):
\begin{multline}
t^{-\frac{1}{2}+i\rho}_{mn}(\mathfrak{g})=
e^{-m(\epsilon+i\varphi)-n(\varepsilon+i\psi)}
Z^{-\frac{1}{2}+i\rho}_{mn}=
e^{-m(\epsilon+i\varphi)-n(\varepsilon+i\psi)}\times\\[0.2cm]
\sum^{\ld\frac{\lambda}{2}\rd}_{k=-\frac{\lambda}{2}}i^{m-k}
\sqrt{\Gamma(\frac{\lambda}{2}-m+1)
\Gamma(\frac{\lambda}{2}+m+1)\Gamma(\frac{\lambda}{2}-k+1)
\Gamma(\frac{\lambda}{2}+k+1)}\times\\
\cos^{\lambda}\frac{\theta}{2}\tg^{m-k}\frac{\theta}{2}\times\\[0.2cm]
\sum^{\ld\min\left(\frac{\lambda}{2}-m,\frac{\lambda}{2}+k\right)
\rd}_{j=\max(0,k-m)}
\frac{i^{2j}\tg^{2j}\dfrac{\theta}{2}}
{\Gamma(j+1)\Gamma(\frac{\lambda}{2}-m-j+1)
\Gamma(\frac{\lambda}{2}+k-j+1)\Gamma(m-k+j+1)}\times\\[0.2cm]
\sqrt{\Gamma(\frac{1}{2}+i\rho-n)
\Gamma(\frac{1}{2}+i\rho+n)\Gamma(\frac{1}{2}+i\rho-k)
\Gamma(\frac{1}{2}+i\rho+k)}
\ch^{-1+2i\rho}\frac{\tau}{2}\tnh^{n-k}\frac{\tau}{2}\times\\[0.2cm]
\sum^{\infty}_{s=\max(0,k-n)}
\frac{\tnh^{2s}\dfrac{\tau}{2}}
{\Gamma(s+1)\Gamma(\frac{1}{2}+i\rho-n-s)
\Gamma(\frac{1}{2}+i\rho+k-s)\Gamma(n-k+s+1)}.\label{MPrincip}
\end{multline}
Thus, the matrix elements of the principal series representations of the
group $\fG_+$ are expressed via the function
\begin{equation}\label{MEPS}
\fM^{-\frac{1}{2}+i\rho}_{mn}(\fg)=e^{-m(\epsilon+i\varphi)}
Z^{-\frac{1}{2}+i\rho}_{mn}(\cos\theta^c)e^{-n(\varepsilon+i\psi)},
\end{equation}
where
\[
Z^{-\frac{1}{2}+i\rho}_{mn}(\cos\theta^c)=
\sum^{\ld\frac{\lambda}{2}\rd}_{k=-\frac{\lambda}{2}}
P^{\frac{\lambda}{2}}_{mk}(\cos\theta)\fP^{-\frac{1}{2}+i\rho}_{kn}(\ch\tau).
\]
In the case of associated functions ($n=0$) we obtain
\begin{equation}\label{AHPS}
Z^m_{-\frac{1}{2}+i\rho}(\cos\theta^c)=
\sum^{\ld\frac{\lambda}{2}\rd}_{k=-\frac{\lambda}{2}}
P^{\frac{\lambda}{2}}_{mk}(\cos\theta)\fP^k_{-\frac{1}{2}+i\rho}(\ch\tau),
\end{equation}
where $\fP^k_{-\frac{1}{2}+i\rho}(\ch\tau)$ are {\it conical functions}
(see \cite{Bat}). In this case our result agrees with the paper \cite{FNR66},
where matrix elements (eigenfunctions of Casimir operators) of noncompact
rotation groups are expressed in terms of conical and spherical functions
(see also \cite{Vil68}).

Further, at $\lambda=0$ and $\rho=i\sigma$ from (\ref{Principal}) it follows
that
\[
V_af(z)=|a_{12}z+a_{22}|^{-2-\sigma}f
\left(\frac{a_{11}z+a_{21}}{a_{12}z+a_{22}}\right).
\]
This formula defines an unitary representation $a\rightarrow V_a$ of
supplementary series
$\fD_\sigma$ of the group $SL(2,\C)$. At this point, for the supplementary
series the relations
\[
l_0=0,\quad l_1=\pm\frac{\sigma}{2}
\]
hold. In turn, the representation $S_l$ of the group $SU(2)$ is contained
in the representation $\fD_\sigma$ of supplementary series when
$l$ is an integer number. In this case $S_l$ is contained in $\fD_\sigma$
exactly one time and the number $\frac{\lambda}{2}=0$ is one from the set
$-l,-l+1,\ldots,l$ \cite{Nai58}.

Thus, matrix elements of supplementary series appear as a particular case
of the matrix elements of principal series at $\lambda=0$
and $\rho=i\sigma$:
\begin{multline}
t^{-\frac{1}{2}-\sigma}_{mn}(\mathfrak{g})=
e^{-m(\epsilon+i\varphi)-n(\varepsilon+i\psi)}
Z^{-\frac{1}{2}-\sigma}_{mn}=
e^{-m(\epsilon+i\varphi)-n(\varepsilon+i\psi)}\times\\[0.2cm]
\sqrt{\Gamma(\tfrac{1}{2}-\sigma-n)
\Gamma(\tfrac{1}{2}-\sigma+n)\Gamma(\tfrac{1}{2}-\sigma-k)
\Gamma(\tfrac{1}{2}-\sigma+k)}
\ch^{-1-2\sigma}\frac{\tau}{2}\tnh^{n-k}\frac{\tau}{2}\times\\[0.2cm]
\sum^{\infty}_{s=\max(0,k-n)}
\frac{\tnh^{2s}\dfrac{\tau}{2}}
{\Gamma(s+1)\Gamma(\tfrac{1}{2}-\sigma-n-s)
\Gamma(\tfrac{1}{2}-\sigma+k-s)\Gamma(n-k+s+1)}.\label{MSupl}
\end{multline}
Or
\[
\fM^{-\frac{1}{2}-\sigma}_{mn}(\fg)=e^{-m(\epsilon+i\varphi)}
\fP^{-\frac{1}{2}-\sigma}_{mn}(\ch\tau)e^{-n(\varepsilon+i\psi)},
\]
that is, the hyperspherical function $Z^{-\frac{1}{2}+i\rho}_{mn}(\cos\theta^c)$
in the case of supplementary series is degenerated to the Jacobi function
$\fP^{-\frac{1}{2}-\sigma}_{mn}(\ch\tau)$. For the associated functions of
supplementary series we obtain
\[
\fP^m_{-\frac{1}{2}-\sigma}(\fg)=e^{-m(\epsilon+i\varphi)}
\fP^m_{-\frac{1}{2}-\sigma}(\ch\tau).
\]

\section{Infinitesimal operators of $SU(2)\otimes SU(2)$}
\begin{sloppypar}\noindent
Let $\omega^c(t)$ be the one--parameter subgroup of $SL(2,\C)$. The operators
of the right regular representation of $SU(2)\otimes SU(2)$, 
corresponded to the
elements of this subgroup, transfer complex functions $f(\mathfrak{g})$
into $R(\omega^c(t))f(\mathfrak{g})=f(\mathfrak{g}\omega^c(t))$.
By this reason the infinitesimal operator of the right regular
representation\index{representation!right regular}
$R(\mathfrak{g})$, associated with one--parameter subgroup
$\omega^c(t)$, transfers the function $f(\mathfrak{g})$ into 
$\frac{df(\mathfrak{g}\omega^c(t))}{dt}$ at $t=0$.\end{sloppypar}

Let us denote Euler angles of the element $\mathfrak{g}\omega^c(t)$ via
$\varphi^c(t),\theta^c(t),\psi^c(t)$. Then there is an equality
\[
\left.\frac{df(\mathfrak{g}\omega^c(t))}{dt}\right|_{t=0}=
\frac{\partial f}{\partial\varphi^c}\left(\varphi^c(0)\right)^\prime+
\frac{\partial f}{\partial\theta^c}\left(\theta^c(0)\right)^\prime+
\frac{\partial f}{\partial\psi^c}\left(\psi^c(0)\right)^\prime.
\]
The infinitesimal operator\index{operator!infinitesimal}
$\sA^c_\omega$, corresponded to the subgroup
$\omega^c(t)$, has a form
\[
\sA^c_\omega=\sA_\omega-i\sB_\omega=
\left(\varphi^c(0)\right)^\prime\frac{\partial}{\partial\varphi^c}+
\left(\theta^c(0)\right)^\prime\frac{\partial}{\partial\theta^c}+
\left(\psi^c(0)\right)^\prime\frac{\partial}{\partial\psi^c}.
\]

Let us calculate infinitesimal operators $\sA^c_1$, $\sA^c_2$,
$\sA^c_3$ corresponding the complex subgroups\index{subgroup!complex}
$\Omega^c_1$, $\Omega^c_2$,
$\Omega^c_3$. The subgroup $\Omega^c_3$ consists of the matrices
\[
\omega_3(t^c)=
\begin{pmatrix}
e^{\frac{it^c}{2}} & 0\\
0 & e^{-\frac{it^c}{2}}
\end{pmatrix}.
\]
Let $\mathfrak{g}=\mathfrak{g}(\varphi^c,\theta^c,\psi^c)$ be a matrix with
complex Euler angles $\varphi^c=\varphi-i\epsilon$, $\theta^c=\theta-i\tau$,
$\psi^c=\psi-i\varepsilon$. Therefore, Euler angles of the matrix
$\mathfrak{g}\omega_3(t^c)$ equal to $\varphi^c$, $\theta^c$,
$\psi^c+t-it$. Hence it follows that
\[
\varphi^\prime(0)=0,\;\;
\epsilon^\prime(0)=0,\;\;
\theta^\prime(0)=0,\;\;
\tau^\prime(0)=0,\;\;
\psi^\prime(0)=1,\;\;
\varepsilon^\prime(0)=-i.
\]
So, the operator $\sA^c_3$, corresponded to the subgroup $\Omega^c_3$,
has a form
\[
\sA^c_3=\frac{\partial}{\partial\psi}-
i\frac{\partial}{\partial\varepsilon}.
\]
Whence
\begin{eqnarray}
\sA_3&=&\frac{\partial}{\partial\psi},\label{SL8}\\
\sB_3&=&\frac{\partial}{\partial\varepsilon}.\label{SL8'}
\end{eqnarray}

Let us calculate the infinitesimal operator $\sA^c_1$ corresponded 
the complex subgroup $\Omega^c_1$. The subgroup $\Omega^c_1$ consists of the
following matrices
\[
\omega_1(t^c)=
{\renewcommand{\arraystretch}{1.3}
\begin{pmatrix}
\cos\frac{t^c}{2} & i\sin\frac{t^c}{2}\\
i\sin\frac{t^c}{2} & \cos\frac{t^c}{2}
\end{pmatrix}}.
\]
The Euler angles of these matrices equal to $0,\,t^c=t-it,\,0$. Let us
represent the matrix $\mathfrak{g}\omega_1(t^c)$ by the product
$\mathfrak{g}_1\mathfrak{g}_2$, the Euler angles of which are described by
the formulae (\ref{SL7}). Then the Euler angles of the matrix
$\omega_1(t^c)$ equal to $\varphi^c_2=0$, $\theta^c_2=t-it$, $\psi^c_2=0$,
and the Euler angles of the matrix $\mathfrak{g}$ equal to
$\varphi^c_1=\varphi^c$, $\theta^c_1=\theta^c$, $\psi^c_1=\psi^c$. Thus,
from the general formulae (\ref{SL7}) we obtain that Euler
angles $\varphi^c(t)$, $\theta^c(t)$, $\psi^c(t)$ of the matrix
$\mathfrak{g}\omega_1(t^c)$ are defined by the following relations:
\begin{eqnarray}
\cos\theta^c(t)&=&\cos\theta^c\cos t^c-\sin\theta^c\sin t^c\cos\psi^c,
\label{SL9}\\[0.2cm]
e^{i\varphi^c(t)}&=&e^{i\varphi^c}\frac{\sin\theta^c\cos t^c+
\cos\theta^c\sin t^c\cos\psi^c+i\sin t^c\sin\psi^c}
{\sin\theta^c(t)},\label{SL10}\\[0.2cm]
e^{\frac{i[\varphi^c(t)+\psi^c(t)]}{2}}&=&e^{\frac{i\varphi^c}{2}}
\frac{\cos\frac{\theta^c}{2}\cos\frac{t^c}{2}e^{\frac{i\psi^c}{2}}-
\sin\frac{\theta^c}{2}\sin\frac{t^c}{2}e^{-\frac{i\psi^c}{2}}}
{\cos\frac{\theta^c(t)}{2}}.\label{SL11}
\end{eqnarray}
For calculation of derivatives $\varphi^\prime(t)$, $\epsilon^\prime(t)$,
$\theta^\prime(t)$, $\tau^\prime(t)$, $\psi^\prime(t)$ , 
$\varepsilon^\prime(t)$ at $t=0$ we differentiate on $t$ the both parts
of the each equality from (\ref{SL9})--(\ref{SL11}) and take $t=0$.
At this point we have $\varphi(0)=\varphi$, $\epsilon(0)=\epsilon$,
$\theta(0)=\theta$, $\tau(0)=\tau$, $\psi(0)=\psi$, 
$\varepsilon(0)=\varepsilon$.

So, let us differentiate the both parts of (\ref{SL9}).
In the result we obtain
\[
\theta^\prime(t)-i\tau^\prime(t)=(1-i)\cos\psi^c.
\]
Taking $t=0$, we find that
\[
\theta^\prime(0)=\cos\psi^c,\quad\tau^\prime(0)=\cos\psi^c.
\]
Differentiating now the both parts of (\ref{SL10}), we obtain
\[
\varphi^\prime(0)-i\epsilon^\prime(0)=
\frac{(1-i)\sin\psi^c}{\sin\theta^c}.
\]
Therefore,
\[
\varphi^\prime(0)=\frac{\sin\psi^c}{\sin\theta^c},\quad
\epsilon^\prime(0)=\frac{\sin\psi^c}{\sin\theta^c}.
\]
Further, differentiating the both parts of (\ref{SL11}), we find that
\[
\psi^\prime(0)-i\varepsilon^\prime(0)=-(1+i)\ctg\theta^c\sin\psi^c
\]
and
\[
\psi^\prime(0)=-\ctg\theta^c\sin\psi^c,\quad
\varepsilon^\prime(0)=-\ctg\theta^c\sin\psi^c.
\]
In such a way, we obtain the following infinitesimal operators:
\begin{eqnarray}
\sA_1&=&\cos\psi^c\frac{\partial}{\partial\theta}+
\frac{\sin\psi^c}{\sin\theta^c}\frac{\partial}{\partial\varphi}-
\ctg\theta^c\sin\psi^c\frac{\partial}{\partial\psi},\label{SL12}\\
\sB_1&=&\cos\psi^c\frac{\partial}{\partial\tau}+
\frac{\sin\psi^c}{\sin\theta^c}\frac{\partial}{\partial\epsilon}-
\ctg\theta^c\sin\psi^c\frac{\partial}{\partial\varepsilon}.\label{SL13}
\end{eqnarray}

Let us calculate now an infinitesimal operator $\sA^c_2$ corresponded
to the complex subgroup $\Omega^c_2$. The subgroup $\Omega^c_2$ consists of
the following matrices
\[
\omega_2(t^c)=
{\renewcommand{\arraystretch}{1.3}
\begin{pmatrix}
\cos\frac{t^c}{2} & -\sin\frac{t^c}{2}\\
\sin\frac{t^c}{2} & \cos\frac{t^c}{2}
\end{pmatrix}},
\]
where the Euler angles equal correspondingly to $0,\,t^c=t-it,\,0$.
It is obvious that the matrix $\mathfrak{g}\omega_2(t^c)$ can be represented
by the product
\[
\mathfrak{g}_1\mathfrak{g}_2=
{\renewcommand{\arraystretch}{1.3}
\begin{pmatrix}
\cos\frac{\theta^c_1}{2} & i\sin\frac{\theta^c_1}{2}\\
i\sin\frac{\theta^c_1}{2} & \cos\frac{\theta^c_1}{2}
\end{pmatrix}}\!\!\!{\renewcommand{\arraystretch}{1.3}
\begin{pmatrix}
\cos\frac{\theta^c_2}{2}e^{\frac{i\varphi^c_2}{2}} &
-\sin\frac{\theta^c_2}{2}e^{\frac{i\varphi^c_2}{2}}\\
\sin\frac{\theta^c_2}{2}e^{-\frac{i\varphi^c_2}{2}} &
\cos\frac{\theta^c_2}{2}e^{-\frac{i\varphi^c_2}{2}}
\end{pmatrix}
}
.
\]
Multiplying the matrices in the right part of this equality, we obtain
that Euler angles of the product $\mathfrak{g}_1\mathfrak{g}_2$ are related
by the formulae
\begin{eqnarray}
\cos\theta^c&=&\cos\theta^c_1\cos\theta^c_2+
\sin\theta^c_1\sin\theta^c_2\sin\varphi^c_2,\label{SL16}\\[0.2cm]
e^{i\varphi^c}&=&\frac{\sin\theta^c_1\cos\theta^c_2-
\cos\theta^c_1\sin\theta^c_2\sin\varphi^c_2+i\sin\theta^c_2\cos\varphi^c_2}
{\sin\theta^c},\label{SL17}\\[0.2cm]
e^{\frac{i(\varphi^c+\psi^c)}{2}}&=&
\frac{\cos\frac{\theta^c_1}{2}\cos\frac{\theta^c_2}{2}
e^{\frac{i\varphi^c_2}{2}}+
i\sin\frac{\theta^c_1}{2}\sin\frac{\theta^c_2}{2}e^{-\frac{i\varphi^c_2}{2}}}
{\cos\frac{\theta^c}{2}}.\label{SL18}
\end{eqnarray}
Or, repeating the calculations as in the case of (\ref{SL7}),
we obtain in general case
\begin{eqnarray}
\cos\theta^c&=&\cos\theta^c_1\cos\theta^c_2+
\sin\theta^c_1\sin\theta^c_2\sin(\varphi^c_2+\psi^c_1),\nonumber\\
e^{i(\varphi^c-\varphi^c_1)}&=&
\frac{\sin\theta^c_1\cos\theta^c_2-
\cos\theta^c_1\sin\theta^c_2\sin(\varphi^c_2+\psi^c_1)+
i\sin\theta^c_2\cos(\varphi^c_2+\psi^c_1)}{\sin\theta^c},\nonumber\\
e^{\frac{i(\varphi^c+\psi^c-\varphi^c_1-\psi^c_2)}{2}}&=&
\frac{\cos\frac{\theta^c_1}{2}\cos\frac{\theta^c_2}{2}
e^{\frac{i(\varphi^c_2+\psi^c_1)}{2}}+\sin\frac{\theta^c_1}{2}
\sin\frac{\theta^c_2}{2}e^{-\frac{i(\varphi^c_2+\psi^c_1)}{2}}}
{\cos\frac{\theta^c}{2}}.\label{SL21}
\end{eqnarray}
Therefore, Euler angles of the matrix $\omega_2(t^c)$ equal to
$\varphi^c_2=0$, $\theta^c_2=t-it$, $\psi^c_2=0$, and Euler angles of the
matrix $\mathfrak{g}$ equal to $\varphi^c_1=\varphi^c$, 
$\theta^c_1=\theta^c$, $\psi^c_1=\psi^c$. Then from the formulae
(\ref{SL21}) we obtain that Euler angles $\varphi^c(t)$,
$\theta^c(t)$, $\psi^c(t)$ of the matrix $\mathfrak{g}\omega_2(t^c)$ are
defined by relations
\begin{eqnarray}
\cos\theta^c(t)&=&\cos\theta^c\cos t^c+\sin\theta^c\sin t^c\sin\psi^c,
\label{SL22}\\[0.2cm]
e^{i\varphi^c(t)}&=&e^{i\varphi^c}\frac{\sin\theta^c\cos t^c-
\cos\theta^c\sin t^c\sin\psi^c+i\sin t^c\cos\psi^c}
{\sin\theta^c(t)},\label{SL23}\\[0.2cm]
e^{\frac{i[\varphi^c(t)+\psi^c(t)]}{2}}&=&e^{\frac{i\varphi^c}{2}}
\frac{\cos\frac{\theta^c}{2}\cos\frac{t^c}{2}e^{\frac{i\psi^c}{2}}+
i\sin\frac{\theta^c}{2}\sin\frac{t^c}{2}e^{-\frac{i\psi^c}{2}}}
{\cos\frac{\theta^c(t)}{2}}.\label{SL24}
\end{eqnarray}
Differentiating on $t$ the both parts of the each equalities
(\ref{SL22})--(\ref{SL24}) and taking $t=0$, we obtain
\begin{eqnarray}
&&\theta^\prime(0)=\tau^\prime(0)=-\sin\psi^c,\nonumber\\
&&\varphi^\prime(0)=\epsilon^\prime(0)=
\frac{\cos\psi^c}{\sin\theta^c},\nonumber\\
&&\psi^\prime(0)=\varepsilon^\prime(0)=-\ctg\theta^c\cos\psi^c.
\end{eqnarray}
Therefore, for the subgroup $\Omega^c_2$ we have the following
infinitesimal operators:
\begin{eqnarray}
\sA_2&=&-\sin\psi^c\frac{\partial}{\partial\theta}+
\frac{\cos\psi^c}{\sin\theta^c}\frac{\partial}{\partial\varphi}-
\ctg\theta^c\cos\psi^c\frac{\partial}{\partial\psi},\label{SL24'}\\
\sB_2&=&-\sin\psi^c\frac{\partial}{\partial\tau}+
\frac{\cos\psi^c}{\sin\theta^c}\frac{\partial}{\partial\epsilon}-
\ctg\theta^c\cos\psi^c\frac{\partial}{\partial\varepsilon}.\label{SL24''}
\end{eqnarray}
It is easy to verify that operators $\sA_i$, $\sB_i$,
defined by the formulae (\ref{SL8}),
(\ref{SL8'}), (\ref{SL12}), (\ref{SL13}) and (\ref{SL24'}), (\ref{SL24''}),
are satisfy the commutation relations (\ref{Com1}).
\subsection{Casimir operators and differential equations for hyperspherical
functions}
Taking into account the expressions (\ref{SL8}), (\ref{SL8'}),
(\ref{SL12}), (\ref{SL13}) and (\ref{SL24'}), (\ref{SL24''}) we can write
the operators (\ref{SL25}) in the form
\begin{eqnarray}
\sX_1&=&\cos\psi^c\frac{\partial}{\partial\theta^c}+
\frac{\sin\psi^c}{\sin\theta^c}\frac{\partial}{\partial\varphi^c}-
\ctg\theta^c\sin\psi^c\frac{\partial}{\partial\psi^c},\label{X1}\\
\sX_2&=&-\sin\psi^c\frac{\partial}{\partial\theta^c}+
\frac{\cos\psi^c}{\sin\theta^c}\frac{\partial}{\partial\varphi^c}-
\ctg\theta^c\cos\psi^c\frac{\partial}{\partial\psi^c},\label{X2}\\
\sX_3&=&\frac{\partial}{\partial\psi^c},\label{X3}\\
\sY_1&=&\cos\dot{\psi}^c\frac{\partial}{\partial\dot{\theta}^c}+
\frac{\sin\dot{\psi}^c}{\sin\dot{\theta}^c}
\frac{\partial}{\partial\dot{\varphi}^c}-
\ctg\dot{\theta}^c\sin\dot{\psi}^c\frac{\partial}{\partial\dot{\psi}^c},
\label{Y1}\\
\sY_2&=&-\sin\dot{\psi}^c\frac{\partial}{\partial\dot{\theta}^c}+
\frac{\cos\dot{\psi}^c}{\sin\dot{\theta}^c}
\frac{\partial}{\partial\dot{\varphi}^c}-
\ctg\dot{\theta}^c\cos\dot{\psi}^c\frac{\partial}{\partial\dot{\psi}^c},
\label{Y2}\\
\sY_3&=&\frac{\partial}{\partial\dot{\psi}^c},\label{Y3}
\end{eqnarray}
where
\begin{gather}
\frac{\partial}{\partial\theta^c}=\frac{1}{2}\left(
\frac{\partial}{\partial\theta}+i\frac{\partial}{\partial\tau}\right),\;\;
\frac{\partial}{\partial\varphi^c}=\frac{1}{2}\left(
\frac{\partial}{\partial\varphi}+i\frac{\partial}{\partial\epsilon}\right),\;\;
\frac{\partial}{\partial\psi^c}=\frac{1}{2}\left(
\frac{\partial}{\partial\psi}+i\frac{\partial}{\partial\varepsilon}\right),
\nonumber\\
\frac{\partial}{\partial\dot{\theta}^c}=\frac{1}{2}\left(
\frac{\partial}{\partial\theta}-i\frac{\partial}{\partial\tau}\right),\;\;
\frac{\partial}{\partial\dot{\varphi}^c}=\frac{1}{2}\left(
\frac{\partial}{\partial\varphi}-i\frac{\partial}{\partial\epsilon}\right),\;\;
\frac{\partial}{\partial\dot{\psi}^c}=\frac{1}{2}\left(
\frac{\partial}{\partial\psi}-i\frac{\partial}{\partial\varepsilon}\right),
\nonumber
\end{gather}
As is known, for the Lorentz group there are two independent 
Casimir operators\index{operator!Casimir}
\begin{eqnarray}
\sX^2&=&\sX^2_1+\sX^2_2+\sX^2_3=\frac{1}{4}(\sA^2-\sB^2+2i\sA\sB),\nonumber\\
\sY^2&=&\sY^2_1+\sY^2_2+\sY^2_3=
\frac{1}{4}(\widetilde{\sA}^2-\widetilde{\sB}^2-
2i\widetilde{\sA}\widetilde{\sB}).\label{KO}
\end{eqnarray}
Substituting (\ref{X1})-(\ref{Y3}) into (\ref{KO}), we obtain 
for the Casimir operators the following expressions
\begin{eqnarray}
\sX^2&=&\frac{\partial^2}{\partial\theta^c{}^2}+
\ctg\theta^c\frac{\partial}{\partial\theta^c}+\frac{1}{\sin^2\theta^c}\left[
\frac{\partial^2}{\partial\varphi^c{}^2}-
2\cos\theta^c\frac{\partial}{\partial\varphi^c}
\frac{\partial}{\partial\psi^c}+
\frac{\partial^2}{\partial\psi^c{}^2}\right],\nonumber\\
\sY^2&=&\frac{\partial^2}{\partial\dot{\theta}^c{}^2}+
\ctg\dot{\theta}^c\frac{\partial}{\partial\dot{\theta}^c}+
\frac{1}{\sin^2\dot{\theta}^c}\left[
\frac{\partial^2}{\partial\dot{\varphi}^c{}^2}-
2\cos\dot{\theta}^c\frac{\partial}{\partial\dot{\varphi}^c}
\frac{\partial}{\partial\dot{\psi}^c}+
\frac{\partial^2}{\partial\dot{\psi}^c{}^2}\right].\label{KO2}
\end{eqnarray}
Matrix elements of 
unitary irreducible representations\index{representation!unitary irreducible}
of the Lorentz
group are eigenfunctions of the operators (\ref{KO2}):
\begin{eqnarray}
\left[\sX^2+l(l+1)\right]\fM^l_{mn}(\varphi^c,\theta^c,\psi^c)&=&0,\nonumber\\
\left[\sY^2+\dot{l}(\dot{l}+1)\right]\fM^{\dot{l}}_{\dot{m}\dot{n}}
(\dot{\varphi}^c,\dot{\theta}^c,\dot{\psi}^c)&=&0,\label{EQ}
\end{eqnarray}
where
\begin{eqnarray}
\fM^l_{mn}(\varphi^c,\theta^c,\psi^c)&=&e^{-i(m\varphi^c+n\psi^c)}
Z^l_{mn}(\theta^c),\nonumber\\
\fM^{\dot{l}}_{\dot{m}\dot{n}}(\dot{\varphi}^c,\dot{\theta}^c,\dot{\psi}^c)&=&
e^{-i(\dot{m}\dot{\varphi}^c+\dot{n}\dot{\varphi}^c)}
Z^{\dot{l}}_{\dot{m}\dot{n}}(\dot{\theta}^c).\label{HF3'}
\end{eqnarray}
Substituting the hyperspherical functions (\ref{HF3'}) into (\ref{EQ}) and
taking into account the operators (\ref{KO2}), we obtain 
\begin{eqnarray}
\left[\frac{d^2}{d\theta^c{}^2}+\ctg\theta^c\frac{d}{d\theta^c}-
\frac{m^2+n^2-2mn\cos\theta^c}{\sin^2\theta^c}+l(l+1)\right]
Z^l_{mn}(\theta^c)&=&0,\nonumber\\
\left[\frac{d^2}{d\dot{\theta}^c{}^2}+\ctg\dot{\theta}^c
\frac{d}{d\dot{\theta}^c}-\frac{\dot{m}^2+\dot{n}^2-
2\dot{m}\dot{n}\cos\dot{\theta}^c}{\sin^2\dot{\theta}^c}+
\dot{l}(\dot{l}+1)\right]Z^{\dot{l}}_{\dot{m}\dot{n}}(\dot{\theta}^c)
&=&0.\nonumber
\end{eqnarray}
Finally, after substitutions $z=\cos\theta^c$ and 
$\overset{\ast}{z}=\cos\dot{\theta}^c$, we come to the following
differential equations
\begin{eqnarray}
\left[(1-z^2)\frac{d^2}{dz^2}-2z\frac{d}{dz}-
\frac{m^2+n^2-2mnz}{1-z^2}+l(l+1)\right]Z^l_{mn}(\arc z)&=&0,\nonumber\\
\left[(1-\overset{\ast}{z}{}^2)\frac{d^2}{d\overset{\ast}{z}{}^2}-
2\overset{\ast}{z}\frac{d}{d\overset{\ast}{z}}-
\frac{\dot{m}^2+\dot{n}^2-2\dot{m}\dot{n}\overset{\ast}{z}}
{1-\overset{\ast}{z}{}^2}+\dot{l}(\dot{l}+1)\right]
Z^{\dot{l}}_{\dot{m}\dot{n}}(\arc\overset{\ast}{z})&=&0.\nonumber
\end{eqnarray}
The latter equations have three singular points $-1$, $+1$, $\infty$.

Further, Casimir operators on the 2-dimensional complex sphere (correspondingly,
on the dual sphere) have the form
\begin{eqnarray}
\sX^2&=&\frac{\partial^2}{\partial\theta^c{}^2}+
\ctg\theta^c\frac{\partial}{\partial\theta^c}+\frac{1}{\sin^2\theta^c}
\frac{\partial^2}{\partial\varphi^c{}^2},\nonumber\\
\sY^2&=&\frac{\partial^2}{\partial\dot{\theta}^c{}^2}+
\ctg\dot{\theta}^c\frac{\partial}{\partial\dot{\theta}^c}+
\frac{1}{\sin^2\dot{\theta}^c{}^2}
\frac{\partial^2}{\partial\dot{\varphi}^c{}^2}.\label{KO3}
\end{eqnarray}
Associated hyperspherical functions $\fM^m_{l}(\varphi^c,\theta^c,0)$ and
$\fM^{\dot{m}}_{\dot{l}}(\dot{\varphi}^c,\dot{\theta}^c,0)$,
defined on the surface of the two-dimensional complex sphere, 
are eigenfunctions
of the operators $\sX^2$ and $\sY^2$:
\begin{eqnarray}
\left[\sX^2+l(l+1)\right]\fM^m_{l}(\varphi^c,\theta^c,0)&=&0,\nonumber\\
\left[\sY^2+\dot{l}(\dot{l}+1)\right]
\fM^{\dot{m}}_{\dot{l}}(\dot{\varphi}^c,\dot{\theta}^c,0)&=&0
\label{EQ2}
\end{eqnarray}
\begin{sloppypar}\noindent
Substituting the functions $\fM^m_{l}(\varphi^c,\theta^c,0)=
e^{-im\varphi^c}Z^m_{l}(\theta^c)$, $\fM^{\dot{m}}_{\dot{l}}
(\dot{\varphi}^c,\dot{\theta}^c,0)=e^{-i\dot{m}\dot{\varphi}^c}
Z^{\dot{m}}_{\dot{l}}(\dot{\theta}^c)$ into (\ref{EQ2}) and
taking into account the operators (\ref{KO3}), we obtain the following
equations\end{sloppypar}
\begin{eqnarray}
\left[\frac{d^2}{d\theta^c{}^2}+\ctg\theta^c\frac{d}{d\theta^c}-
\frac{m^2}{\sin^2\theta^c}+l(l+1)\right]Z^m_{l}(\theta^c)&=&0,\nonumber\\
\left[\frac{d^2}{d\dot{\theta}^c{}^2}+\ctg\dot{\theta}^c
\frac{d}{d\dot{\theta}^c}-\frac{m^2}{\sin^2\dot{\theta}^c}+
\dot{l}(\dot{l}+1)\right]Z^{\dot{m}}_{\dot{l}}(\dot{\theta}^c)&=&0.
\nonumber
\end{eqnarray}
Or, introducing the substitutions $z=\cos\theta^c$,
$\overset{\ast}{z}=\cos\dot{\theta}^c$, we find that
\begin{eqnarray}
\left[(1-z^2)\frac{d^2}{dz^2}-2z\frac{d}{dz}-\frac{m^2}{1-z^2}
+l(l+1)\right]Z^m_{l}(\arc z)&=&0,\nonumber\\
\left[(1-\overset{\ast}{z}{}^2)\frac{d^2}{d\overset{\ast}{z}{}^2}-
2\overset{\ast}{z}\frac{d}{d\overset{\ast}{z}}-
\frac{\dot{m}^2}{1-\overset{\ast}{z}{}^2}+\dot{l}(\dot{l}+1)\right]
Z^{\dot{m}}_{\dot{l}}(\arc\overset{\ast}{z})&=&0.\nonumber
\end{eqnarray}
Analogously, Casimir operators, corresponding to zonal hyperspherical
functions $Z_l(\theta^c)$, have the form
\begin{eqnarray}
\sX^2&=&\frac{\partial^2}{\partial\theta^c{}^2}+
\ctg\theta^c\frac{\partial}{\partial\theta^c},\nonumber\\
\sY^2&=&\frac{\partial^2}{\partial\dot{\theta}^c{}^2}+
\ctg\dot{\theta}^c\frac{\partial}{\partial\dot{\theta}^c}.\label{KO4}
\end{eqnarray}
And the equations for $Z_l$ are
\begin{eqnarray}
\left[(1-z^2)\frac{d^2}{dz^2}-2z\frac{d}{dz}
+l(l+1)\right]Z_{l}(\arc z)&=&0,\nonumber\\
\left[(1-\overset{\ast}{z}{}^2)\frac{d^2}{d\overset{\ast}{z}{}^2}-
2\overset{\ast}{z}\frac{d}{d\overset{\ast}{z}}
+\dot{l}(\dot{l}+1)\right]
Z_{\dot{l}}(\arc\overset{\ast}{z})&=&0.\nonumber
\end{eqnarray}

\section{Recurrence relations between hyperspherical functions}
Between generalized hyperspherical 
functions\index{function!hyperspherical!generalized} 
$\fT^l_{mn}$ (and also the
hyperspherical functions $Z^l_{mn}$) there exists a wide variety of
recurrence relations. Part of them relates the hyperspherical functions
of one and the same order (with identical $l$), other part relates the
functions of different orders.

In virtue of the Van der Waerden 
representation\index{representation!Van der Waerden} (\ref{Waerden}) the
recurrence formulae for the hyperspherical functions of one and the same
order follow from the equalities
\begin{eqnarray}
&&\sX_-\fT^l_{mn}=\boldsymbol{\alpha}_n\fT^l_{m,n-1},\quad
\sX_+\fT^l_{mn}=\boldsymbol{\alpha}_{n+1}
\fT^l_{m,n+1},\label{SL26'}\\
&&\sY_-\fT^{\dot{l}}_{\dot{m}\dot{n}}=
\boldsymbol{\alpha}_{\dot{n}}\fT^{\dot{l}}_{\dot{m},\dot{n}-1},\quad
\sY_+\fT^{\dot{l}}_{\dot{m}\dot{n}}=\boldsymbol{\alpha}_{\dot{n}+1}
\fT^{\dot{l}}_{\dot{m},\dot{n}+1},\label{SL26''}
\end{eqnarray}
where
\[
\boldsymbol{\alpha}_n=\sqrt{(l+n)(l-n+1)},\quad
\boldsymbol{\alpha}_{\dot{n}}=\sqrt{(\dot{l}+\dot{n})(\dot{l}-\dot{n}+1)}.
\]
From (\ref{SL25}) and (\ref{SL26}) it follows that
\begin{eqnarray}
\sX_+&=&\frac{1}{2}\left(i\sA_1-\sA_2-\sB_1-i\sB_2\right),\nonumber\\
\sX_-&=&\frac{1}{2}\left(i\sA_1+\sA_2-\sB_1+i\sB_2\right),\nonumber\\
\sY_+&=&\frac{1}{2}\left(i\widetilde{\sA}_1-
\widetilde{\sA}_2+\widetilde{\sB}_1+i\widetilde{\sB}_2\right),\nonumber\\
\sY_-&=&\frac{1}{2}\left(i\widetilde{\sA}_1+
\widetilde{\sA}_2+\widetilde{\sB}_1-i\widetilde{\sB}_2\right).\label{GenH}
\end{eqnarray}
Using the formulae (\ref{SL12}), (\ref{SL13}) and (\ref{SL24'}), (\ref{SL24''}),
we obtain
\begin{eqnarray}
\sX_+&=&\frac{e^{-i\psi^c}}{2}\left[i\frac{\partial}{\partial\theta}-
\frac{1}{\sin\theta^c}\frac{\partial}{\partial\varphi}
+\ctg\theta^c\frac{\partial}{\partial\psi}-\frac{\partial}{\partial\tau}-
\frac{i}{\sin\theta^c}\frac{\partial}{\partial\epsilon}+
i\ctg\theta^c\frac{\partial}{\partial\varepsilon}\right],\label{SL27}\\
\sX_-&=&\frac{e^{i\psi^c}}{2}\left[i\frac{\partial}{\partial\theta}+
\frac{1}{\sin\theta^c}\frac{\partial}{\partial\varphi}
-\ctg\theta^c\frac{\partial}{\partial\psi}-\frac{\partial}{\partial\tau}+
\frac{i}{\sin\theta^c}\frac{\partial}{\partial\epsilon}-
i\ctg\theta^c\frac{\partial}{\partial\varepsilon}\right],\label{SL28}\\
\sY_+&=&\frac{e^{-i\dot{\psi}^c}}{2}\left[i\frac{\partial}{\partial\theta}-
\frac{1}{\sin\dot{\theta}^c}\frac{\partial}{\partial\varphi}+
\ctg\dot{\theta}^c\frac{\partial}{\partial\psi}+\frac{\partial}{\partial\tau}+
\frac{i}{\sin\dot{\theta}^c}\frac{\partial}{\partial\epsilon}-
i\ctg\dot{\theta}^c\frac{\partial}{\partial\varepsilon}\right],\label{SL29}\\
\sY_-&=&\frac{e^{i\dot{\psi}^c}}{2}\left[i\frac{\partial}{\partial\theta}+
\frac{1}{\sin\dot{\theta}^c}\frac{\partial}{\partial\varphi}
-\ctg\dot{\theta}^c\frac{\partial}{\partial\psi}+\frac{\partial}{\partial\tau}-
\frac{i}{\sin\dot{\theta}^c}\frac{\partial}{\partial\epsilon}+
i\ctg\dot{\theta}^c\frac{\partial}{\partial\varepsilon}\right].\label{SL30}
\end{eqnarray}
Further, substituting the function $\fT^l_{mn}=e^{-m(\epsilon-i\varphi)}
\do{Z}^l_{mn}(\theta,\tau)e^{-n(\varepsilon-i\psi)}$ 
into the relations (\ref{SL26'})
and taking into account the operators (\ref{SL27}) and (\ref{SL28}),
we find that
\begin{eqnarray}
i\frac{\partial\dot{Z}^l_{mn}}{\partial\theta}-
\frac{\partial\dot{Z}^l_{mn}}{\partial\tau}-
\frac{2i(m-n\cos\theta^c)}{\sin\theta^c}\dot{Z}^l_{mn}&=&
2\boldsymbol{\alpha}_n\dot{Z}^l_{m,n-1},\label{SL31}\\
i\frac{\partial\dot{Z}^l_{mn}}{\partial\theta}-
\frac{\partial\dot{Z}^l_{mn}}{\partial\tau}+
\frac{2i(m-n\cos\theta^c)}{\sin\theta^c}\dot{Z}^l_{mn}&=&
2\boldsymbol{\alpha}_{n+1}\dot{Z}^l_{m,n+1}.\label{SL32}
\end{eqnarray}
Since the functions $\dot{Z}^l_{mn}(\theta,\tau)$ are symmetric, that is,
$\dot{Z}^l_{mn}(\theta,\tau)=\dot{Z}^l_{nm}(\theta,\tau)$, then substituting
$\dot{Z}^l_{nm}(\theta,\tau)$ in lieu of $\dot{Z}^l_{mn}$ into the formulae
(\ref{SL31})--(\ref{SL32}) and replacing $m$ by $n$, and $n$ by $m$,
we obtain
\begin{eqnarray}
i\frac{\partial Z^l_{mn}}{\partial\theta}-
\frac{\partial Z^l_{mn}}{\partial\tau}-
\frac{2i(n-m\cos\theta^c)}{\sin\theta^c}Z^l_{mn}&=&
2\boldsymbol{\alpha}_mZ^l_{m-1,n},\label{SL33}\\
i\frac{\partial Z^l_{mn}}{\partial\theta}-
\frac{\partial Z^l_{mn}}{\partial\tau}+
\frac{2i(n-m\cos\theta^c)}{\sin\theta^c}Z^l_{mn}&=&
2\boldsymbol{\alpha}_{m+1}Z^l_{m+1,n}.\label{SL34}
\end{eqnarray}
Analogously, 
substituting the function
$\fT^{\dot{l}}_{\dot{m}\dot{n}}=e^{-\dot{m}(\epsilon-i\varphi)}
Z^{\dot{l}}_{\dot{m}\dot{n}}(\theta,\tau)e^{-\dot{n}(\varepsilon-i\psi)}$
into the relations (\ref{SL26''}), we obtain
\begin{eqnarray}
i\frac{\partial Z^{\dot{l}}_{\dot{m}\dot{n}}}{\partial\theta}
+\frac{\partial Z^{\dot{l}}_{\dot{m}\dot{n}}}{\partial\tau}+
\frac{2i(\dot{m}-\dot{n}\cos\dot{\theta}^c)}{\sin\dot{\theta}^c}
Z^{\dot{l}}_{\dot{m}\dot{n}}&=&
2\boldsymbol{\alpha}_{\dot{n}} Z^{\dot{l}}_{\dot{m},\dot{n}-1},\label{SL37}\\
i\frac{\partial Z^{\dot{l}}_{\dot{m}\dot{n}}}{\partial\theta}
+\frac{\partial Z^{\dot{l}}_{\dot{m}\dot{n}}}{\partial\tau}-
\frac{2i(\dot{m}-\dot{n}\cos\dot{\theta}^c)}{\sin\dot{\theta}^c} 
Z^{\dot{l}}_{\dot{m}\dot{n}}&=&
2\boldsymbol{\alpha}_{\dot{n}+1} 
Z^{\dot{l}}_{\dot{m},\dot{n}+1}.\label{SL38}
\end{eqnarray}
Further, using the symmetry of the functions 
$Z^{\dot{l}}_{\dot{m}\dot{n}}$, we obtain
\begin{eqnarray}
i\frac{\partial Z^{\dot{l}}_{\dot{m}\dot{n}}}{\partial\theta}
+\frac{\partial Z^{\dot{l}}_{\dot{m}\dot{n}}}{\partial\tau}+
\frac{2i(\dot{n}-\dot{m}\cos\dot{\theta}^c)}{\sin\dot{\theta}^c}
Z^{\dot{l}}_{\dot{m}\dot{n}}&=&
2\boldsymbol{\alpha}_{\dot{m}}Z^{\dot{l}}_{\dot{m}-1,\dot{n}},\label{SL39}\\
i\frac{\partial Z^{\dot{l}}_{\dot{m}\dot{n}}}{\partial\theta}
+\frac{\partial Z^{\dot{l}}_{\dot{m}\dot{n}}}{\partial\tau}-
\frac{2i(\dot{n}-\dot{m}\cos\dot{\theta}^c)}{\sin\dot{\theta}^c}
Z^{\dot{l}}_{\dot{m}\dot{n}}&=&
2\boldsymbol{\alpha}_{\dot{m}+1}Z^{\dot{l}}_{\dot{m}+1,\dot{n}}.\label{SL40}
\end{eqnarray}
Supposing $n=0$ ($\dot{n}=0$) in the formulas (\ref{SL33})--(\ref{SL34})
and (\ref{SL39})--(\ref{SL40}), we obtain recurrence relations for the
associated hyperspherical functions:
\begin{eqnarray}
i\frac{\partial Z^m_{l}}{\partial\theta}-
\frac{\partial Z^m_{l}}{\partial\tau}+
2im\cot\theta^cZ^m_{l}&=&
2\boldsymbol{\alpha}_mZ^{m-1}_l,\nonumber\\
i\frac{\partial Z^m_{l}}{\partial\theta}-
\frac{\partial Z^m_{l}}{\partial\tau}-
2im\cot\theta^cZ^m_{l}&=&
2\boldsymbol{\alpha}_{m+1}Z^{m+1}_l,\nonumber
\end{eqnarray}
and
\begin{eqnarray}
i\frac{\partial Z_{\dot{l}}^{\dot{m}}}{\partial\theta}
+\frac{\partial Z_{\dot{l}}^{\dot{m}}}{\partial\tau}-
2i\dot{m}\cot\dot{\theta}^c
Z_{\dot{l}}^{\dot{m}}&=&
2\boldsymbol{\alpha}_{\dot{m}}Z_{\dot{l}}^{\dot{m}-1},\nonumber\\
i\frac{\partial Z_{\dot{l}}^{\dot{m}}}{\partial\theta}
+\frac{\partial Z_{\dot{l}}^{\dot{m}}}{\partial\tau}+
2i\dot{m}\cot\dot{\theta}^c
Z_{\dot{l}}^{\dot{m}}&=&
2\boldsymbol{\alpha}_{\dot{m}+1}Z_{\dot{l}}^{\dot{m}+1}.\nonumber
\end{eqnarray}

Let us consider now recurrence relations between hyperspherical functions
with different order. These recurrence formulae are related with
the tensor products of irreducible representations of the Lorentz group.
Indeed, in accordance with Van der Waerden representation (\ref{Waerden})
an arbitrary finite--dimensional representation of the group $\fG_+$ has
a form $\boldsymbol{\tau}_{l0}\otimes\boldsymbol{\tau}_{0\dot{l}}\sim
\boldsymbol{\tau}_{l\dot{l}}$, where $\boldsymbol{\tau}_{l0}$ and
$\boldsymbol{\tau}_{0\dot{l}}$ are representations of the group $SU(2)$.
Then a product of the two representations $\boldsymbol{\tau}_{l_1\dot{l}_1}$
and $\boldsymbol{\tau}_{l_2\dot{l}_2}$ of the Lorentz group is defined by
an expression
\begin{equation}\label{CGF}
\boldsymbol{\tau}_{l_1\dot{l}_1}\otimes\boldsymbol{\tau}_{l_2\dot{l}_2}=
\sum_{|l_1-l_2|\leq m\leq l_1+l_2;|\dot{l}_1-\dot{l}_2|\leq \dot{m}\leq
\dot{l}_1+\dot{l}_2}\boldsymbol{\tau}_{m\dot{m}}.
\end{equation}
The vectors $\zeta_{lm;\dot{l}\dot{m}}=\mid lm;\dot{l}\dot{m}\rangle$ 
of the helicity basis have the form
\begin{equation}\label{CG}
\zeta_{lm;\dot{l}\dot{m}}=
\sum_{\substack{m_1+m_2=m,\\ \dot{m}_1+\dot{m}_2=\dot{m}}}
C(l_1,l_2,l;m_1,m_2,m)C(\dot{l}_1,\dot{l}_2,\dot{l};\dot{m}_1,\dot{m}_2,\dot{m})
\zeta_{l_1 m_1;\dot{l}_1\dot{m}_1}\otimes 
\zeta_{l_2 m_2;\dot{l}_2\dot{m}_2},
\end{equation}
where
\[
C(l_1,l_2,l;m_1,m_2,m)C(\dot{l}_1,\dot{l}_2,\dot{l};\dot{m}_1,\dot{m}_2,\dot{m})=
B^{l_1,l_2,l;m_1,m_2,m}_{\dot{l}_1,\dot{l}_2,
\dot{l};\dot{m}_1,\dot{m}_2,\dot{m}}
\]
are the Clebsch--Gordan coefficients\index{Clebsch-Gordan coefficients}
of the group $SL(2,\C)$. Expressing
the Clebsch--Gordan coefficients $C(l_1,l_2,l;m_1,m_2,m_1+m_2)$ of the group
$SU(2)$ via a generalized hypergeometric 
function\index{function!hypergeometric!generalized} 
${}_3F_2$ (see, for example,
\cite{Ros55,VK,VKS75}), we see that CG--coefficients of $SL(2,\C)$ have the form
\begin{multline}
B^{l_1,l_2,l;m_1,m_2,m}_{\dot{l}_1,\dot{l}_2,\dot{l};\dot{m}_1,\dot{m}_2,\dot{m}}=
(-1)^{l_1+\dot{l}_1-m_1-\dot{m}_1}\times\\
\frac{\Gamma(l_1+l_2-m+1)\Gamma(\dot{l}_1+\dot{l}_2-\dot{m}+1)}
{\Gamma(l_2-l_1+m+1)\Gamma(\dot{l}_2-\dot{l}_1+\dot{m}+1)}\times\\
\left(\frac{(l-m)!(l+l_2-l_1)!(l_1-m_1)!(l_2+m_2)!
(l+m)!(2l+1)}{(l_1-l_2+l)!(l_1+l_2-l)!
(l_1+l_2+l)!(l_1-m_1)!(l_2-m_2)!}\right)^{\frac{1}{2}}\times\\
\left(\frac{(\dot{l}-\dot{m})!(\dot{l}+\dot{l}_2-\dot{l}_1)!
(\dot{l}_1-\dot{m}_1)!(\dot{l}_2+\dot{m}_2)!
(\dot{l}+\dot{m})!(2\dot{l}+1)}
{(\dot{l}_1-\dot{l}_2+\dot{l})!
(\dot{l}_1+\dot{l}_2-\dot{l})!
(\dot{l}_1+\dot{l}_2+\dot{l})!
(\dot{l}_1-\dot{m}_1)!(\dot{l}_2-\dot{m}_2)!}
\right)^{\frac{1}{2}}\times\\
\hypergeom{3}{2}{l+m+1,-l+m,-l_1+m_1}{-l_1-l_2+m,l_2-l_1+m+1}{1}
\hypergeom{3}{2}{\dot{l}+\dot{m}+1,-\dot{l}+\dot{m},-\dot{l}_1+\dot{m}_1}
{-\dot{l}_1-\dot{l}_2+\dot{m},\dot{l}_2-\dot{l}_1+\dot{m}+1}{1},
\label{CG'}
\end{multline}
where $m=m_1+m_2$, $\dot{m}=\dot{m}_1+\dot{m}_2$. In virtue of the
orthogonality of the Clebsch--Gordan coefficients from (\ref{CG}) it
follows that
\begin{eqnarray}
\zeta_{l_1 m_1;\dot{l}_1\dot{m}_2}\otimes
\zeta_{l_2 m_2;\dot{l}_2\dot{m}_2}
&=&\sum_{\substack{m_1+m_2=m,\\ \dot{m}_1+\dot{m}_2=\dot{m}}}
\overline{
C(l_1,l_2,l;m_1,m_2,m)C(\dot{l}_1,\dot{l}_2,\dot{l};\dot{m}_1,\dot{m}_2,\dot{m})}
\zeta_{l m;\dot{l}\dot{m}}\nonumber\\
&=&\sum_{\substack{m_1+m_2=m,\\ \dot{m}_1+\dot{m}_2=\dot{m}}}
\bar{B}^{l_1,l_2,l;m_1,m_2,m}_{\dot{l}_1,\dot{l}_2,\dot{l};\dot{m}_1,\dot{m}_2,
\dot{m}}\zeta_{l m;\dot{l}\dot{m}}.\label{CG2}
\end{eqnarray}
Further, assume that $l_1=1$ and $\dot{l}_1=1$, then at $l_2\ge 1$ and
$\dot{l}_2\ge 1$ the numbers $l$ and $m$ (correspondingly $\dot{l}$ and
$\dot{m}$) take the values $l=l_2-1,\,l_2,\,l_2+1$, $m=m_2-1,\,m_2,\,m_2+1$
(correspondingly $\dot{l}=\dot{l}_2-1,\,\dot{l}_2,\,\dot{l}_2+1$,
$\dot{m}=\dot{m}_2-1,\,\dot{m}_2,\,\dot{m}_2+1$). In this case the system
(\ref{CG2}) can be rewritten as follows
\begin{eqnarray}
\zeta_{1,-1;1,-1}\otimes\zeta_{l_2,m_2+1;\dot{l}_2,\dot{m}_2+1}&=&
\bar{b}^m_{11}\zeta_{l+1,m;\dot{l}+1,\dot{m}}+
\bar{b}^m_{12}\zeta_{lm;\dot{l}\dot{m}}+
\bar{b}^m_{13}\zeta_{l-1,m;\dot{l}-1,\dot{m}},\nonumber\\
\zeta_{1,0;1,0}\otimes\zeta_{l_2m_2;\dot{l}_2\dot{m}_2}&=&
\bar{b}^m_{21}\zeta_{l+1,m;\dot{l}+1,\dot{m}}+
\bar{b}^m_{22}\zeta_{lm;\dot{l}\dot{m}}+
\bar{b}^m_{23}\zeta_{l-1,m;\dot{l}-1,\dot{m}},\nonumber\\
\zeta_{1,1;1,1}\otimes\zeta_{l_2,m_2-1;\dot{l}_2,\dot{m}_2-1}&=&
\bar{b}^m_{31}\zeta_{l+1,m;\dot{l}+1,\dot{m}}+
\bar{b}^m_{32}\zeta_{lm\dot{l}\dot{m}}+
\bar{b}^m_{33}\zeta_{l-1,m;\dot{l}-1,\dot{m}},
\label{RR1t}
\end{eqnarray}
where
\[
\bar{b}^{(m)}={\renewcommand{\arraystretch}{1.4}\begin{pmatrix}
\bar{B}^{1,l_2,l_2-1;1,m_2,m}_{1,\dot{l}_2,\dot{l}_2-1;1,\dot{m}_2,\dot{m}} &
\bar{B}^{1,l_2,l_2-1;0,m_2,m}_{1,\dot{l}_2,\dot{l}_2-1;1,\dot{m}_2,\dot{m}} &
\bar{B}^{1,l_2,l_2-1;-1,m_2,m}_{1,\dot{l}_2,\dot{l}_2-1;-1,\dot{m}_2,\dot{m}}\\
\bar{B}^{1,l_2,l_2;1,m_2,m}_{1,\dot{l}_2,\dot{l}_2;1,\dot{m}_2,\dot{m}} &
\bar{B}^{1,l_2,l_2;0,m_2,m}_{1,\dot{l}_2,\dot{l}_2;0,\dot{m}_2,\dot{m}} &
\bar{B}^{1,l_2,l_2;-1,m_2,m}_{1,\dot{l}_2,\dot{l}_2;-1,\dot{m}_2,\dot{m}}\\
\bar{B}^{1,l_2,l_2+1;1,m_2,m}_{1,\dot{l}_2,\dot{l}_2+1;1,\dot{m}_2,\dot{m}} &
\bar{B}^{1,l_2,l_2+1;0,m_2,m}_{1,\dot{l}_2,\dot{l}_2+1;0,\dot{m}_2,\dot{m}} &
\bar{B}^{1,l_2,l_2+1;-1,m_2,m}_{1,\dot{l}_2,\dot{l}_2+1;-1,\dot{m}_2,\dot{m}}
\end{pmatrix}}=
\]
\begin{multline}
{\renewcommand{\arraystretch}{1.4}\left(\begin{array}{cc}
\sqrt{\frac{(l_2-m)(l_2-m+1)(\dot{l}_2-\dot{m})(\dot{l}_2-\dot{m}+1)}
{(2l_2+1)(2l_2+2)(2\dot{l}_2+1)(2\dot{l}_2+2)}} &
\sqrt{\frac{(l_2+m+1)(l_2-m)(\dot{l}_2+\dot{m}+1)(\dot{l}_2-\dot{m}}
{4l_2(l_2+1)\dot{l}_2(\dot{l}_2+1)}} \\
\sqrt{\frac{(l_2+m+1)(l_2-m+1)(\dot{l}_2+\dot{m}+1)(\dot{l}_2-\dot{m}+1)}
{(2l_2+1)(l_2+1)(2\dot{l}_2+1)(\dot{l}_2+1)}} &
\sqrt{\frac{m\dot{m}}{l_2(l_2+1)\dot{l}_2(\dot{l}_2+1)}} \\
\sqrt{\frac{(l_2+m)(l_2+m+1)(\dot{l}_2+\dot{m})(\dot{l}_2+\dot{m}+1)}
{(2l_2+1)(2l_2+2)(2\dot{l}_2+1)(2\dot{l}_2+2)}} &
\sqrt{\frac{(l_2+m)(l_2-m+1)(\dot{l}_2+\dot{m})(\dot{l}_2-\dot{m}+1)}
{4l_2(l_2+1)\dot{l}_2(\dot{l}_2+1)}} 
\end{array}\right.}\\
{\renewcommand{\arraystretch}{1.4}
\left.\begin{array}{c}
\sqrt{\frac{(l_2+m)(l_2+m+1)(\dot{l}_2+\dot{m})(\dot{l}_2+\dot{m}+1)}
{4l_2(2l_2+1)\dot{l}_2(\dot{l}_2+1)}}\\
\sqrt{\frac{(l_2+m_2)(l_2-m_2)(\dot{l}_2+\dot{m}_2)(\dot{l}_2-\dot{m}_2)}
{l_2(2l_2+1)\dot{l}_2(2\dot{l}_2+1)}}\\
\sqrt{\frac{(l_2-m)(l_2-m+1)(\dot{l}_2-\dot{m})(\dot{l}_2-\dot{m}+1)}
{4l_2(2l_2+1)\dot{l}_2(\dot{l}_2+1)}}
\end{array}\right).
}
\end{multline}
Let $T^l_{\mathfrak{g}}$ be a matrix of the irreducible representation of
the weight $l$ in the helicity basis. Let us apply the transformation
$T_{\mathfrak{g}}$ to the left and right parts of the each equalities
(\ref{RR1t}). In the left part we have
\begin{multline}
T_{\mathfrak{g}}\zeta_{1,k;1,\dot{k}}\otimes 
\zeta_{l_2,m_2-k;\dot{l}_2-\dot{k}}=T^1_{\mathfrak{g}}
\zeta_{1,k;1,\dot{k}}\otimes
T^{l_2}_{\mathfrak{g}}\zeta_{l_2,m_2-k;\dot{l}_2-\dot{k}}=\\
\bar{b}^m_{k+2,1}T^{l+1}_{\mathfrak{g}}\zeta_{l+1,m;\dot{l}+1}+
\bar{b}^m_{k+2,2}T^l_{\mathfrak{g}}\zeta_{lm;\dot{l}\dot{m}}+
\bar{b}^m_{k+2,3}T^{l-1}_{\mathfrak{g}}\zeta_{l-1,m;\dot{l}-1,\dot{m}},
\nonumber
\end{multline}
where $k,\dot{k}=-1,0,1$. Denoting the elements of $T^l_{\mathfrak{g}}$ via
$\fT^l_{mn}$ (generalized hyperspherical functions), we find
\begin{multline}
\left(\fT^1_{-1,k}\zeta_{1,-1;1,-1}+\fT^1_{0,k}\zeta_{1,0;1,0}+
\fT^1_{1,k}\zeta_{1,1;1,1}\right)
\sum\fT^l_{j,m-k}\zeta_{l_2,m_2-j;\dot{l}_2-\dot{j}}=\\
\sum\left(\bar{b}^m_{k+2,1}\fT^{l+1}_{jm}\zeta_{l+1,j;\dot{l}+1,\dot{j}}+
\bar{b}^m_{k+2,2}\fT^l_{jm}\zeta_{lj;\dot{l}\dot{j}}+
\bar{b}^m_{k+2,3}\fT^{l-1}_{jm}\zeta_{l-1,j;\dot{l}-1;\dot{j}}\right).\nonumber
\end{multline}
Replacing in the right part the vectors 
$\zeta_{l+1,j;\dot{l}+1,\dot{j}}$, $\zeta_{lj;\dot{l}\dot{j}}$, 
$\zeta_{l-1,j;\dot{l}-1,\dot{j}}$
via 
$\zeta_{1,-1;1,-1}\otimes\zeta_{l_2,j+1;\dot{l}_2,\dot{j}+1}$, 
$\zeta_{1,0;1,0}\otimes\zeta_{l_2j;\dot{l}_2\dot{j}}$, 
$\zeta_{1,1;1,1}\otimes\zeta_{l_2,j-1;\dot{l}_2,\dot{j}-1}$ 
and comparising the coefficients at 
$\zeta_{1,-1;1,-1}\otimes\zeta_{l_2,j+1;\dot{l}_2,\dot{j}+1}$, 
$\zeta_{1,0;1,0}\otimes\zeta_{l_2j;\dot{l}_2\dot{j}}$, 
$\zeta_{1,1;1,1}\otimes\zeta_{l_2,j-1;\dot{l}_2,\dot{j}-1}$ 
in the left and right parts, we
obtain three relations depending on $k$. Giving in these relations
three possible values $-1,0,1$ to the number $k$ and substituting
the functions $\fT^1_{mn}(\varphi,\epsilon,\theta,\tau,\psi,\varepsilon)$
(the matrix (\ref{T2})), we find the following nine recurrence relations:
\begin{multline}
\bar{b}^m_{11}\fT^{l+1}_{jm}\bar{b}^j_{11}+
\bar{b}^m_{12}\fT^l_{jm}\bar{b}^j_{12}+
\bar{b}^m_{13}\fT^{l-1}_{jm}\bar{b}^j_{13}=\\
\shoveright{
\left(\cos^2\frac{\theta}{2}\ch^2\frac{\tau}{2}+
\frac{i\sin\theta\sh\tau}{2}-\sin^2\frac{\theta}{2}\sh^2\frac{\tau}{2}\right)
e^{\epsilon+i\varphi+\varepsilon+i\psi}\fT^l_{j+1,m+1},}\nonumber\\
\shoveleft{
\bar{b}^m_{11}\fT^{l+1}_{jm}\bar{b}^j_{21}+
\bar{b}^m_{12}\fT^l_{jm}\bar{b}^j_{22}+
\bar{b}^m_{13}\fT^{l-1}_{jm}\bar{b}^j_{23}=}\\
\shoveright{
\frac{1}{\sqrt{2}}(\cos\theta\sh\tau+i\sin\theta\ch\tau)
e^{-\varepsilon-i\psi}\fT^l_{j,m+1},}\nonumber\\
\shoveleft{
\bar{b}^m_{11}\fT^{l+1}_{jm}\bar{b}^j_{31}+
\bar{b}^m_{12}\fT^l_{jm}\bar{b}^j_{32}+
\bar{b}^m_{13}\fT^{l-1}_{jm}\bar{b}^j_{33}=}\\
\shoveright{
\left(\cos^2\frac{\theta}{2}\sh^2\frac{\tau}{2}+
\frac{i\sin\theta\sh\tau}{2}-\sin^2\frac{\theta}{2}\ch^2\frac{\tau}{2}\right)
e^{\epsilon+i\varphi-\varepsilon-i\psi}\fT^l_{j-1,m+1},}\nonumber\\
\shoveleft{
\bar{b}^m_{21}\fT^{l+1}_{jm}\bar{b}^j_{11}+
\bar{b}^m_{22}\fT^l_{jm}\bar{b}^j_{12}+
\bar{b}^m_{23}\fT^{l-1}_{jm}\bar{b}^j_{13}=}\\
\shoveright{
\frac{1}{\sqrt{2}}(\cos\theta\sh\tau+i\sin\theta\ch\tau)
e^{\varepsilon+i\psi}\fT^l_{j+1,m},}\nonumber\\
\bar{b}^m_{21}\fT^{l+1}_{jm}\bar{b}^j_{21}+
\bar{b}^m_{22}\fT^l_{jm}\bar{b}^j_{22}+
\bar{b}^m_{23}\fT^{l-1}_{jm}\bar{b}^j_{23}=
(\cos\theta\ch\tau+i\sin\theta\sh\tau)\fT^l_{j,m},
\nonumber\\
\shoveleft{
\bar{b}^m_{21}\fT^{l+1}_{jm}\bar{b}^j_{31}+
\bar{b}^m_{22}\fT^l_{jm}\bar{b}^j_{32}+
\bar{b}^m_{23}\fT^{l-1}_{jm}\bar{b}^j_{33}=}\\
\shoveright{
\frac{1}{\sqrt{2}}(\cos\theta\sh\tau+i\sin\theta\ch\tau)
e^{-\varepsilon-i\psi}\fT^l_{j-1,m},}\nonumber\\
\shoveleft{
\bar{b}^m_{31}\fT^{l+1}_{jm}\bar{b}^j_{11}+
\bar{b}^m_{32}\fT^l_{jm}\bar{b}^j_{12}+
\bar{b}^m_{33}\fT^{l-1}_{jm}\bar{b}^i_{13}=}\\
\shoveright{
\left(\cos^2\frac{\theta}{2}\sh^2\frac{\tau}{2}+
\frac{i\sin\theta\sh\tau}{2}-\sin^2\frac{\theta}{2}\ch^2\frac{\tau}{2}\right)
e^{-\epsilon-i\varphi+\epsilon+i\psi}\fT^l_{j+1,m-1},}\nonumber\\
\shoveleft{
\bar{b}^m_{31}\fT^{l+1}_{jm}\bar{b}^j_{21}+
\bar{b}^m_{32}\fT^l_{jm}\bar{b}^j_{22}+
\bar{b}^m_{33}\fT^{l-1}_{jm}\bar{b}^j_{23}=}\\
\shoveright{
\frac{1}{\sqrt{2}}(\cos\theta\sh\tau+i\sin\theta\ch\tau)
e^{-\epsilon-i\varphi}\fT^l_{j,m-1},}\nonumber\\
\shoveleft{
\bar{b}^m_{31}\fT^{l+1}_{jm}\bar{b}^j_{31}+
\bar{b}^m_{32}\fT^l_{jm}\bar{b}^j_{32}+
\bar{b}^m_{33}\fT^{l-1}_{jm}\bar{b}^j_{33}=}\\
\left(\cos^2\frac{\theta}{2}\ch^2\frac{\tau}{2}+
\frac{i\sin\theta\sh\tau}{2}-\sin^2\frac{\theta}{2}\sh^2\frac{\tau}{2}\right)
e^{-\epsilon-i\varphi-\varepsilon-i\psi}\fT^l_{j-1,m-1}.\nonumber
\end{multline}
Let us find recurrence relations between the functions
$\fT^l_{mn}$, where the weight $l$ changed by $\frac{1}{2}$. 
Thus, at $l_1=1/2,\,\dot{l}_2=1/2$ and $l_1\ge 1/2$, $\dot{l}_2\ge 1/2$
by analogy with (\ref{RR1t}) we obtain
\begin{eqnarray}
\zeta_{\frac{1}{2},\frac{1}{2};\frac{1}{2},\frac{1}{2}}\otimes 
\zeta_{l_2,m_2-\frac{1}{2};\dot{l}_2,\dot{m}_2-\frac{1}{2}}&=&
\bar{b}^m_{00}\zeta_{l+\frac{1}{2},m;\dot{l}+\frac{1}{2},\dot{m}}+
\bar{b}^m_{01}\zeta_{l-\frac{1}{2},m;\dot{l}-\frac{1}{2},\dot{m}},\nonumber\\
\zeta_{\frac{1}{2},-\frac{1}{2};\frac{1}{2},-\frac{1}{2}}\otimes 
\zeta_{l_2,m_2+\frac{1}{2};\dot{l}_2,\dot{m}_2+\frac{1}{2}}&=&
\bar{b}^m_{10}\zeta_{l+\frac{1}{2},m;\dot{l}+\frac{1}{2},\dot{m}}+
\bar{b}^m_{11}\zeta_{l-\frac{1}{2},m;\dot{l}-\frac{1}{2},\dot{m}},\nonumber
\end{eqnarray}
where
\begin{gather}
\bar{b}^{(m)}=\begin{pmatrix}
\bar{B}^{\frac{1}{2},l_2,l_2+\frac{1}{2};-\frac{1}{2},m_2,m}_{\frac{1}{2},
\dot{l}_2,\dot{l}_2+\frac{1}{2};-\frac{1}{2},\dot{m}_2,\dot{m}} &
\bar{B}^{\frac{1}{2},l_2,l_2-\frac{1}{2};-\frac{1}{2},m_2,m}_{\frac{1}{2},
\dot{l}_2,\dot{l}_2-\frac{1}{2};-\frac{1}{2},\dot{m}_2,\dot{m}} \\
\bar{B}^{\frac{1}{2},l_2,l_2+\frac{1}{2};\frac{1}{2},m_2,m}_{\frac{1}{2},
\dot{l}_2,\dot{l}_2+\frac{1}{2};\frac{1}{2},\dot{m}_2,\dot{m}} &
\bar{B}^{\frac{1}{2},l_2,l_2-\frac{1}{2};\frac{1}{2},m_2,m}_{\frac{1}{2},
\dot{l}_2,\dot{l}_2-\frac{1}{2};\frac{1}{2},\dot{m}_2,\dot{m}} 
\end{pmatrix}=\nonumber\\
\begin{pmatrix}
\sqrt{\frac{(l_2-m+\frac{1}{2})(\dot{l}_2-\dot{m}+\frac{1}{2})}
{(2l_2+1)(2\dot{l}_2+1)}} & 
\sqrt{\frac{(l_2+m+\frac{1}{2})(\dot{l}_2+\dot{m}+\frac{1}{2})}
{(2l_2+1)(2\dot{l}_2+1)}}\\
\sqrt{\frac{(l_2+m+\frac{1}{2})(\dot{l}_2+\dot{m}+\frac{1}{2})}
{(2l_2+1)(2\dot{l}_2+1)}} & 
\sqrt{\frac{(l_2-m+\frac{1}{2})(\dot{l}_2-\dot{m}+\frac{1}{2})}
{(2l_2+1)(2\dot{l}_2+1)}}
\end{pmatrix}.\nonumber
\end{gather}
Carrying out the analogous calculations as for the case $l=1$ and
using the matrix (\ref{T1}), we come to the following recurrence
relations
\begin{eqnarray}
\bar{b}^m_{00}\fT^{l+\frac{1}{2}}_{jm}\bar{b}^j_{00}+
\bar{b}^m_{01}\fT^{l-\frac{1}{2}}_{jm}\bar{b}^j_{01}&=&
\left(\cos\frac{\theta}{2}\ch\frac{\tau}{2}+i\sin\frac{\theta}{2}
\sh\frac{\tau}{2}\right)e^{\frac{\epsilon+i\varphi+\varepsilon+i\psi}{2}}
\fT^l_{j+\frac{1}{2},m+\frac{1}{2}},
\nonumber\\
\bar{b}^m_{00}\fT^{l+\frac{1}{2}}_{jm}\bar{b}^j_{10}+
\bar{b}^m_{01}\fT^{l-\frac{1}{2}}_{jm}\bar{b}^j_{11}&=&
\left(\cos\frac{\theta}{2}\sh\frac{\tau}{2}+i\sin\frac{\theta}{2}
\ch\frac{\tau}{2}\right)e^{\frac{\epsilon+i\varphi-\varepsilon-i\psi}{2}}
\fT^l_{j-\frac{1}{2},m+\frac{1}{2}},
\nonumber\\
\bar{b}^m_{10}\fT^{l+\frac{1}{2}}_{jm}\bar{b}^j_{00}+
\bar{b}^m_{11}\fT^{l-\frac{1}{2}}_{jm}\bar{b}^j_{01}&=&
\left(\cos\frac{\theta}{2}\sh\frac{\tau}{2}+i\sin\frac{\theta}{2}
\ch\frac{\tau}{2}\right)e^{\frac{-\epsilon-i\varphi+\varepsilon+i\psi}{2}}
\fT^l_{j+\frac{1}{2},m-\frac{1}{2}},\nonumber\\
\bar{b}^m_{10}\fT^{l+\frac{1}{2}}_{jm}\bar{b}^j_{10}+
\bar{b}^m_{11}\fT^{l-\frac{1}{2}}_{jm}\bar{b}^j_{11}&=&
\left(\cos\frac{\theta}{2}\ch\frac{\tau}{2}+i\sin\frac{\theta}{2}
\sh\frac{\tau}{2}\right)e^{\frac{-\epsilon-i\varphi-\varepsilon-i\psi}{2}}
\fT^l_{j-\frac{1}{2},m-\frac{1}{2}}.\nonumber
\end{eqnarray}

\section{Harmonic analysis on the group $SU(2)\otimes SU(2)$}
Since $SU(2)\otimes SU(2)$ is locally compact, then there exists an
invariant measure $d\fg$ (Haar measure) on this group, that is, such a
measure that for any finite continuous function
$f(\fg)$ we have
\[
\int f(\fg)d\fg=\int f(\fg_0\fg)d\fg=\int f(\fg\fg_0)d\fg=\int f(\fg^{-1})d\fg.
\]
Applying the equations (\ref{X1})--(\ref{Y3}), we come to a following
expression for the Haar measure in terms of the parameters (\ref{CEA}):
\begin{equation}\label{HA1}
d\fg=\sin\theta^c\sin\dot{\theta}^cd\theta d\varphi d\psi d\tau d\epsilon
d\varepsilon.
\end{equation}
Thus, an invariant integration on the group $SU(2)\otimes SU(2)$
is defined by the formula
\[
\int\limits_{SU(2)\otimes SU(2)}f(g)d\fg=\frac{1}{32\pi^4}
\int\limits^{+\infty}_{-\infty}
\int\limits^{+\infty}_{-\infty}
\int\limits^{+\infty}_{-\infty}
\int\limits^{2\pi}_{-2\pi}
\int\limits^{2\pi}_0
\int\limits^\pi_0
f(\theta,\varphi,\psi,\tau,\epsilon,\varepsilon)
\sin\theta^c\sin\dot{\theta}^cd\theta d\varphi d\psi d\tau d\epsilon
d\varepsilon.
\]
Since a dimension of the spinor representation $T_{\fg}$ of
$SU(2)\otimes SU(2)$ is equal to $(2l+1)(2\dot{l}+1)$, then the functions
$\sqrt{(2l+1)(2\dot{l}+1)}t^l_{mn}(\fg)$ form a full orthogonal normalized
system on this group with respect to the invariant measure $d\fg$. At this point,
the index $l$ runs all possible integer or half-integer non-negative values,
and the indices $m$ and $n$ run the values
$-l,-l+1,\ldots,l-1,l$. In virtue of (\ref{HS}) the matrix elements
$t^l_{mn}$ are expressed via the generalized hyperspherical function
$t^l_{mn}(\fg)=\fM^l_{mn}(\varphi,\epsilon,\theta,\tau,\psi,\varepsilon)$.
Therefore,
\begin{equation}\label{HA2}
\int\limits_{SU(2)\otimes SU(2)}\fM^l_{mn}(\fg)
\overline{\fM^l_{mn}(\fg)}d\fg=\frac{32\pi^4}{(2l+1)(2\dot{l}+1)}
\delta(\fg^\prime-\fg),
\end{equation}
where $\delta(\fg^\prime-\fg)$ is a $\delta$-function on the group
$SU(2)\otimes SU(2)$. An explicit form of $\delta$-function is
\begin{multline}
\delta(\fg^\prime-\fg)=\delta(\varphi^\prime-\varphi)
\delta(\epsilon^\prime-\epsilon)\delta(\cos\theta^\prime\ch\tau^\prime-
\cos\theta\ch\tau)\times\\
\times\delta(\sin\theta^\prime\sh\tau^\prime-\sin\theta\sh\tau)
\delta(\psi^\prime-\psi)\delta(\varepsilon^\prime-\varepsilon).\nonumber
\end{multline}
Substituting into (\ref{HA2}) the expression
\[
\fM^l_{mn}(\fg)=e^{-m(\epsilon+i\varphi)}Z^l_{mn}e^{-n(\varepsilon+i\psi)}
\]
and taking into account (\ref{HA1}), we obtain
\begin{multline}
\int\limits^{+\infty}_{-\infty}
\int\limits^{+\infty}_{-\infty}
\int\limits^{+\infty}_{-\infty}
\int\limits^{2\pi}_{-2\pi}
\int\limits^{2\pi}_0
\int\limits^\pi_0
Z^l_{mn}\overline{Z^s_{pq}}e^{-(m+p)\epsilon}e^{-i(m-p)\varphi}\times\\
\times e^{-(n+q)\varepsilon}e^{-i(n-q)\psi}
\sin\theta^c\sin\dot{\theta}^cd\theta d\varphi d\psi d\tau d\epsilon
d\varepsilon=\frac{32\pi^4\delta_{ls}\delta_{mp}\delta_{nq}
\delta(\fg^\prime-\fg)}{(2l+1)(2\dot{l}+1)}.\nonumber
\end{multline}
Since the hyperspherical function of the principal series has the form
\[
Z^{-\frac{1}{2}+i\rho}_{mn}(\cos\theta^c)=
\sum^{\ld\frac{\lambda}{2}\rd}_{k=-\frac{\lambda}{2}}
P^{\frac{\lambda}{2}}_{mk}(\cos\theta)\fP^{-\frac{1}{2}+i\rho}_{kn}(\ch\tau),
\]
then any function $f(\fg)$ on the group $SU(2)\otimes SU(2)$, such that
$\int|f(\fg)|^2d\fg<+\infty$, is expanded on the matrix elements
of the principal series
\[
\fM^{-\frac{1}{2}+i\rho}_{mn}(\fg)=
e^{-m(\epsilon+i\varphi)-n(\varepsilon+i\psi)}
\sum^{\ld\frac{\lambda}{2}\rd}_{k=-l}
P^{\frac{\lambda}{2}}_{mk}(\cos\theta)
\fP^{-\frac{1}{2}+i\rho}_{kn}(\ch\tau)
\]
Since the Lorentz group is noncompact, then an expansion should be run on the
conical functions
$\fP^{-\frac{1}{2}+i\rho}_{kn}(\ch\tau)$, where the parameter $\tau$ changes
in the limits $0\leq\tau<\infty$. In other words, this problem can be
formulated as follows.

Let $m$ and $n$ be simultaneously integer or half-integer numbers,
and let $F(x)$ be the function such that
\[
\int\limits^\infty_1|F(x)|^2dx<+\infty.
\]
It takes to expand the function $F(x)$ via the functions $\fP^l_{mn}(x)$,
where $l$ is an expansion parameter and $1\leq x<\infty$.

As it shown in \cite{Vil68}, the functions
$\fP^l_{mn}(x)$ satisfy the self-conjugate differential equation
\[
\frac{d}{dx}(x^2-1)\frac{du}{dx}-\frac{m^2+n^2-2mnx}{x^2-1}=l(l+1)u.
\]
The functions $\fP^l_{mn}(x)$ are continuous at the point $x=1$ 
corresponding to $\tau=0$. In other words,
$\fP^l_{mn}(x)$ are eigenfunctions of the self-conjugate operator
\[
\frac{d}{dx}(x^2-1)\frac{du}{dx}-\frac{m^2+n^2-2mnx}{x^2-1},\quad
1\leq x<\infty.
\]

Using the standard expansion technique on eigenfunctions of self-conjugate
operators (see, for example, \cite{Lev50,Nai54}), Vilenkin
\cite{Vil68} derived the following result.
\begin{theorem}[{\rm Vilenkin \cite{Vil68}}]
If $m$ and $n$ are integer numbers, then any function $F(x)$, such that
$\int\limits^\infty_1|F(x)|^2dx<+\infty$, has an expansion
\[
F(x)=\frac{1}{4\pi^2}\int\limits^\infty_0a(\rho)
\fP^{-\frac{1}{2}+i\rho}_{mn}(x)\rho\tnh\pi\rho d\rho
+\frac{1}{4\pi^2}\sum^M_{l=1}\left(l-\frac{1}{2}\right)b(l)\fP^l_{mn}(x),
\]
where
\[
M=
\begin{cases}
\min(|m|,|n|) & \text{if\;$mn\ge 0$},\\
0 & \text{if $mn<0$}.
\end{cases}
\]
The summation is produced via integer values $l$. The coefficients
$a(\rho)$ and $b(l)$ in this expansion are expressed by the formulas:
\[
a(\rho)=\int\limits^\infty_1F(x)\fP^{-\frac{1}{2}-i\rho}_{mn}(x)dx\equiv
\int\limits^\infty_1F(x)\overline{\fP^{-\frac{1}{2}+i\rho}_{mn}(x)}dx
\]
and
\begin{multline}
b(l)=(-1)^{m-n}\int\limits^\infty_1F(x)\fP^{-l-1}_{mn}(x)dx\equiv\\
\equiv(-1)^{m-n}\frac{\Gamma(l+m+1)\Gamma(l-m+1)}
{\Gamma(l+n+1)\Gamma(l-n+1)}\int\limits^\infty_1F(x)
\overline{\fP^l_{mn}(x)}dx.\nonumber
\end{multline}
There is an analog of the Plancherel formula:
\begin{multline}
\int\limits^\infty_1|F(x)|^2dx=\frac{1}{4\pi^2}\int\limits^\infty_0
|a(\rho)|^2\rho\tnh\pi\rho d\rho+\\
+\frac{(-1)^{n-m}}{4\pi^2}\sum^M_{l=1}\left(l-\frac{1}{2}\right)
\frac{\Gamma(l+n+1)\Gamma(l-n+1)}{\Gamma(l+m+1)\Gamma(l-m+1)}|b(l)|^2.
\nonumber
\end{multline}
Analogously, if $m$ and $n$ are half-integer numbers, then
\[
F(x)=\frac{1}{4\pi^2}\int\limits^\infty_0a(\rho)
\fP^{-\frac{1}{2}+i\rho}_{mn}(x)\rho\cth\pi\rho d\rho
+\frac{1}{4\pi^2}\sum^M_{l=\frac{1}{2}}
\left(l-\frac{1}{2}\right)b(l)\fP^l_{mn}(x),
\]
where
\[
M=
\begin{cases}
\min(|m|,|n|) & \text{if\;$mn > 0$},\\
0 & \text{if $mn<0$}.
\end{cases}
\]
The summation is produced via half-integer values $l$. 
The coefficients $a(\rho)$ and
$b(l)$ are expressed by the same formulas that take place 
for integer $m$ and $n$. In this case, the Plancherel formula has a
following form
\begin{multline}
\int\limits^\infty_1|F(x)|^2dx=\frac{1}{4\pi^2}\int\limits^\infty_0
|a(\rho)|^2\rho\cth\pi\rho d\rho+\\
+\frac{(-1)^{m-n}}{4\pi^2}\sum^M_{l=\frac{1}{2}}\left(l-\frac{1}{2}\right)
\frac{\Gamma(l+n+1)\Gamma(l-n+1)}{\Gamma(l+m+1)\Gamma(l-m+1)}|b(l)|^2.
\nonumber
\end{multline}
\end{theorem}
Let us apply the Vilenkin Theorem to an expansion of the functions $f(\fg)$ 
on the Lorentz group, such that
$\int|f(\fg)|^2d\fg<+\infty$. Then, taking into account that matrix elements
of the principal series representations have the form
\[
\fM^{-\frac{1}{2}+i\rho}_{mn}(\fg)=e^{-m(\epsilon+i\varphi)-n(\varepsilon+
i\psi)}\sum^{\ld\frac{\lambda}{2}\rd}_{k=-\frac{\lambda}{2}}
P^{\frac{\lambda}{2}}_{mk}(\cos\theta)\fP^{-\frac{1}{2}+i\rho}_{kn}(\ch\tau),
\]
we obtain
\begin{multline}
f(\fg)=\frac{(2l_0+1)(2\dot{l}_0+1)}{32\pi^4}\sum_{m,n}
\sum^{\ld\frac{\lambda}{2}\rd}_{k=-\frac{\lambda}{2}}
\left[\int\limits^\infty_0a_{mn}(\rho)
e^{-m(\epsilon+i\varphi)-n(\varepsilon+i\psi)}\times\right.\\
\times P^{\frac{\lambda}{2}}_{mk}(\cos\theta)
\fP^{-\frac{1}{2}+i\rho}_{kn}(\ch\tau)
\rho\tnh\pi\rho d\rho+\\
+\left.\sum^\infty_{l_0=1}\left(l_0-\frac{1}{2}\right)b_{mn}(l_0)
e^{-m(\epsilon+i\varphi)-n(\varepsilon+i\psi)}P^{l_0}_{mk}(\cos\theta)
\fP^{l_0}_{kn}(\ch\tau)\right].\label{Decomp1}
\end{multline}
The values of the coefficients $a_{mn}(\rho)$ and $b_{mn}(l_0)$ are expressed
by the formulas
\begin{gather}
a_{mn}(\rho)=\sum^{\ld\frac{\lambda}{2}\rd}_{k=-\frac{\lambda}{2}}
\int f(\fg)e^{-m(\epsilon+i\varphi)-n(\varepsilon+
i\psi)}P^{\frac{\lambda}{2}}_{mk}(\cos\theta)
\fP^{-\frac{1}{2}-i\rho}_{kn}(\ch\tau)d\fg,
\label{Coefa}\\
b_{mn}(l_0)=\frac{(-1)^{m-n}\Gamma(l_0+m+1)\Gamma(l_0-m+1)}
{\Gamma(l_0+n+1)\Gamma(l_0-n+1)}
\int f(\fg)\fM^{l_0}_{mn}(\fg)d\fg,\label{Coefb}
\end{gather}
where $d\fg$ is the Haar measure on the Lorentz group in Euler parameters:
\[
d\fg=\sin\theta^c\sin\dot{\theta}^cd\theta d\varphi d\psi d\tau d\epsilon
d\varepsilon.
\]

In the case of discrete series of representations, the expansion takes
a form
\begin{multline}
f(\fg)=\frac{(2l_0+1)(2\dot{l}_0+1)}{32\pi^4}\sum_{m,n}
\sum^{\ld\frac{\lambda}{2}\rd}_{k=-\frac{\lambda}{2}}
\left[\int\limits^\infty_0a_{mn}(\rho)
e^{-m(\epsilon+i\varphi)-n(\varepsilon+i\psi)}\times\right.\\
\times P^{\frac{\lambda}{2}}_{mk}(\cos\theta)
\fP^{-\frac{1}{2}+i\rho}_{kn}(\ch\tau)
\rho\tnh\pi\rho d\rho+\\
+\left.\sum^M_{l_0=1}\left(l_0-\frac{1}{2}\right)b_{mn}(l_0)
e^{-m(\epsilon+i\varphi)-n(\varepsilon+i\psi)}P^{l_0}_{mk}(\cos\theta)
\fP^{l_0}_{kn}(\ch\tau)\right].\label{Decomp2}
\end{multline}
where
\[
M=
\begin{cases}
\min(|m|,|n|) & \text{if\;$mn\ge 0$},\\
0 & \text{if $mn<0$}.
\end{cases}
\]
The coefficients $a_{mn}(\rho)$ and $b_{mn}(l)$ have the form
(\ref{Coefa}) and (\ref{Coefb}). Analogously, if
$m$ and $n$ are half-integer numbers, then
\begin{multline}
f(\fg)=\frac{(2l_0+1)(2\dot{l}_0+1)}{32\pi^4}\sum_{m,n}
\sum^{\ld\frac{\lambda}{2}\rd}_{k=-\frac{\lambda}{2}}
\left[\int\limits^\infty_0a_{mn}(\rho)
e^{-m(\epsilon+i\varphi)-n(\varepsilon+i\psi)}\times\right.\\
\times P^{\frac{\lambda}{2}}_{mk}(\cos\theta)
\fP^{-\frac{1}{2}+i\rho}_{kn}(\ch\tau)
\rho\cth\pi\rho d\rho+\\
+\left.\sum^M_{l_0=\frac{1}{2}}\left(l_0-\frac{1}{2}\right)b_{mn}(l_0)
e^{-m(\epsilon+i\varphi)-n(\varepsilon+i\psi)}P^{l_0}_{mk}(\cos\theta)
\fP^{l_0}_{kn}(\ch\tau)\right].\label{Decomp3}
\end{multline}
where
\[
M=
\begin{cases}
\min(|m|,|n|) & \text{if\;$mn> 0$},\\
0 & \text{if $mn<0$}.
\end{cases}
\]
The coefficients $a_{mn}(\rho)$ and $b_{mn}(l_0)$ have the same form
as in the case of integer $m$ and $n$.

Since
\[
\cth\pi\rho=\tnh\pi\left(\rho+\frac{i}{2}\right),
\]
then integral terms in the expansions (\ref{Decomp2}) and (\ref{Decomp3})
can be written uniformly if we replace $\tnh\pi\rho$ and
$\cth\pi\rho$ by $\tnh\pi(\rho+oi)$, where $o=0$ in the integer case and
$o=1/2$ in the half-integer case.
The following expansion is an unification of the expansions (\ref{Decomp2})
and (\ref{Decomp3}):
\begin{multline}
f(\fg)=\frac{(2l_0+1)(2\dot{l}_0+1)}{32\pi^4}\sum_{m,n}
\sum^{\ld\frac{\lambda}{2}\rd}_{k=-\frac{\lambda}{2}}
\left[\int\limits^\infty_0a^o_{mn}(\rho)
e^{-m(\epsilon+i\varphi)-n(\varepsilon+i\psi)}\right.\times\\
\times P^{\frac{\lambda}{2}}_{mk}(\cos\theta)
\fP^{\left(-\frac{1}{2}+i\rho,o\right)}_{kn}(\ch\tau)
\rho\tnh\pi(\rho+oi)
d\rho+\\
+\left.\sum^M_{l_0=1-o}\left(l_0-\frac{1}{2}\right)b^o_{mn}(l_0)
e^{-m(\epsilon+i\varphi)-n(\varepsilon+i\psi)}P^{l_0}_{mk}(\cos\theta)
\fP^{l_0}_{kn}(\ch\tau)\right],\label{Decomp4}
\end{multline}
where
\begin{gather}
a^o_{mn}(\rho)=\sum^{\ld\frac{\lambda}{2}\rd}_{k=-\frac{\lambda}{2}}
\int f(\fg)e^{-m(\epsilon+i\varphi)-n(\varepsilon+
i\psi)}P^{\frac{\lambda}{2}}_{mk}(\cos\theta)
\fP^{\left(-\frac{1}{2}+i\rho,o\right)}_{kn}(\ch\tau)d\fg,
\label{Coefa2}\\
b^o_{mn}(l_0)=\frac{(-1)^{m-n}\Gamma(l_0+m+o+1)\Gamma(l_0-m-o+1)}
{\Gamma(l_0+n+o+1)\Gamma(l_0-n-o+1)}\int f(\fg)
\fM^{l_0}_{mn}(\fg)d\fg,\label{Coefb2}
\end{gather}
\[
d\fg=\sin\theta^c\sin\dot{\theta}^cd\theta d\varphi d\psi d\tau d\epsilon
d\varepsilon.
\]
There is an analog of the Plancherel formula:
\begin{multline}
\int|f(\fg)|^2d\fg=\frac{(2l_0+1)(2\dot{l}_0+1)}{32\pi^4}\sum_{m,n,o}
\left[\int\limits^\infty_0|a^o_{mn}(\rho)|^2\rho
\tnh\pi(\rho+oi)d\rho+\right.\\
+\left.(-1)^{m-n}\sum^M_{l_0=1-o}\frac{\Gamma(l_0+n+o+1)\Gamma(l_0-n-o+1)}
{\Gamma(l_0+m+o+1)\Gamma(l_0-m-o+1)}\left(l_0-\frac{1}{2}\right)
|b^o_{mn}(l_0)|^2\right].\nonumber
\end{multline}
Thus, the expansion of square integrable functions $f(\fg)$ on the group
$SL(2,\C)$ has only matrix elements of the principal series of representations.
Representations of supplementary series do not participate in the expansion.

An expansion of the functions $f(\fg)$ on the Lorentz group in terms
of associated hyperspherical functions, that is, an expansion on the
two-dimensional complex sphere, has an important meaning in physical
applications. Using the explicit expression
(\ref{AHPS}) for the associated hyperspherical functions of the principal
series, we find the following expansion:
\begin{multline}
f(\fg)=\frac{(2l_0+1)(2\dot{l}_0+1)}{32\pi^4}\sum_{m}
\sum^{\ld\frac{\lambda}{2}\rd}_{k=-\frac{\lambda}{2}}
\left[\int\limits^\infty_0a^o_{m}(\rho)
e^{-m(\epsilon+i\varphi)}\right.\times\\
\times P^{\frac{\lambda}{2}}_{mk}(\cos\theta)
\fP_{\left(-\frac{1}{2}+i\rho,o\right)}^{k}(\ch\tau)
\rho\tnh\pi(\rho+oi)
d\rho+\\
+\left.\sum^M_{l_0=1-o}\left(l_0-\frac{1}{2}\right)b^o_{m}(l_0)
e^{-m(\epsilon+i\varphi)}P^{l_0}_{mk}(\cos\theta)
\fP_{l_0}^{k}(\ch\tau)\right],\label{Decomp5}
\end{multline}
where
\begin{gather}
a^o_{m}(\rho)=\sum^{\ld\frac{\lambda}{2}\rd}_{k=-\frac{\lambda}{2}}
\int f(\fg)e^{-m(\epsilon+i\varphi)}P^{\frac{\lambda}{2}}_{mk}(\cos\theta)
\fP_{\left(-\frac{1}{2}+i\rho,o\right)}^{k}(\ch\tau)d\fg,
\label{Coefa3}\\
b^o_{m}(l_0)=\frac{(-1)^{m}\Gamma(l_0+m+o+1)\Gamma(l_0-m-o+1)}
{\Gamma(l_0+o+1)\Gamma(l_0-o+1)}\int f(\fg)
\fM_{l_0}^{m}(\fg)d\fg\label{Coefb3}
\end{gather}
and
\[
d\fg=\sin\theta^c\sin\dot{\theta}^cd\theta d\varphi d\tau d\epsilon.
\]
is a Haar measure on the surface of the two-dimensional complex sphere.

In turn, physical fields, describing particles with an arbitrary spin
(both integer and half-integer), are defined in terms of finite-dimensional
(spinor) representations of the group
$\fG_+$ via the expansions on associated hyperspherical functions realized
on the surface of the two-dimensional complex sphere.
At this point, the expansion
(\ref{Decomp5}) takes a form
\begin{equation}\label{Decomp6}
f(\fg)=\frac{(2l+1)(2\dot{l}+1)}{32\pi^4}\sum^{l}_{k=-l}
\sum^M_{l=1-o}\left(l-\frac{1}{2}\right)b^o_m(l)
e^{-m(\epsilon+i\varphi)}P^l_{mk}(\cos\theta)\fP^k_l(\ch\tau),
\end{equation}
where
\begin{equation}\label{Coefb4}
b^o_m(l)=\frac{(-1)^m\Gamma(l+m+o+1)\Gamma(l-m-o+1)}
{\Gamma(l+o+1)\Gamma(l-o+1)}\int f(\fg)\fM^m_l(\fg)d\fg.
\end{equation}

An expansion on the cone for the square integrable function (the wave
function of free particle with the null spin) was first given by
I. S. Shapiro \cite{Sha56,Sha62}. Later on, Chou Kuang-chao and Zastavenko
\cite{CZ58} refined the Shapiro expansion for the particles with non-null
spin (see also \cite{DT59,Pop59}). From mathematical viewpoint the functions
obtained in \cite{Sha56,CZ58} present an integral transformation, which allows one
to obtain an expansion of the function, defined on the hyperboloid, via the
conical functions. Such expansions with the use of integral geometry have
been further studied in the Gel'fand-Graev works
\cite{GGV62} and later have been used by Vilenkin and Smorodinsky
\cite{VS64} for a definition of the formulas of direct and inverse expansions
on degenerate representations of the Lorentz group in different coordinate
systems. Further, in sequel of the work \cite{VS64} the authors of
\cite{WF65,KS67} studied the questions about relation between different
subgroups of the Lorentz group and coordinate systems on the hyperboloid and
also the questions concerning convergence and asymptotic expansions. The
following step in this direction was became in
\cite{LSS68,KLMS69}, where the authors constructed irreducible unitary
representations  of the Lorentz group, realized on the space of functions
defined on the direct product of two spaces -- the hyperboloid (with the
infinite-dimensional representation) and the sphere (with the 
finite-dimensional spinor representation). These functions, forming the
basis for integral representations of scattering amplitudes, are represented
by the product of spherical functions (Legendre functions) and conical
functions (see also\cite{PW70}). In the work
\cite{KLMS69} it has been shown that a realization of the Lorentz group
representations on the cone is closely related with the formulas of 
helicity expansions \cite{JW59}. Basis functions on the cone and
relativistically-invariant expansions of the spiral scattering amplitudes
have been obtained by Verdiev \cite{Ver69,Ver68}.

{\bf Example}. By way of example let us consider a decomposition of the
functions on the group $SU(2)\otimes SU(2)$, defining solutions for the
Dirac field $(1/2,0)\oplus(0,1/2)$. In \cite{Var031,Var033} it has been
shown that these solutions are defined in terms of associated hyperspherical
functions:
\begin{eqnarray}
\psi_1(r,\varphi^c,\theta^c)&=&\boldsymbol{f}^l_{\frac{1}{2},\frac{1}{2}}
(\re r)\fM^{\frac{1}{2}}_l(\varphi,\epsilon,\theta,\tau,0,0),\nonumber\\
\psi_2(r,\varphi^c,\theta^c)&=&\pm\boldsymbol{f}^l_{\frac{1}{2},\frac{1}{2}}
(\re r)\fM^{-\frac{1}{2}}_l(\varphi,\epsilon,\theta,\tau,0,0),\nonumber\\
\dot{\psi}_1(r^\ast,\dot{\varphi}^c,\dot{\theta}^c)&=&
\mp\boldsymbol{f}^{\dot{l}}_{\frac{1}{2},-\frac{1}{2}}
(\re r^\ast)\fM_{\dot{l}}^{\frac{1}{2}}
(\varphi,\epsilon,\theta,\tau,0,0),\nonumber\\
\dot{\psi}_2(r^\ast,\dot{\varphi}^c,\dot{\theta}^c)&=&
\boldsymbol{f}^{\dot{l}}_{\frac{1}{2},-\frac{1}{2}}
(\re r^\ast)\fM_{\dot{l}}^{-\frac{1}{2}}(\varphi,\epsilon,\theta,\tau,0,0),
\nonumber
\end{eqnarray}
where
\begin{eqnarray}
&&l=\frac{1}{2},\;\frac{3}{2},\;\frac{5}{2},\ldots;\nonumber\\
&&\dot{l}=\frac{1}{2},\;\frac{3}{2},\;\frac{5}{2},\ldots;\nonumber
\end{eqnarray}
\[
\fM^{\pm\frac{1}{2}}_l(\varphi,\epsilon,\theta,\tau,0,0)=
e^{\mp\frac{1}{2}(\epsilon+i\varphi)}Z^{\pm\frac{1}{2}}_l(\theta,\tau),
\]
\begin{multline}
Z^{\pm\frac{1}{2}}_l(\theta,\tau)=\cos^{2l}\frac{\theta}{2}
\ch^{2l}\frac{\tau}{2}\sum^l_{k=-l}i^{\pm\frac{1}{2}-k}
\tg^{\pm\frac{1}{2}-k}\frac{\theta}{2}\tnh^{-k}\frac{\tau}{2}\times\\
\hypergeom{2}{1}{\pm\frac{1}{2}-l+1,1-l-k}{\pm\frac{1}{2}-k+1}
{i^2\tg^2\frac{\theta}{2}}
\hypergeom{2}{1}{-l+1,1-l-k}{-k+1}{\tnh^2\frac{\tau}{2}},\nonumber
\end{multline}
\[
\fM_{\dot{l}}^{\pm\frac{1}{2}}(\varphi,\epsilon,\theta,\tau,0,0)=
e^{\mp\frac{1}{2}(\epsilon-i\varphi)}
Z_{\dot{l}}^{\pm\frac{1}{2}}(\theta,\tau),
\]
\begin{multline}
Z_{\dot{l}}^{\pm\frac{1}{2}}(\theta,\tau)=
\cos^{2\dot{l}}\frac{\theta}{2}
\ch^{2\dot{l}}\frac{\tau}{2}
\sum^{\dot{l}}_{\dot{k}=-\dot{l}}i^{\pm\frac{1}{2}-\dot{k}}
\tg^{\pm\frac{1}{2}-\dot{k}}\frac{\theta}{2}
\tnh^{-\dot{k}}\frac{\tau}{2}\times\\
\hypergeom{2}{1}{\pm\frac{1}{2}-\dot{l}+1,1-\dot{l}-\dot{k}}
{\pm\frac{1}{2}-\dot{k}+1}
{i^2\tg^2\frac{\theta}{2}}
\hypergeom{2}{1}{-\dot{l}+1,1-\dot{l}-\dot{k}}
{-\dot{k}+1}{\tnh^2\frac{\tau}{2}}.\nonumber
\end{multline}
The radial components $\boldsymbol{f}^l_{\frac{1}{2},\frac{1}{2}}(\re r)$ and
$\boldsymbol{f}^{\dot{l}}_{\frac{1}{2},-\frac{1}{2}}(\re r^\ast)$, where
$r$ is the radius of the two-dimensional complex sphere, are expressed via
the Bessel functions of half-integer order:
\begin{eqnarray}
\boldsymbol{f}^l_{\frac{1}{2},\frac{1}{2}}(\re r)&=&
C_1\sqrt{\kappa^c\dot{\kappa}^c}\re r 
J_l\left(\sqrt{\kappa^c\dot{\kappa}^c}\re r\right)+
C_2\sqrt{\kappa^c\dot{\kappa}^c}\re r 
J_{-l}\left(\sqrt{\kappa^c\dot{\kappa}^c}\re r\right),\nonumber\\
\boldsymbol{f}^{\dot{l}}_{\frac{1}{2},-\frac{1}{2}}(\re r^\ast)&=&
\frac{C_1}{2}\sqrt{\frac{\dot{\kappa}^c}{\kappa^c}}\re r^\ast
J_{l+1}\left(\sqrt{\kappa^c\dot{\kappa}^c}\re r^\ast\right)-
\frac{C_2}{2}\sqrt{\frac{\dot{\kappa}^c}{\kappa^c}}\re r^\ast
J_{-l-1}\left(\sqrt{\kappa^c\dot{\kappa}^c}\re r^\ast\right).\nonumber
\end{eqnarray}
Then the decomposition for the field $(1/2,0)\oplus(0,1/2)$ takes a form:
\[
\psi_1=\frac{(2l+1)(2\dot{l}+1)}{32\pi^4}
\boldsymbol{f}^l_{\frac{1}{2},\frac{1}{2}}(\re r)\sum^l_{k=-l}
\sum^M_{l=\frac{1}{2}}\left(l-\frac{1}{2}\right)b_{\frac{1}{2}}(l)
e^{-\frac{1}{2}(\epsilon+i\varphi)}P^l_{\frac{1}{2},k}(\cos\theta)
\fP^k_{l}(\ch\tau),
\]
\[
\psi_2=\pm\frac{(2l+1)(2\dot{l}+1)}{32\pi^4}
\boldsymbol{f}^l_{\frac{1}{2},\frac{1}{2}}(\re r)\sum^l_{k=-l}
\sum^M_{l=\frac{1}{2}}\left(l-\frac{1}{2}\right)b_{-\frac{1}{2}}(l)
e^{\frac{1}{2}(\epsilon+i\varphi)}P^l_{-\frac{1}{2},k}(\cos\theta)
\fP^k_{l}(\ch\tau),
\]
\[
\dot{\psi}_1=\mp\frac{(2l+1)(2\dot{l}+1)}{32\pi^4}
\boldsymbol{f}^{\dot{l}}_{\frac{1}{2},-\frac{1}{2}}(\re r^\ast)
\sum^{\dot{l}}_{\dot{k}=-\dot{l}}
\sum^{\dot{M}}_{\dot{l}=\frac{1}{2}}\left(\dot{l}-\frac{1}{2}\right)
b_{\frac{1}{2}}(\dot{l})e^{-\frac{1}{2}(\epsilon-i\varphi)}
P^{\dot{l}}_{\frac{1}{2},\dot{k}}(\cos\theta)
\fP_{\dot{l}}^{\dot{k}}(\ch\tau),
\]
\[
\dot{\psi}_2=\frac{(2l+1)(2\dot{l}+1)}{32\pi^4}
\boldsymbol{f}^{\dot{l}}_{\frac{1}{2},-\frac{1}{2}}(\re r^\ast)
\sum^{\dot{l}}_{\dot{k}=-\dot{l}}
\sum^{\dot{M}}_{\dot{l}=\frac{1}{2}}\left(\dot{l}-\frac{1}{2}\right)
b_{-\frac{1}{2}}(\dot{l})e^{\frac{1}{2}(\epsilon-i\varphi)}
P^{\dot{l}}_{-\frac{1}{2},\dot{k}}(\cos\theta)
\fP_{\dot{l}}^{\dot{k}}(\ch\tau),
\]
where
\[
b_{\pm\frac{1}{2}}(l)=\frac{(-1)^{\pm\frac{1}{2}}\Gamma(l\pm\frac{1}{2}+1)
\Gamma(l\mp\frac{1}{2}+1)}{\Gamma(l+1)\Gamma(l+1)}
\int\ld\fM^{\pm\frac{1}{2}}_l(\varphi,\epsilon,\theta,\tau,0,0)\rd^2d\fg,
\]
\[
b_{\pm\frac{1}{2}}(\dot{l})=
\frac{(-1)^{\pm\frac{1}{2}}\Gamma(\dot{l}\pm\frac{1}{2}+1)
\Gamma(\dot{l}\mp\frac{1}{2}+1)}{\Gamma(\dot{l}+1)
\Gamma(\dot{l}+1)}\int\ld\fM_{\dot{l}}^{\pm\frac{1}{2}}
(\varphi,\epsilon,\theta,\tau,0,0)\rd^2d\fg,
\]
here $d\fg$ is a Haar measure on the surface of the two-dimensional complex
sphere:
\[
d\fg=\sin\theta^c\sin\dot{\theta}^cd\theta d\varphi d\tau d\epsilon.
\]
In conclusion, it should be noted that such a description corresponds
to a quantum field theory on the Poincar\'{e} group introduced by
Lur\c{c}at \cite{Lur64} 
(see also \cite{BK69,Kih70,BF74,Aro76,Tol96,Dre97,GS01} and 
references therein). Moreover, it allows us to widely 
use a harmonic analysis on the
Poincar\'{e} group \cite{Rid66,Hai69,Hai71} 
(or, on the product $SU(2)\otimes SU(2)$)
at the study of relativistic amplitudes. Harmonic analysis on the
Poincar\'{e} group in terms of hyperspherical functions will be considered
in the following work.

\end{document}